\documentclass[fleqn,usenatbib]{mnras} 
\usepackage[T1]{fontenc}
\usepackage[normalem]{ulem}
\usepackage{xcolor}
\usepackage{ae,aecompl}

\DeclareRobustCommand{\VAN}[3]{#2}
\let\VANthebibliography\thebibliography
\def\thebibliography{\DeclareRobustCommand{\VAN}[3]{##3}\VANthebibliography}

\usepackage{graphicx}	% Including figure files
\usepackage{amsmath}	% Advanced maths commands
\usepackage{amssymb}	% Extra maths symbols
\usepackage{multicol}
\usepackage{caption}
\usepackage{subcaption}
\usepackage{orcidlink}
\usepackage{newtxtext,newtxmath}

\newcommand{\Msun}{M$_{\odot}$} 
\newcommand{\Rsun}{R$_{\odot}$}

\newcommand{\review}{}
\newcommand{\phant}{{\sc phantom}}

\newcommand{\astrobear}{{\sc astrobear}}
\newcommand{\arepo}{{\sc arepo}}
\newcommand{\mesa}{{\sc mesa}}

\title[Common envelope with a late phase AGB star donor]{Common envelope binary interaction simulations between a thermally-pulsating AGB star and a low mass companion}

\author[M. González-Bolívar et al.]{Miguel González-Bolívar \orcidlink{0000-0002-5939-9269 },$^{1,2}$ \thanks{E-mail: miguel-angel.gonzalez-boliv@hdr.mq.edu.au}
    Orsola De Marco\orcidlink{0000-0002-1126-869X},$^{1,2}$ 
    Mike Y. M. Lau\orcidlink{0000-0002-6592-2036},$^{3,4}$
    \newauthor
    Ryosuke Hirai\orcidlink{0000-0002-8032-8174}$^{3,4}$ and
	Daniel J. Price\orcidlink{0000-0002-4716-4235}$^{3}$
	\\
$^{1}$ Department of Physics and Astronomy, Macquarie University, Sydney, NSW 2109, Australia\\
$^{2}$ Astronomy, Astrophysics and Astrophotonics Research Centre, Macquarie University, Sydney, NSW 2109, Australia\\
$^{3}$ School of Physics and Astronomy, Monash University, Clayton, Victoria 3800, Australia\\
$^{4}$ OzGrav: The ARC Centre of Excellence for Gravitational Wave Discovery, Australia\\
}

\date{Accepted XXX. Received YYY; in original form ZZZ}

\pubyear{2022}

\begin{document}
\label{firstpage}
\pagerange{\pageref{firstpage}--\pageref{lastpage}}
\maketitle

% Abstract of the paper

\begin{abstract}
    At least one in five of all planetary nebulae are the product of a common envelope (CE) interaction, where the companion in-spirals into the envelope of an asymptotic giant branch (AGB) star ejecting the nebula and leaving behind a compact binary. In this work we carry out 3D smoothed particle hydrodynamics simulations of the CE interaction between a 1.7~\Msun\ AGB star and a 0.6~\Msun\ companion. \review{We model the AGB structure using a 1D stellar model taken at the seventh thermal pulse.} The interaction takes place when the giant is on the expanding phase of the seventh thermal pulse and has a radius of 250~\Rsun. The post-CE orbital separations varies between 20--31~\Rsun, with the inclusion of recombination energy resulting in wider separations. Based on the observed short in-spiral timescales, we suggest that thermal pulses can trigger CEs, extending the ability of AGB stars to capture companions into CEs, that would lead to the prediction of a larger population of post-AGB post-CE binaries. Simulations that include a tabulated equation of state unbind a great deal more gas, likely unbinding the entire envelope on short timescales. The shape of the CE after the in-spiral is more spherical for AGB than RGB stars, and even more so if recombination energy is included. We expect that the planetary nebula formed from this CE will have different features from those observed by \review{Zou} et al. 2020.
\end{abstract}

\begin{keywords}
hydrodynamics --- methods: numerical --- Stars: AGB --- binaries: close
\end{keywords}

%%%%%%%%%%%%%%%%%%%%%%%%%%%%%%%%%%%%%%%%%%%%%%%%%%

%%%%%%%%%%%%%%%%% BODY OF PAPER %%%%%%%%%%%%%%%%%%

\section{Introduction}

Common envelope (CE) interactions occur in binary stellar systems when one of the stars, whether because of radial expansion or changes in the binary orbit, engulfs the companion. This interaction is a critical step in the formation of compact binaries, such as cataclysmic variables, the progenitors of type Ia supernovae, at least one in five of all planetary nebulae (PN) and even of binaries that later become gravitational waves sources. Since such interactions have timescales of the order of months to years, the number of systems observed during a CE is low \citep{Ivanova2013,DeMarco2017}.

At least one in five PN have a post-CE central star binary \citep[e.g.,][]{Miszalski2009,Jacoby2021}. These close binary stars (typical periods $\lesssim$3 days) owe their short periods to a CE in-spiral following unstable mass transfer between an asymptotic giant branch (AGB) star and a companion (main sequence or white dwarf). The PN around these binaries tend to have axial symmetry or no apparent symmetry. PN in general tend to have short lifetimes in virtue of their expansion, which leads to visibility times of $\sim 50\,000$~years \citep{Schoenberner1983}. When the central star of a PN is a post-CE binary, this ensures that the PN is the ejected CE\footnote{Some of the PN material could technically have been ejected before the CE interaction. This would be the case in those instances when the AGB star has entered the short superwind phase just ahead of the CE in-spiral, something that would require some degree of fine tuning, but is not impossible.}. This in turn means \review{that the observed nebular morphology and dynamics, as well as the central stars' and binary parameters may provide observational constraints} on CE interaction energies and timescales. 
%\review{When the central star of a PN is a post-CE binary, this ensures that the PN itself is the result of that previous CE interaction, which provides an observational constraint on CE interaction energies and timescales.} 
Additionally, the post-CE central stars of PN provide an observational data point for the end state of a CE interaction.
 
Most CE simulations in the literature have used red giant branch (RGB), instead of AGB donors, because they exhibit greater 3D hydrostatic stability in the numerical domain. All giant stars are in an evolutionary phase with relatively short timescales. RGB and AGB stars expand on timescales of millions of years. Thermally-pulsating AGB stars, the last phase of AGB evolution, expand significantly on timescales of decades-to-centuries. These thermodynamically unstable objects tend to expand or contract on dynamical timescales once they are mapped on the 3D computational domain, because of typical numerical shortcuts,  \review{such as a lack of adequate resolution}, or the use of a simplistic equation of state. The problem is worse at the end of the AGB phase, where even 1D stellar structure and evolution codes find it increasingly difficult to find a hydrostatic solution for the envelope \citep{Ohlmann2017}.

%\ryo{Do you want to explain what ``stability'' means? Maybe add a paragraph before this one to briefly remind the reader how stars behave in the later stages (RGB$\rightarrow$AGB$\rightarrow$TP-AGB) and what a ``stable/unstable'' envelope is. I know you explain in more detail in the following sections, but I feel it is important to emphasize this in the Intro as well because it's the main focus of this paper.} 

There are only a handful of AGB primary CE simulations in the literature. \citet{Sandquist1998} and \citet{Staff2016} carried out simulations with relatively massive AGB primaries (3-5~\Msun). More recently, \citet{Chamandy2020} and \citet{Sand2020} have implemented CE simulations using low mass AGB stars as donors ($1.78$ and $0.97$ \Msun, respectively, evolved from 2.0 and 1.2~\Msun\ main sequence stars, respectively). These simulations were carried out using the grid code \astrobear\ \citep{Carroll-Nellenback2013} and moving mesh code \arepo\ \citep{Springel2010}.  \review{Donors} were in the non-thermally-pulsating AGB phase (the stars had radii of 122 and 173~\Msun, respectively), when the star is more dynamically stable as opposed to the later stage. Particularly, there is no previous work done using a thermally pulsating AGB star as the CE donor. 

In order to pursue the link between PN and CE interactions, \cite{GarciaSegura2018,GarciaSegura2020,GarciaSegura2021}, \citet{Frank2018} and \citet{Zou2020} carried out hydrodynamics simulations of the ``fast wind'' (the wind that the post-AGB star emits and that ploughs into the previously emitted, slow AGB wind) ploughing into gas ejected during a CE. The first group used a simulation by \citet{Ricker2012}, which, by virtue of its small computational domain, was halted after $56$ days. The PN simulation was in 2.5D ensuring that symmetry would be found. The latter two papers used instead a simulation by \citet{Reichardt2019}, which, having been carried out using a smoothed particle hydrodynamics (SPH) technique, could be followed for a time of 14.5 years after the start, resulting in more expanded ejecta. Their PN simulation was in 3D but covered a much shorter PN evolution time than the one by \citet{GarciaSegura2018}. These initial efforts, however, showed promise to connect PN morphologies with CE events.

Another possible shortcoming of the PN formation work of \citet{GarciaSegura2018} and subsequent papers as well as \citet{Zou2020} is that the CE simulations used to provide initial conditions for the simulation had not fully ejected their envelopes. Hence the velocity of the gas into which the super-wind ploughs was slower than is expected in a real PN formation situation, where the AGB has ejected its envelope. Today, we have simulations that can fully eject the envelope, by adopting a tabulated equation of state \review{including recombination energy that can be released into the envelope}. While these simulations may overestimate the dynamical effect of the recombination energy by assuming an adiabatic gas, they allow one to test its effect on the formation of the PN.

Finally \review{\cite{Iaconi2019b}} found that the observed post-RGB CE systems appear to be systematically more efficient in ejecting the envelope than the post-AGB objects, or, in other words, they have similar final separations despite having considerably more bound envelopes. At the same time they found that the lack of simulations with low mass AGB stars made it impossible to determine whether simulations retrieve this characteristic.  

In this paper we therefore carry out the first 3D hydrodynamical simulation of a CE interaction with a thermally-pulsating AGB star, carried out with an ideal gas equation of state as well as with a tabulated equation of state that includes the effects of radiation pressure and recombination. The initial mass, 2.0~\Msun, was chosen because stars with masses lower value than this tend to grow as large on the RGB as on the AGB, with the implication that most CE interactions would take place on the RGB.

The structure of the paper is as follows: we outline the evolution through the thermally pulsating AGB phase (TP-AGB) of the donor star in Section~\ref{sec2:TP-AGB}. In Section~\ref{sec3:setup}, we discuss the simulation setup: the star relaxation procedure is detailed in Section~\ref{sec3.1:star_relaxation}, the binding energy of the AGB stellar structures is discussed in Section~\ref{sec3.2:binding_energy} and the binary setup is discussed in Section~\ref{sec3.3:bin_setup}. The results are presented in Section~\ref{sec4:results}, starting with the orbital evolution in Section~\ref{sec4.1:orb}, an analysis of the unbound mass in Section~\ref{sec4.2:bound}, a discussion on the envelope rotation in Section~\ref{sec4.3:angmom}, an appraisal of the morphology of the common envelope in Section~\ref{sec4.5:morph} and an projection of the likely envelope dust content in Section~\ref{ssec:dust}. We finally present our discussion in Section~\ref{sec5:discussion} and summarise in Section~\ref{sec6:conclusions}.

\section{The thermally-pulsating AGB primary}
\label{sec2:TP-AGB}
\begin{figure*}
  \centering
  \includegraphics[width=0.45\linewidth]{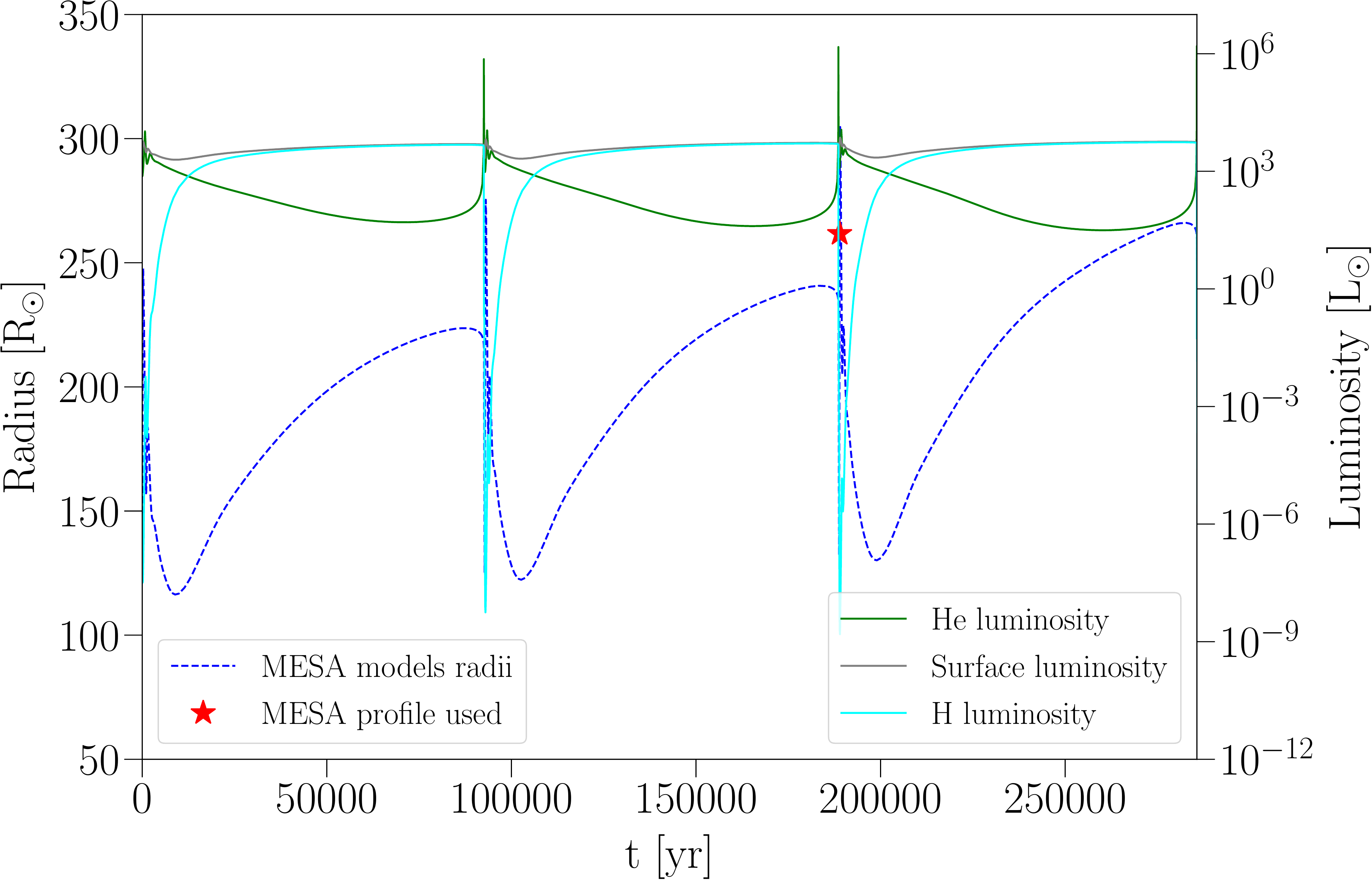}\ \ \ \ \   
    \includegraphics[width=0.45\linewidth]{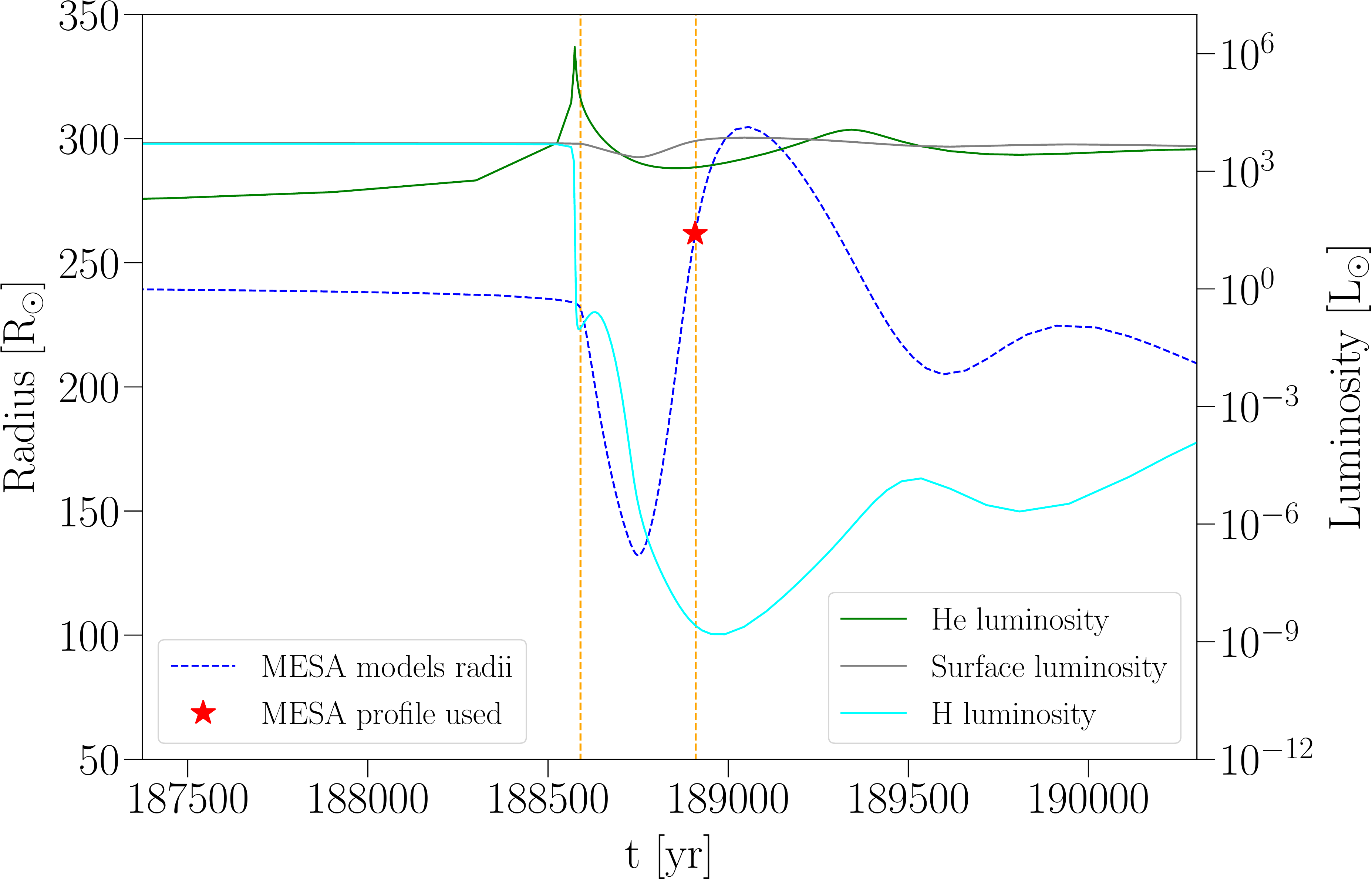}\vspace{0.2in}
    \includegraphics[width=0.45\linewidth]{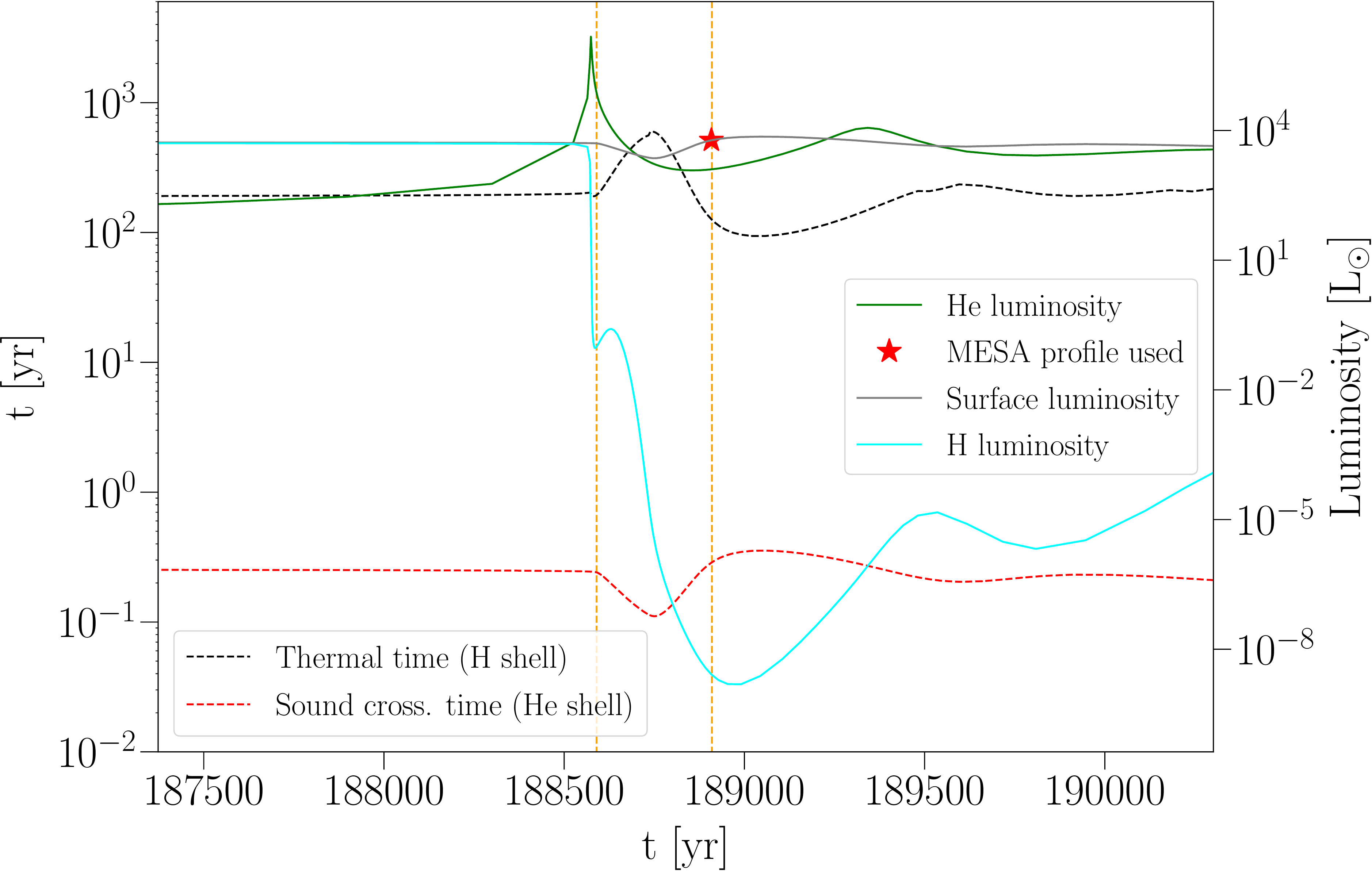}\ \ \ \ \
    \includegraphics[width=0.45\linewidth]{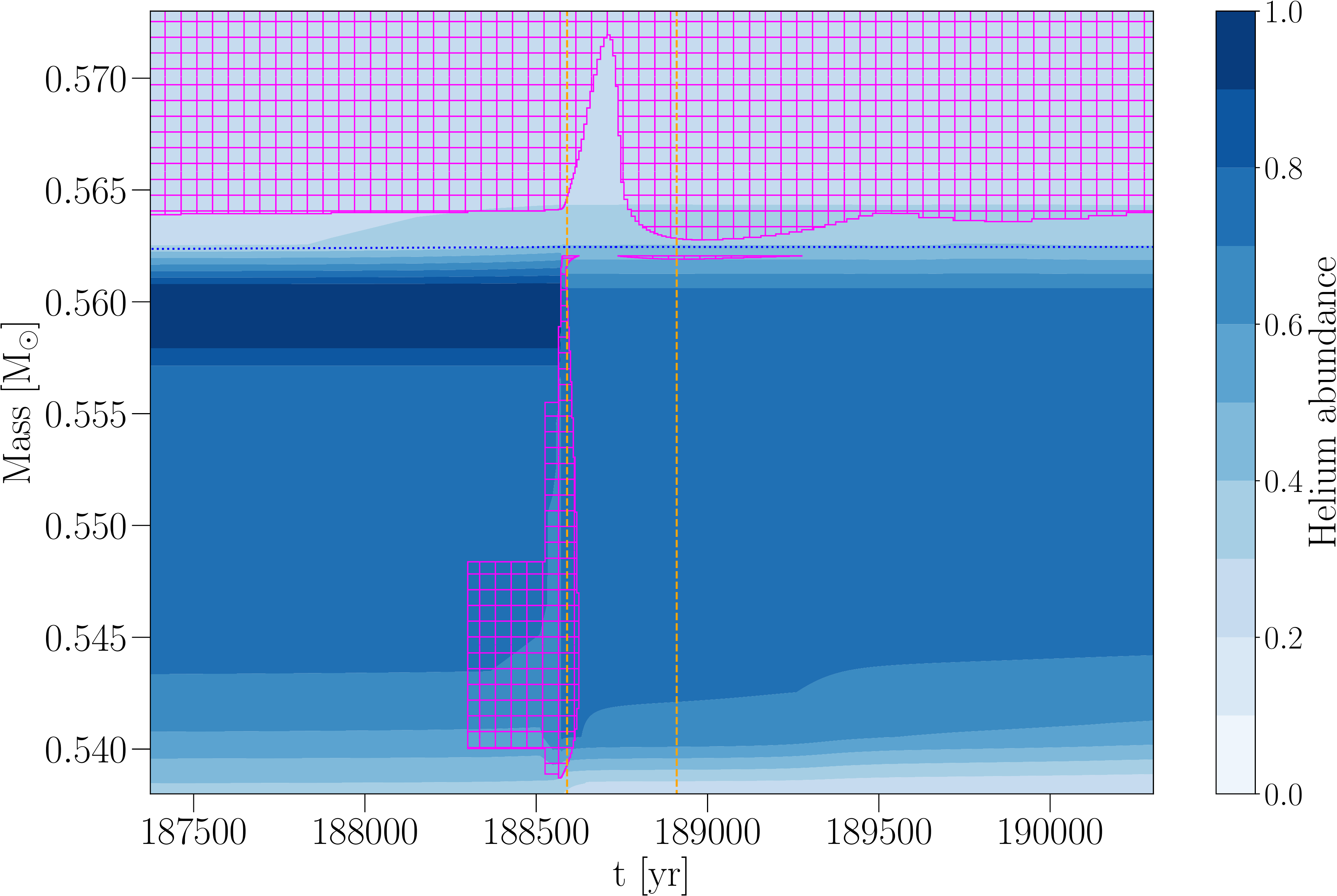}
  \caption{Evolution during the TP-AGB phase. Top left panel: evolution of the radius and luminosity. Top right panel: an enlargement of a thermal pulse; the orange vertical lines indicate the star contraction period right after the helium shell flash. Bottom left panel: evolution of the thermal and dynamical timescales compared to luminosity outputs. The red symbol represents the \mesa\ stellar profile used for the common envelope simulation at $R=260$~\Rsun. Bottom right panel: Kippenhahn diagram of the star during the thermal pulse. The contour represents the helium abundance clearly showing the boundary of the hydrogen and helium shells. The blue dotted line at $\sim 0.562$~\Msun\ is the helium core limit. The square magenta pattern indicates the convective regions. The time $t=0$ is set, arbitrarily, two pulses before the pulse we analyse.}
  \label{fig:TP_mom}
\end{figure*}

The AGB is one of the last evolutionary phases experienced by low-and intermediate-mass stars ($\lesssim$ 8~\Msun). It is divided into an early and late/thermally-pulsating phase. The early AGB phase is characterised by thick helium shell burning above a mostly carbon-oxygen core and a slow and steady radial expansion. Above this layer, there is a helium-rich shell, and above that, a hydrogen-burning shell \citep{Lattanzio2001}. After the helium shell thins out, the late-AGB phase starts, which is underpinned by a series of thermal pulses as the star alternates between explosive, runaway shell helium burning and thermally stable shell hydrogen burning. 

Figure~\ref{fig:TP_mom} illustrates the relevant aspects of stellar evolution through this phase, where we have used the Modules for Experiment in Stellar Astrophysics \citep[\mesa, version 12778; ][]{Paxton2011,Paxton2013,Paxton2015} to calculate the structure of an AGB star with a zero age main sequence mass of 2~\Msun. Here we see a cyclical ignition of the helium shell with concomitant variability of the hydrogen shell burning rate (green and cyan solid lines in Figure~\ref{fig:TP_mom}, top left-panel, respectively). Helium shell flashes repeat approximately every 100\,000 years. 

Thermal flashes are caused by the thermonuclear ignition of a thin layer of helium, at the base of the helium shell just above the CO degenerate core and below a large hydrogen-rich envelope. The instability that leads to the flashes is explained by a combination of the thin-shell instability and partial degeneracy \citep{Kippenhahn1990}. Flashes follow long, quiescent periods of hydrogen burning just above the helium shell. During these periods, hydrogen burning shell accretes onto the helium shell until the base of the shell ignites under partial degeneracy conditions.

These ignitions of the helium layer produce a thermal energy increase of the inner layers, which leads to its expansion and to a decreased burning rate. 
At this point there is relatively low energy input in any part of the star and the surface of the star begins to contract. Eventually, the energy radiated from the helium shell produces a radial expansion of the envelope, which reaches a larger radius compared to the hydrogen shell burning phase before the flash.

For stars with main sequence masses of 2~\Msun, both hydrogen and helium burning layers are at $R<0.1$~\Rsun\ by the time the TP-AGB starts. We define those radii as the (radial) cell with the highest burning rate of hydrogen or helium.

At the time of the flash, the stellar radius is 234~\Rsun. After a period of contraction, the radius returns to its initial value (900 years after the flash peak), expands to 304~\Rsun\ (1050 years after the flash peak) and eventually returns to its initial value 37\,000 years after the peak of the flash.
The interval between the helium flash and the maximum stellar radial expansion is $\sim400$ years. This lag is of the order of the thermal timescale of the layers above the helium shell (bottom plot, black dashed line in Figure \ref{fig:TP_mom}, calculated from the 1D \mesa\ model as the time for the layers above the helium burning shell to radiate all their thermal energy), and this time is about two orders of magnitude slower than the dynamical timescale of these regions.

The upper layers ($R \geq 0.2$~\Rsun) are convective during the TP-AGB phase. Figure \ref{fig:TP_mom}, bottom-right panel, shows the Kippenhahn diagram during the same time span as the previous plot. The orange vertical lines are the same as for the top-right panel and are plotted here for reference. The non-burning hydrogen envelope is convective, while convection extends all the way to the core just preceding the maximum helium shell luminosity output, during the early part of the thermonuclear runaway of the helium shell. This is the convective event responsible for the third dredge-up. 
This combination of events creates a unique situation for the star. The thermally-pulsating star is a mostly convective, rapidly expanding star. Thermal pulses expand the star into volumes never reached before creating new opportunity for interaction. Accounting for tidal interactions between the star and its companion shows that the thermal pulses are able to progressively draw in the companion such that even if one thermal pulse expansion does not result in a CE, the following pulse can generate a CE even before the star expands past the maximum radius of the previous pulse \citep[figure 5]{Madappatt2016}. The next question is whether the short duration of the pulse radial expansion is such that a CE would be stymied by the contraction of the star post-pulse. 

\section{Simulation setup}
\label{sec3:setup}

\begin{figure}
    \centering
    \includegraphics[width=\linewidth]{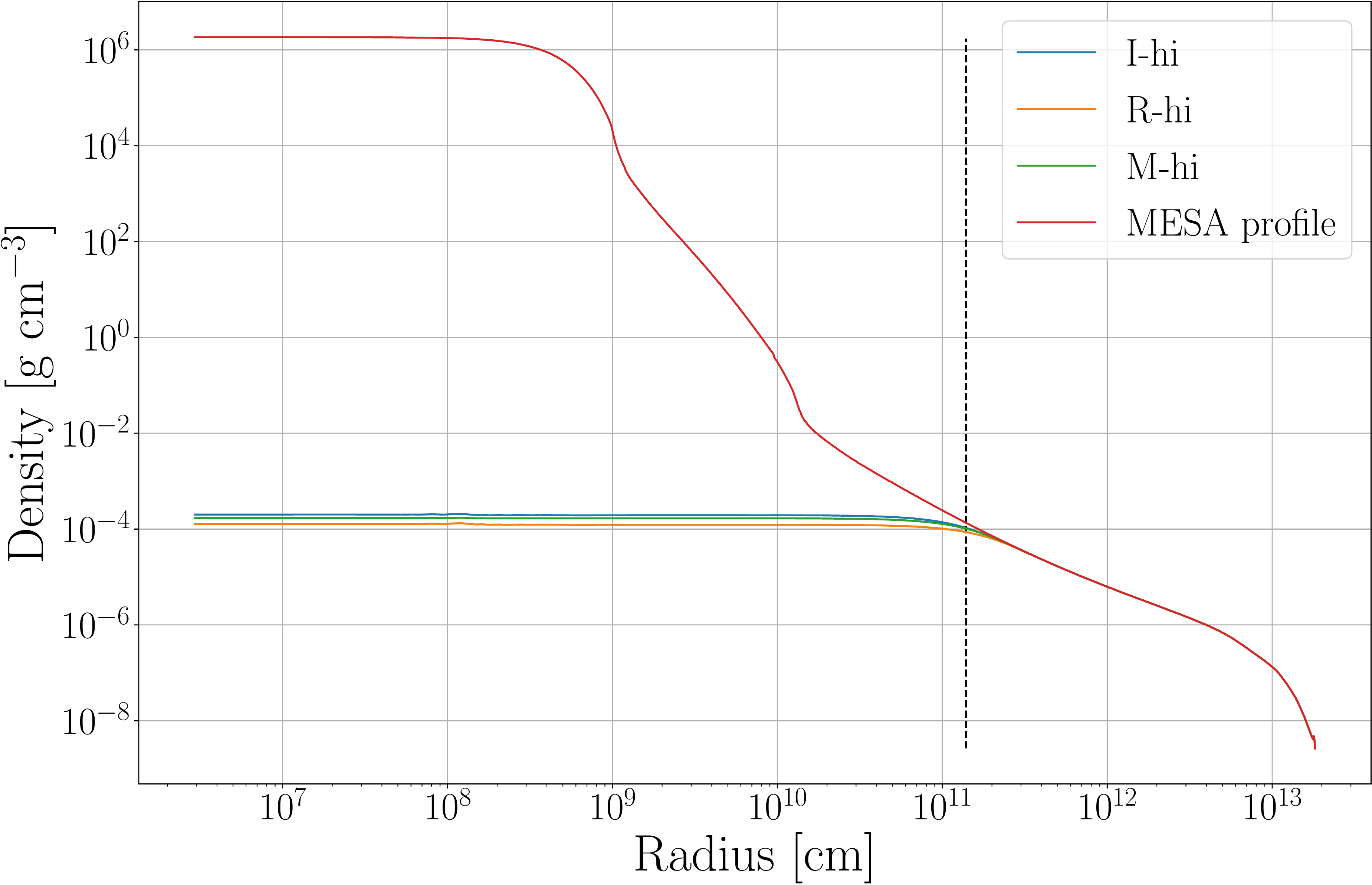}
    \caption{The \mesa\ and softened density profiles of the four models used. The black dashed vertical line is at $R=2$~\Rsun, which is the value set for $r_{\rm soft}$ in all the models except for R-hi. The density and pressure profiles outside the softening length are identical for all three softened models.  I-hi, R-hi and M-hi are the stellar profile mapped in the 3D computational domain using the ideal gas, ideal gas plus radiation and tabulated equations of state, respectively.}
    \label{fig:rho_soft_profile}

\end{figure}

In this section we describe the initial conditions for our 3D CE simulation, starting with the single star 3D mapping method, the relaxation procedure to obtain a stable model of the TP-AGB star and the subsequent binary star setup. We use the \mesa\ model of the TP-AGB star discussed in Section~\ref{sec2:TP-AGB}. The zero age main sequence mass is 2~\Msun\ with solar metallicity ($Y=0.28$; $Z=0.02$). The stellar wind parameter implemented had the code default values: cool wind RGB scheme was set to ``Reimers'', with a Reimers scaling factor of $0.1$; the cool wind AGB scheme was ``Blocker'' with a scaling factor of 0.5 and the RGB to AGB wind switch was set at $10^{-4}$. We let the model evolve until it reaches the seventh thermal pulse expansion. At this point, the stellar radius is 260~\Rsun, the carbon-oxygen core mass is 0.45~\Msun\ and the total stellar mass is 1.7~\Msun. The peak radius of 304~\Rsun\ would be reached $150$ years later (Figure~\ref{fig:TP_mom}).
We took the 1D stellar profile and mapped it into the 3D SPH code~\phant\ \citep{Price2018} using the procedure explained in Section~\ref{sec3.1:star_relaxation}.
We then proceeded to create four different, stable 3D setups of the same \mesa\ model: two with an ideal gas equation of state at low and high resolutions of $1.37 \times10^5$ and $1.37\times10^6$ SPH particles, respectively. The other two models used an equation of state including ideal gas plus radiation pressure and a tabulated equation of state implemented by \citet{Reichardt2020} and taken from \mesa, both with the same number of particles as the high resolution simulation for the ideal gas case. Both equations of state assume local thermodynamic equilibrium, which is a good approximation in the optically thick regions of the star.

\subsection{Core excision and star relaxation procedure}
\label{sec3.1:star_relaxation}
\begin{table*}
\centering
\begin{tabular}{ccccccccccc}
\hline

Model & Equation & $N_{\rm part}$ & Relaxation &  $r_{1, \rm soft}$ & $r_{2, \rm soft}$ & $M_{\rm core}$ &  $t_{\rm model}$ & $E_{\rm bin}$ & Artificial\\
      & of state &                &  technique &  (\Rsun) & (\Rsun) & (\Msun) &  ($t_{\rm dyn}$) & ($10^{46}$ erg) & conductivity\\
\hline
I-lo & ideal gas & $1.37 \times 10^5$ & {\sc damp} &  2 & 2 & 0.5633 &  5.2 & $3.12$ & $1$\\
I-hi & ideal gas & $1.37 \times 10^6$ & {\sc damp} &  2 & 2 & 0.5634 &  33.9 &$3.11$ & $1$\\
R-hi & ideal + rad & $1.37 \times 10^6$ & {\sc relax-o-matic} &  3 & 3 & 0.5642 &  17.3 &$2.92$ & $0.8$\\
M-hi & tabulated   & $1.37 \times 10^6$ & {\sc relax-o-matic} &  2.5 & 2.5 & 0.5639 &  5.2 & $-0.11$ & $0.1$\\
\hline
\multicolumn{6}{l}{$^\dagger$ For our TP-AGB star, $t_{\rm dyn} \approx 42$ days}
\end{tabular}
\caption{Parameters of the stellar models used as starting conditions for our common envelope simulations. The eighth column, $t_{\rm model}$ is the number of dynamical times that the stellar structure was evolved in isolation, before implementing it in the binary setup. These times correspond to the colored squares in Figure~\ref{fig:PD_rad_fracc}.  }
\label{tab1:table}
\end{table*}

A procedure to generate a reasonable stellar structure that is in equilibrium over sufficiently long timescales is an integral and important component of all hydrodynamics simulations of stars. Typically stars generated by 1D implicit codes have high resolution and sophisticated equations of state and are not stable once they are mapped in the 3D computational domain at lower resolution, a different equation of state, grid mismatch and less physics included, such as, for example, the lack of radiation transport. 

Most approaches followed today are similar to those in the early simulations of \citet[][who used an SPH approach]{Rasio1996} and \citet[][who used a grid-based approach]{Sandquist1998}, who replaced the cores of their giant stars with a point mass particle, because the extremely high central densities are orders of magnitude larger than near the surface, resulting in the Courant–Friedrichs–Lewy condition near the centre imposing impractically short time steps. In order to include a central point mass particles, these authors implemented a smoothed gravitational potential over a certain distance from the center. 
This technique solves the time stepping problem in the core region and similar solutions have been adopted by stellar numerical simulations ever since. They also added a velocity-dependent friction term \citep[e.g.][]{Gingold1977} for the gas applied over a certain time, that aims to dampen abnormal speeds due to the mapping process or lack of hydrostatic equilibrium. Running the code with this damping term for a few dynamical timescales relaxes stellar profiles mapped from hydrostatic equilibrium solutions. 

Variations on this technique were used before in CE simulations with the \phant\ code \citep{Iaconi2018,Reichardt2019,Reichardt2020}, when the primary was an RGB star. 
In this work we have used two different techniques to excise the core and relax the stellar profile. The first is a slightly modified version of the technique presented by \citet{Reichardt2019}, which was very similar to what was devised by \citet{Ohlmann2017}. 
The second procedure was introduced by \cite{Lau2022} in order to accurately reproduce the density profile in their work with a more massive star and which resulted to be superior in the stability of our TP-AGB star with an equation of state other than ideal gas. Below we give an account of these procedures. All relevant parameters are listed in Table \ref{tab1:table}. 
The first prescription was used to create an input stellar model that adopted an ideal gas equation of state (in Table~\ref{tab1:table} we nickname this first procedure {\sc damp}). We start with the selection of a softening radius, $r_{\rm 1,soft}$, for the potential of the point mass particle that replaces the core. The potential is softened within a volume that has twice this radius. The \mesa\ density profile is modified inside $2 \times r_{\rm 1,soft}$, such that the density is lower and almost constant but it joins smoothly with the original \mesa\ profile at $2\times r_{\rm 1,soft}$.

The value of $r_{\rm 1,soft}$ needs to be smaller than the final core-companion orbit, but large enough that the time step in the simulation is not impractically short. Once a value of $r_{\rm 1,soft}$ is selected, \phant\ then calculates the mass for the core point mass particle ($M_{\rm core}$ in Table~\ref{tab1:table}). After the softened profile is constructed, a geometric distribution of particles,
with an approximate density similar to that of the softened profile, is generated (see Figure \ref{fig:rho_soft_profile}, where the three profiles mapped using three equations of state are displayed alongside the original 1D \mesa\ profile). 
After finalising the mapping process into the \phant\ computational domain, we evolve the 3D star with the velocity-dependent damping factor implemented by \cite{Reichardt2019} for 5 dynamical times, $5t_{\rm dyn}$. \review{We calculate $t_{\rm dyn}$ as:}

\begin{equation}
    t_{\rm dyn} = \frac{R}{v_{esc}} = \sqrt{\frac{R^3}{2GM}},
\end{equation}

\noindent \review{where G is the gravitational constant and R and M are the photosphere radius and mass of the original \mesa\ stellar profile, respectively.} We then let the simulation evolve for an additional 5 dynamical times with no velocity damping to assess the numerical stability of the model. In Appendix~\ref{app:stellar_relaxation} we give further details of this process and show the particle distributions and density cross sections for the stellar models taken at all stages in the relaxation process just described.

\begin{figure*}
    \centering
    \includegraphics[width=0.45\linewidth]{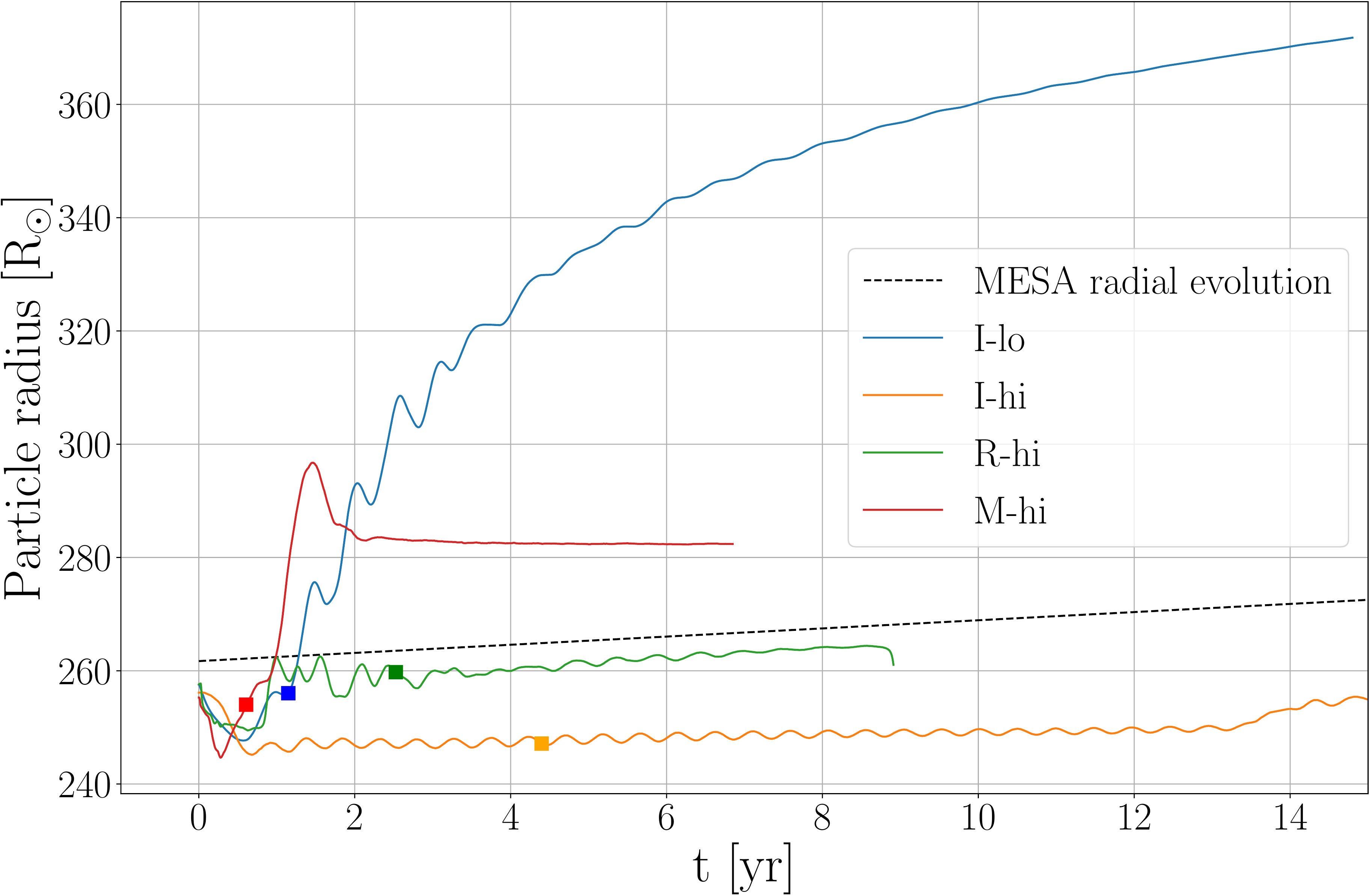}\ \ \ \ \ \ 
    \includegraphics[width=0.45\linewidth]{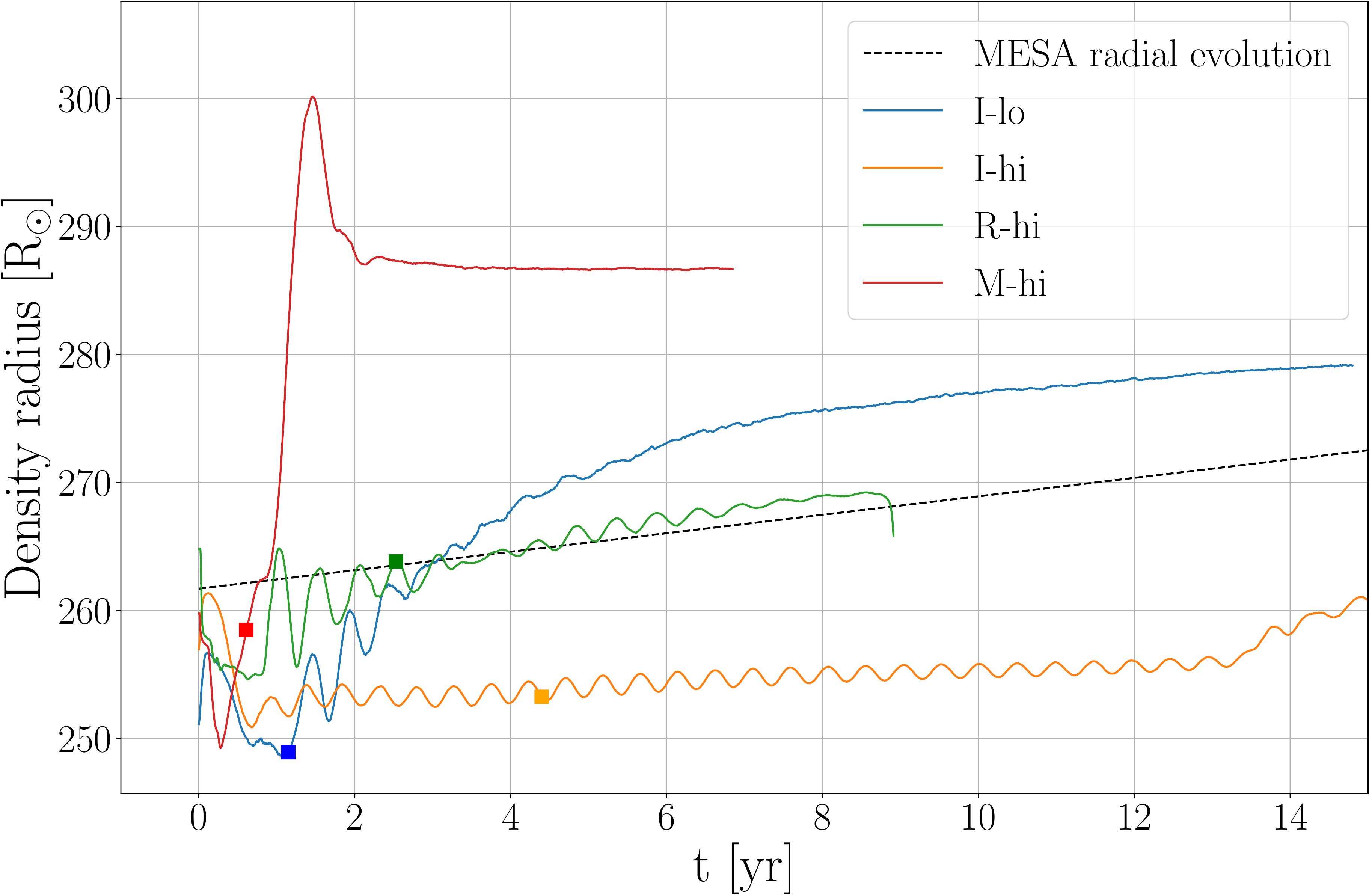}\\
    \caption{Radial evolution of the implemented models during and after relaxation. The ``particle radius'' is shown in the left panel and the ``density radius'' evolution is showing in the right panel. The square symbol represents the profile we have used in the binary simulations. The black dashed line is the radial evolution of the 1D, \mesa\ profiles during the thermal pulse. 
    %\mike{The blue line, the only low-resolution run, looks like it has very poor stability in the left plot. I suggest deleting it and show results at high resolution.} \orsola{I want to leave this. We can use to explain how expansion per se does not lead to an overly fast in-spiral.}
    }
    \label{fig:PD_rad_fracc}
\end{figure*}

The second relaxation technique used in this paper (called {\sc relax-o-matic} in Table~\ref{tab1:table}) similarly follows from mapping the 1D profile and creating a density distribution inside the core region based on the selected softening length. This time, however, we set the density and pressure within $r<r_\mathrm{1,soft}$, such that the entropy is constant. In the case of an ideal gas the specific entropy in the central region is given by:

\begin{equation}
    s = \frac{k_B}{m_h}\frac{1}{\mu} \ln\left( \frac{T_{\rm soft}^{3/2}}{\rho_{\rm soft}}\right),
    \end{equation}

\noindent whereas for ideal gas with radiation pressure and for the tabulated equation of state we use:

\begin{equation}
    s = \frac{k_B}{m_h}\frac{1}{\mu} \ln\left( \frac{T_{\rm soft}^{3/2}}{\rho_{\rm soft}}\right) + \frac{4a_{\rm rad}}{3}\frac{T_{\rm soft}^3}{\rho_{\rm soft}},
\end{equation}

\noindent where $k_B$ is the Boltzmann constant, $a_{\rm rad}$ is the radiation constant, $m_h$ the hydrogen atomic mass, and $T_{\rm soft}$ and $\rho_{\rm soft}$ are the temperature and density values at $r=2\times r_{\rm soft}$, respectively (we only used the constant entropy prescription for R-hi and M-hi; see \citealt{Hirai2020}, appendix B for details of this procedure). At this point, a random distribution of SPH particles is generated. Particles are generated in pairs such that their positions are symmetric with respect to the origin of the domain. This initially removes any net gravitational force on the core. Instead of using velocity damping, we apply an artificial particle shifting method where the thermal energy of the particles is updated after each iteration so that it follows the given $P(m)/\rho(m)$ profile (for a more detailed explanation see appendix C of \citealt{Lau2022}). 
The relaxation procedure iterates until the particle distribution resembles the original softened density profile within a certain tolerance,
the total kinetic energy falls below $10^{-7}$ of the
potential energy, or after 1000 iterations - for our case, after 1000 iterations $E_{\rm kin}/E_{\rm pot}\sim 2 \times 10^{-7}$. Once again, after the relaxation procedure was complete, we evolved the star in isolation for a few dynamical times to test the model's hydrostatic stability. We used the {\sc relax-o-matic} technique to relax the stellar structures with radiation pressure and a tabulated equation of state. 
We adopted default artificial conductivity parameter values \citep[see section 2.2.8 of][]{Price2018} of unity for the ideal gas models (I-hi and I-lo), 0.8 for R-hi and 0.1 for M-hi (see Table~\ref{tab1:table}). These values are the same for the relaxation phase and for the CE simulations. The default value is set to ensure accurate treatment of contact discontinuities \citep{Price2008}. The conductivity is second order in {\phant}, with the effective thermal conduction $\kappa_{\rm cond} \propto \alpha_u h^2 (\nabla\cdot{\bf v})$, where $h$ is the resolution length \citep[c.f.][]{Price2018}. It thus vanishes when velocity \review{gradients} are zero and the resolution is high, but we found that the remaining heat transfer could nevertheless lead to a slight undesirable expansion of the star over the timescales of interest. The simple solution was to lower the pre-factor for the R-hi and M-hi models to $\alpha_u = 0.8$ and $0.1$, respectively.

The fidelity of the stellar structures was further assessed by inspecting the density distributions as well as by assessing the individual velocities of the SPH particles. This analysis leads to the conclusion that the stellar structures are sufficiently in hydrostatic equilibrium to proceed with the binary interaction simulations. This numerical stability analysis is fully presented in Appendix~\ref{app:stellar_relaxation}.

In Figure~\ref{fig:PD_rad_fracc}, we plot the stellar radius over time. Time zero is the time when we start running the simulation of the {\it single} star to check its degree of stability. The square symbols indicate the times used to set up the binary simulations. As can be seen in Figure~\ref{fig:PD_rad_fracc}, we run the single star simulations for longer to determine the stability of the star over a time that is almost as long as the entire CE simulation. 
In Figure~\ref{fig:PD_rad_fracc}, ``particle radius'' is the average distance of the 0.5~per cent of all particles that are farthest from the center of mass, whereas the ``density radius'' is defined as the volume-equivalent radius \citep{Nandez2014} of all the SPH particles whose density is higher than the least dense SPH particle at $t=0$ ($2.29 \times 10^{-9}$, $1.81 \times 10^{-9}$, $3.16 \times 10^{-9}$, and $3.16\times 10^{-9}$~g~cm$^{-3}$ for the I-lo, I-hi, R-hi and M-hi, respectively). 

For the I-lo model, the stellar ``particle radius'' is still growing by 14 years, while the ``density radius'' seems to be asymptotically reaching a value of $\sim$280~\Rsun.  On the other hand, I-hi remains reasonably constant in both radius definitions for the entire time. The R-hi model's radius is gently growing over the 5 years for which the single model was run, while the M-hi model rapidly expands but then just as rapidly decreases and stabilises thereafter at a value of $\sim$285~\Rsun, depending on the radius definition. 

In Figure~\ref{fig:PD_rad_fracc} we also plot the natural radial growth of the 1D {\sc mesa} model which, being at the top of a thermal pulse is still growing on relatively short timescales. The radial growth of our single star models is due to residual motion that we have not been able to eliminate. Later on we discuss the slow expansion of our numerical stellar models in relation to the expected expansion of the actual TP-AGB star during the same time as the CE interaction.

\subsection{The binding energy of the AGB star}
\label{sec3.2:binding_energy}

\review{To calculate the binding energy of our models we perform a numerical integration using two terms: the gravitational potential energy, which is the same for all the models, and the internal energy, which depends on the equation of state implemented. This second term represents one of the sources of energy that will be used to accelerate the gas during the common envelope simulation.}
%\ryo{"gravitational potential energy" is a local quantity but what you are computing is the "binding energy". Rephrase the first sentence. For the second sentence, consider briefly explaining "why" you include internal energy. We include it because it could later become a useful source of energy to accelerate the gas.}

\begin{figure}
    \centering
    \includegraphics[width=\linewidth]{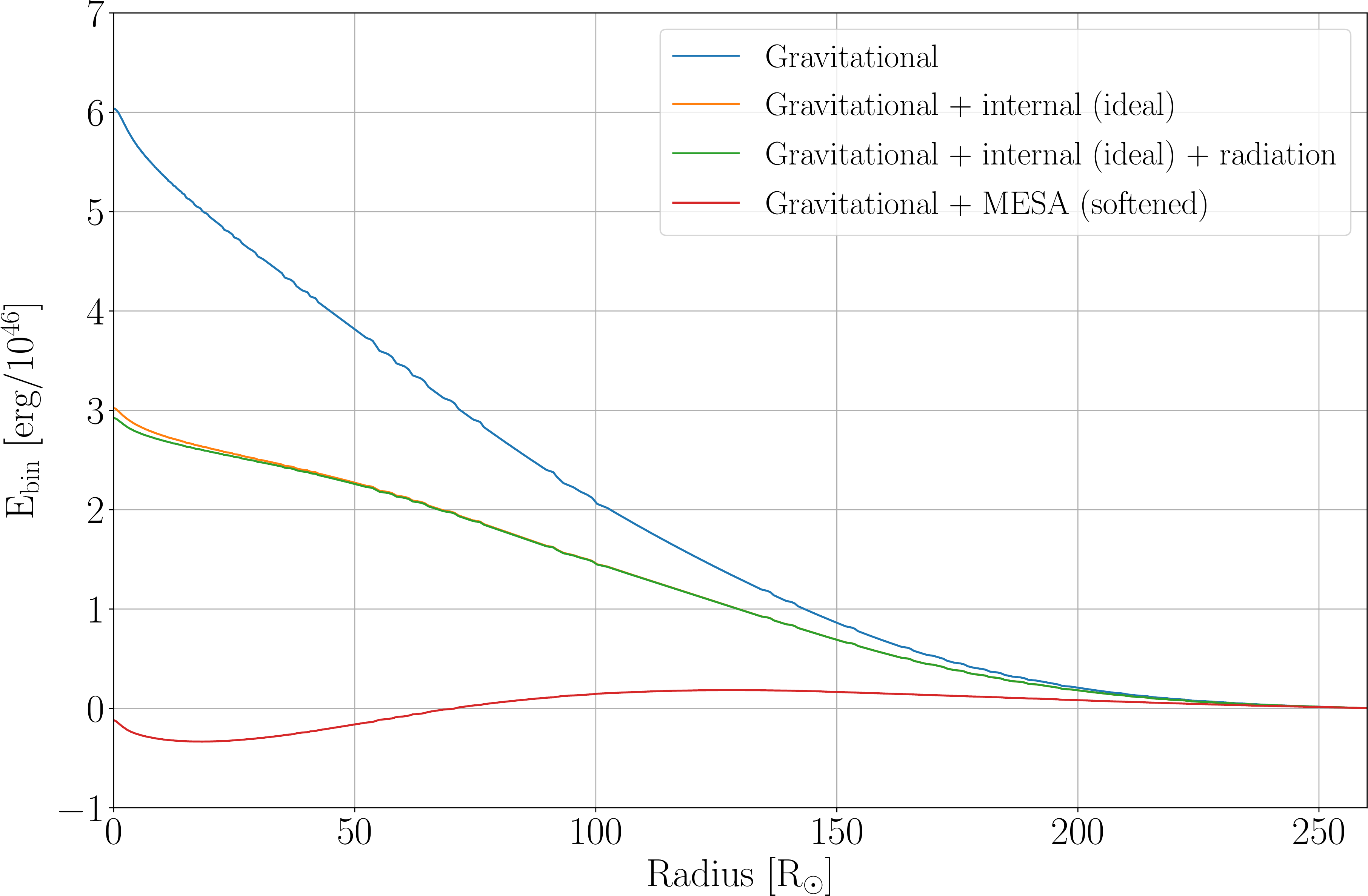}
    \caption{The cumulative binding energy for the TP-AGB star. The blue curve includes only the gravitational potential energy (second term in the integrand in Equation \ref{eq:ebin}). Orange, green and red lines include the $u(r)$ term in Equation \ref{eq:ebin} which is the internal energy for an ideal gas (equivalent to the thermal energy), ideal gas with radiation pressure (equivalent to thermal energy plus radiation energy) and the \mesa\ equations of state (equivalent to thermal, radiation and recombination energies), respectively. All of the quantities correspond to the softened profile generated during the core excision of the original \mesa profile.}
    \label{fig:ebin}
\end{figure}

In Figure~\ref{fig:ebin}, we plot the the cumulative binding energy as a function of radius $r$, defined as:

\begin{equation}
    E_{\rm bind} (r) = \int_M^{m(r)} \left[u(r') - \frac{Gm^\prime}{r'}\right] dm^\prime
    \label{eq:ebin}
\end{equation}

\noindent where M is the total stellar mass, $G$ is the gravitational constant and $m^\prime$ is the mass coordinate. The integration is carried out from the surface inward. 
The internal energy, $u(r)$, depends on the equation of state. As expected, when we only account for the gravitational energy, the integration has the highest binding energy. Adding the internal energy corresponding to an ideal gas with the density of the stellar profile (green and orange curves in Figure \ref{fig:ebin}), the full integration gives a total binding energy equivalent to half of the  gravitational binding energy, satisfying the Virial theorem. This result also applies when adding the radiation pressure, since this does not contribute significantly for low mass stars. Lastly, calculating $u(r)$ with the tabulated equation of state, the contribution of the latent energy of ionised hydrogen and helium decreases the binding energy in the inner layer to such level that the central region ($<75$ \Rsun) has a negative integrated binding energy value. 

\begin{table}
\centering
\begin{tabular}{cccc}
\hline

Model & $E_{\rm bin}$ (\phant) & Equivalent radius & Equivalent mass \\
& ($10^{46}$ erg)  & (\Rsun) & (\Msun)\\
\hline
I-lo & 3.118  &0.084& 0.5603 \\
I-hi & 3.105  &0.084& 0.5603 \\
R-hi & 2.918  &0.086& 0.5604 \\
M-hi & -0.1062  &2.004& 0.5645 \\

\hline
\end{tabular}
\caption{Envelope binding energy of the implemented models. The second column is the binding energy calculated using Eq.~\ref{eq:ebin} and the \phant\ profiles. 
The third column is the radius coordinate in the \mesa\ profile that has the same binding energy value. The fourth column is the mass inside that radial location. The binding energy, including only gravitational potential energy of the model, is $6.036\times 10^{46}$~erg.}
 \label{tab2:table}
\end{table}

This analysis works as a sanity check: in Table~\ref{tab2:table} we see that the mass inside the location with the binding energy calculated using the \phant\ file (``Equivalent mass'', in Table~\ref{tab2:table}), is less than 1~per cent different from the value of the degenerate core mass in the \mesa\ profile ($M_{\rm core}$ in Table~\ref{tab1:table}) or the value of the point mass we used.

\subsection{Binary setup}
\label{sec3.3:bin_setup}

Each of our four stellar models, the two models with ideal gas equation of state at low and high resolutions (I-lo and I-hi), the high resolution model with radiation pressure added (R-hi) and the model with a tabulated equation of state (M-hi) are combined with a companion with a mass of 0.6~\Msun\ at a separation of 550~\Rsun. For simplicity, we set the softening length of the companion, $r_{\rm 2, soft}$, to be the same as that of the primary's core. The initial separation is such that the donor with a radius of {260~\Rsun} fills its Roche lobe, which was calculated using the formula of \cite{Eggleton1983}. The primary and companion are set to orbit one another with zero eccentricity and Keplerian velocities of $35.3$ km~s$^{-1}$. The initial period is 2.7 years. None of our simulations were relaxed in the potential of the binary. For a discussion of the differences of such a setup see \citet{Reichardt2019}.

The primary is not rotating, and we do not relax the primary in the potential of the companion. When the companion is introduced in the computational domain the primary suffers a ``plucking'' effect, where a sudden tide is induced that sets off oscillations. This results in a slight additional expansion of the star. 

Table~\ref{tab3:table} shows the binary setup parameters for the four models and for the simulation conducted by \citet[][hereafter Rei19]{Reichardt2019} that we use as reference. 

\begin{table*}
\begin{tabular}{ccccccccccccccc}
\hline
Model & $q$ & $a_{\rm i}$ & $a_{\rm max,pl}$ & $a_{\rm f,pl}$ & $a_{\rm f}$ & $a_{\rm f,fit}$  & $t_{\rm max,pl}$   &$t_{\rm f, pl}$& $t_{\rm f}$ & $e_{\rm f,pl}$ & $M_{\rm b,PN}$ & $M_{\rm b,PN}$ \\
&  & (\Rsun) & (\Rsun) &(\Rsun)   &(\Rsun)  & (\Rsun) & (yr) & (yr)  & (yr)&& (\Msun) & (\%)\\
\hline
I-lo & $0.35$ & $550$ & $166$ & $54$ & $25.9$ & $25.0$ -- $27.4$   & $13.3$ & $14.2$ & $17.7$ & $0.06$ & $0.97$ & 85\\
I-hi & $0.35$ & $550$ & $161$ & $32$ & $20.6$ & $19.1$ -- $19.4$   & $15.3$ & $16.8$ & $27.4$ & $0.03$ & $0.96$ & 84\\
R-hi & $0.35$ & $550$ & $162$ & $44$ & $25.8$ & $23.3$ -- $24.8$   & $7.5$ & $8.7$  & $15.2$  & $0.09$ & $0.97$ & 85\\
M-hi & $0.35$ & $550$ & $250$ & $35$ & $32.7$ & $31.8$ -- $32.2$   & $4.6$  & $8.7$  & $9.8$  & $0.005$& $0.69$ & 60\\
Rei19 & $0.68$ & $218$ & $61$ & $23$ & $18.5$ & $17.9$ -- $18.6$   & $13.3$ & $14.2$ & $20.1$ & $0.01$ & $0.46$ & 94\\
\hline

\end{tabular}
\caption{Simulation parameters for our simulations and for the simulation of \citet{Reichardt2019}. Columns are as follows: $q$ is the mass ratio $M_2/M_1$; $a_1$, $a_{\rm max, pl}$, $a_{f, pl}$ and $a_f$ are the separations at the start of the simulation, at the time of fastest in-spiral, at the end of the in-spiral, as defined in the text, and at the end of the simulation. \review{Column $a_{\rm f,fit}$ contains the range of the extrapolated final separations (periastron and apastron separations).}  Columns $t_{\rm max, pl}$, $t_{f, pl}$ and $t_f$, are the corresponding times. The last two columns list the bound mass at the time of the morphology comparison described in Subsection \ref{sec4.5:morph}.}
 \label{tab3:table}

\end{table*}

\section{Simulation results}
\label{sec4:results}

\begin{figure*}
    \centering
    \includegraphics[width=0.299\linewidth]{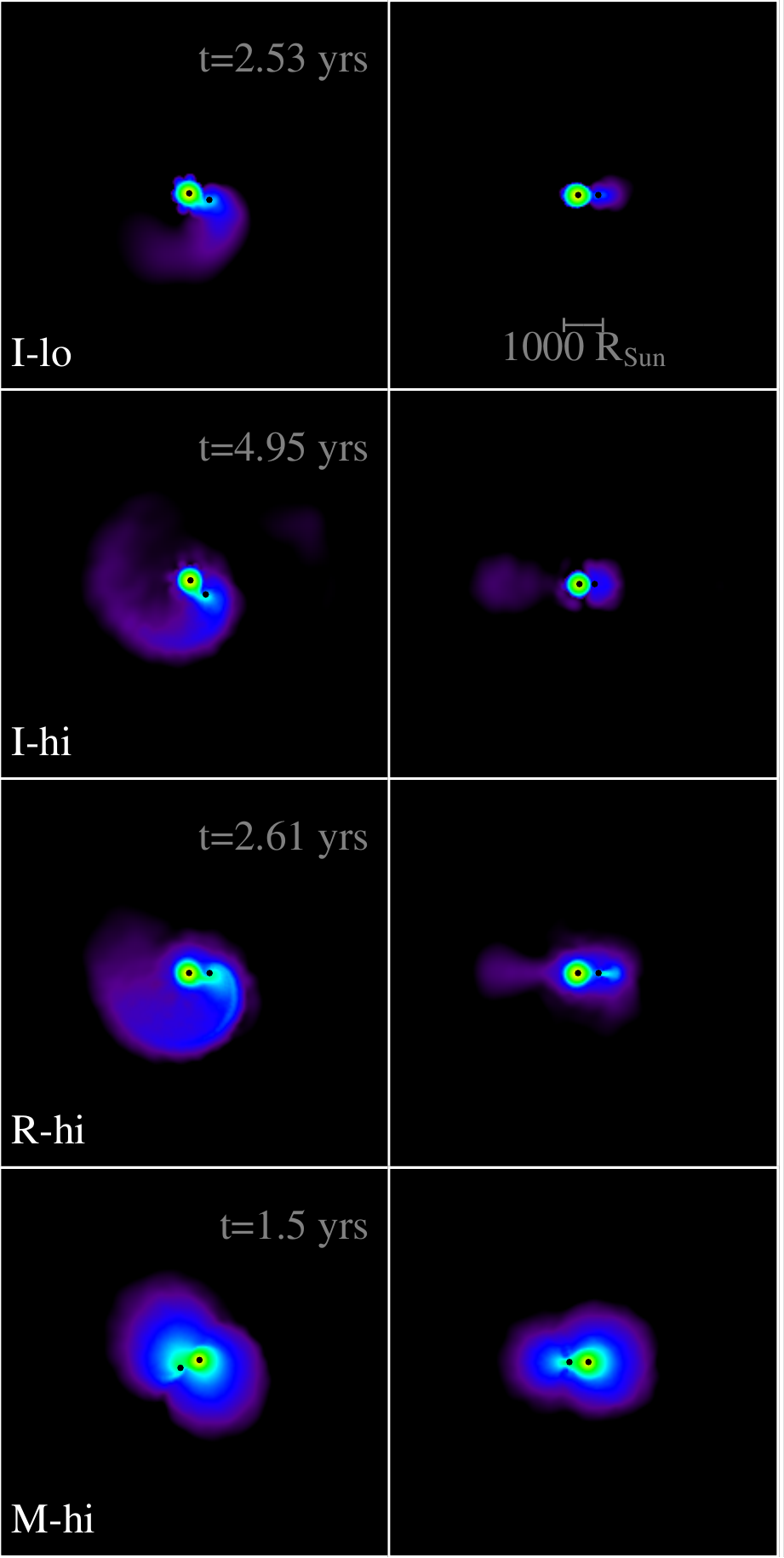}\
    \includegraphics[width=0.299\linewidth]{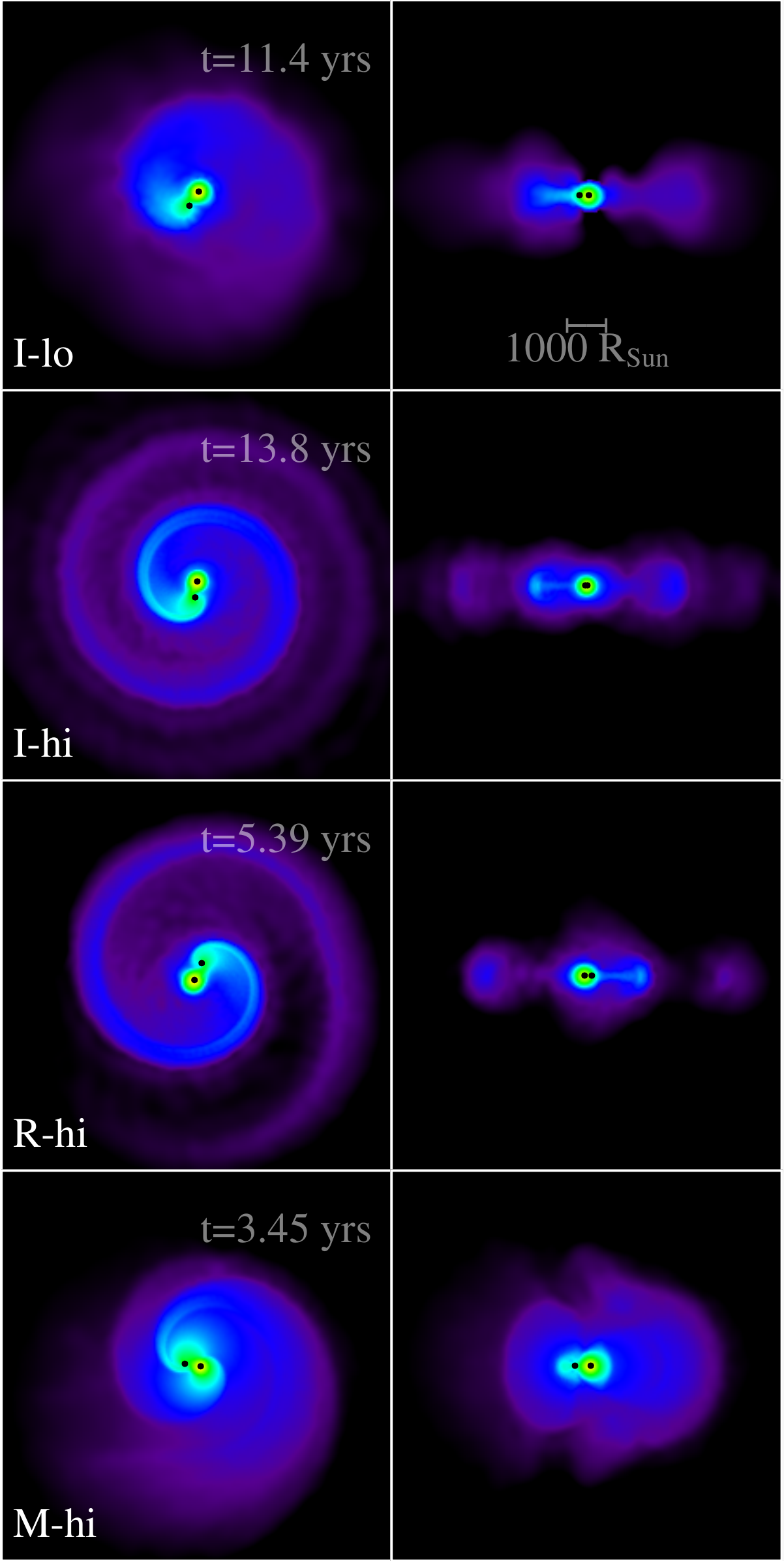}\
    \includegraphics[width=0.39\linewidth]{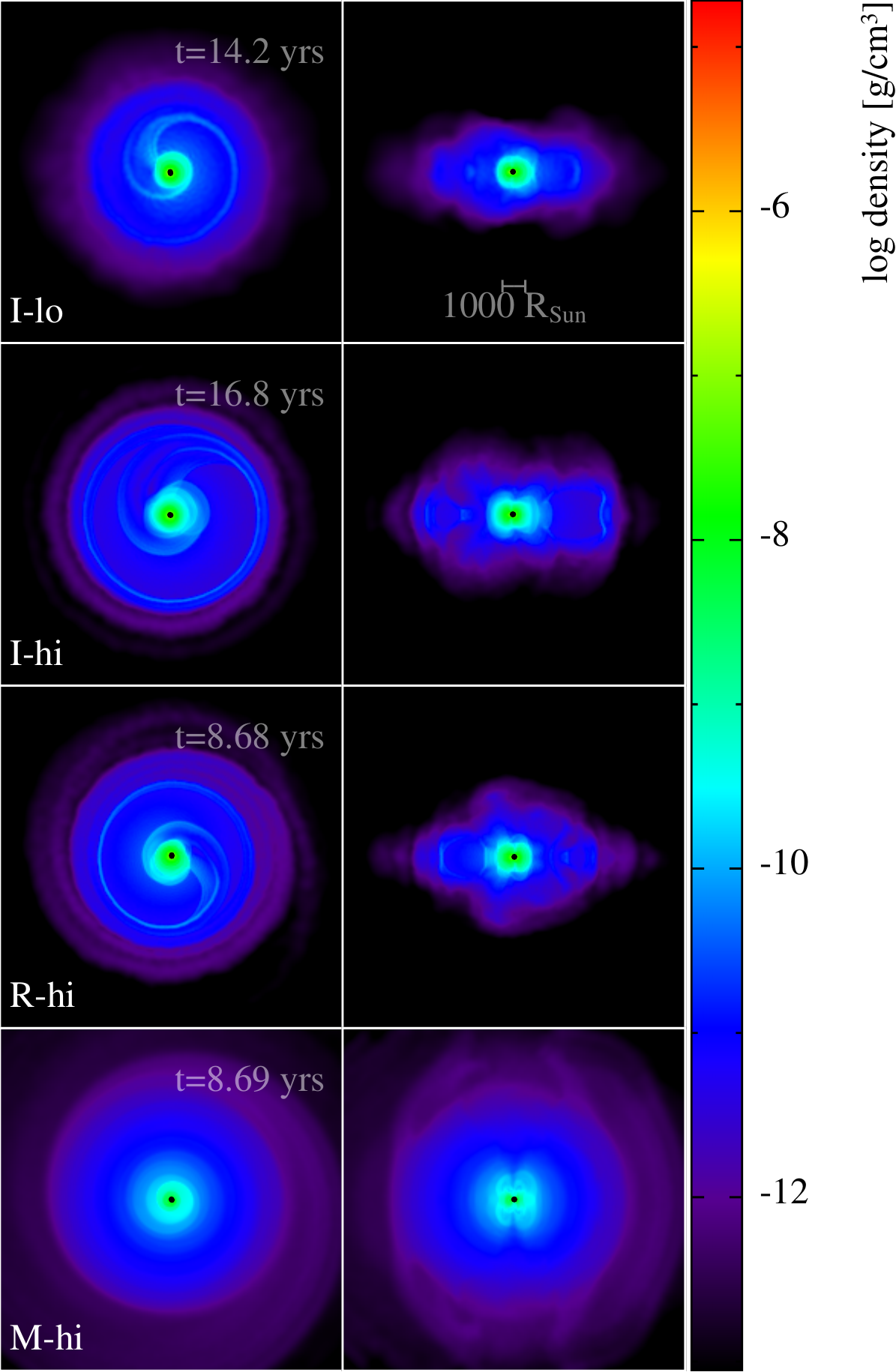}\
    \caption{Density cross sections in the orbital plane (first, third and fifth columns) and the perpendicular plane (second, fourth and last columns) of all models at the start (first two columns) in the middle (third and fourth columns) and at the end (last two columns - note these slices are zoomed out) of the plunge-in phase. Columns 1 to 4 have panels of $10^4$~\Rsun on each side. Columns 5 and 6 have panels of $1.5 \times 10^4$~\Rsun on each side. The rows, from top to bottom, represent the four models: I-lo, I-hi, R-hi and M-hi, respectively. A complete movie of the I-hi, R-hi and M-hi models is available by following the link:~\url{https://miguelglezb.github.io/mgb/simulations/2msun-tp-agb.html}. }
    \label{fig:render_ins}
\end{figure*}

Here we discuss the simulations' results using the high resolution, ideal gas equation of state simulation (I-hi) as our baseline. In Section~\ref{sec4.1:orb} we analyse the orbital evolution of all our simulations, including that of \citet{Reichardt2019}. In Section~\ref{sec4.2:bound} we discuss the mass unbound during the simulations. We analyse the rotational properties of the stars in Section~\ref{sec4.3:angmom}. Finally, the morphology of the post-CE system is explained in Section~\ref{sec4.5:morph}. All our simulations conserve total energy and total angular momentum to 0.1 per cent and 0.05 per cent, at worst, respectively (more information is presented in Appendix~\ref{app:conservation_properties}).

\subsection{Orbital evolution}
\label{sec4.1:orb}

\begin{figure*}
    \centering
    \includegraphics[width=0.45\textwidth]{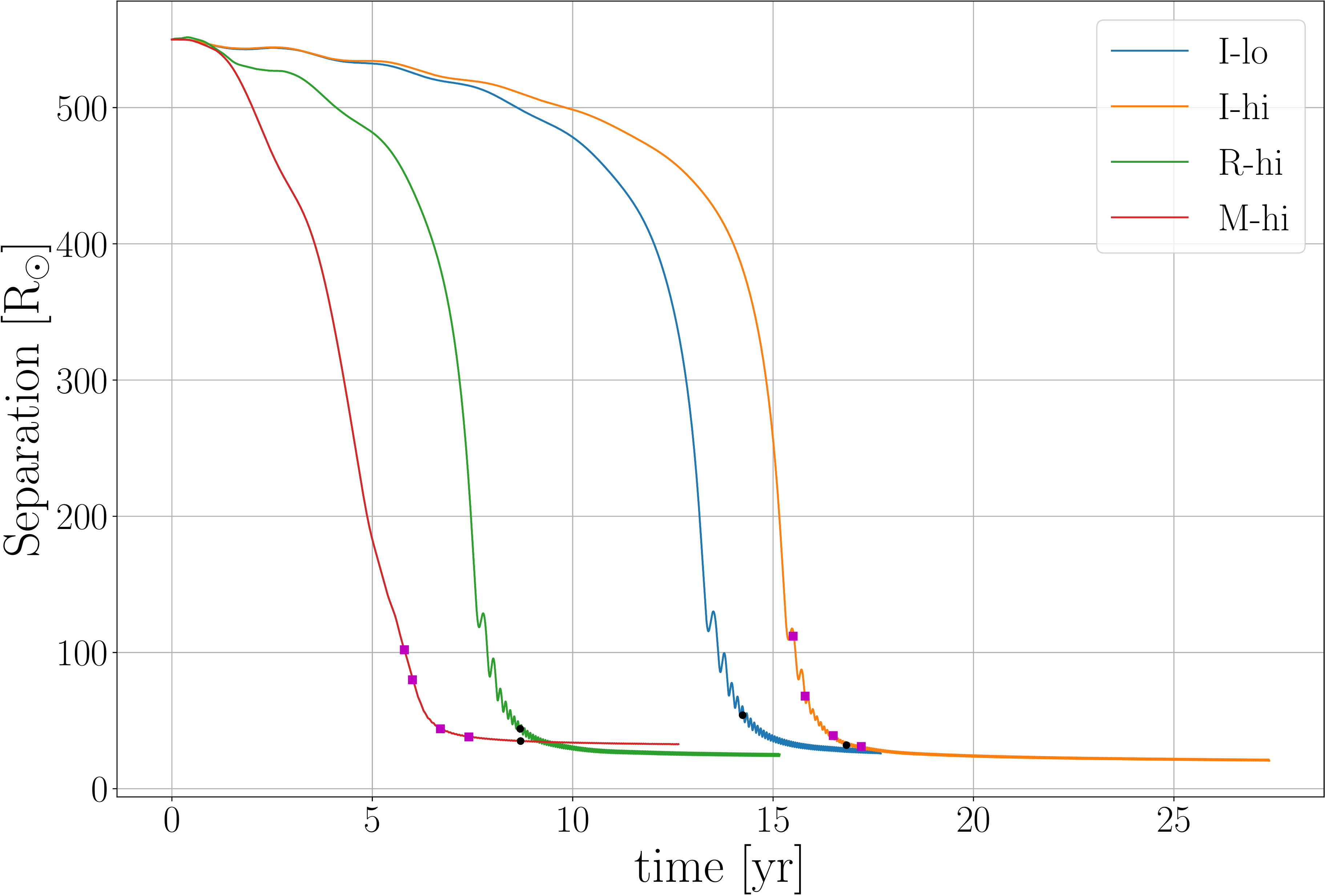} \ \ \ \ \ \ \ 
    \includegraphics[width=0.45\textwidth]{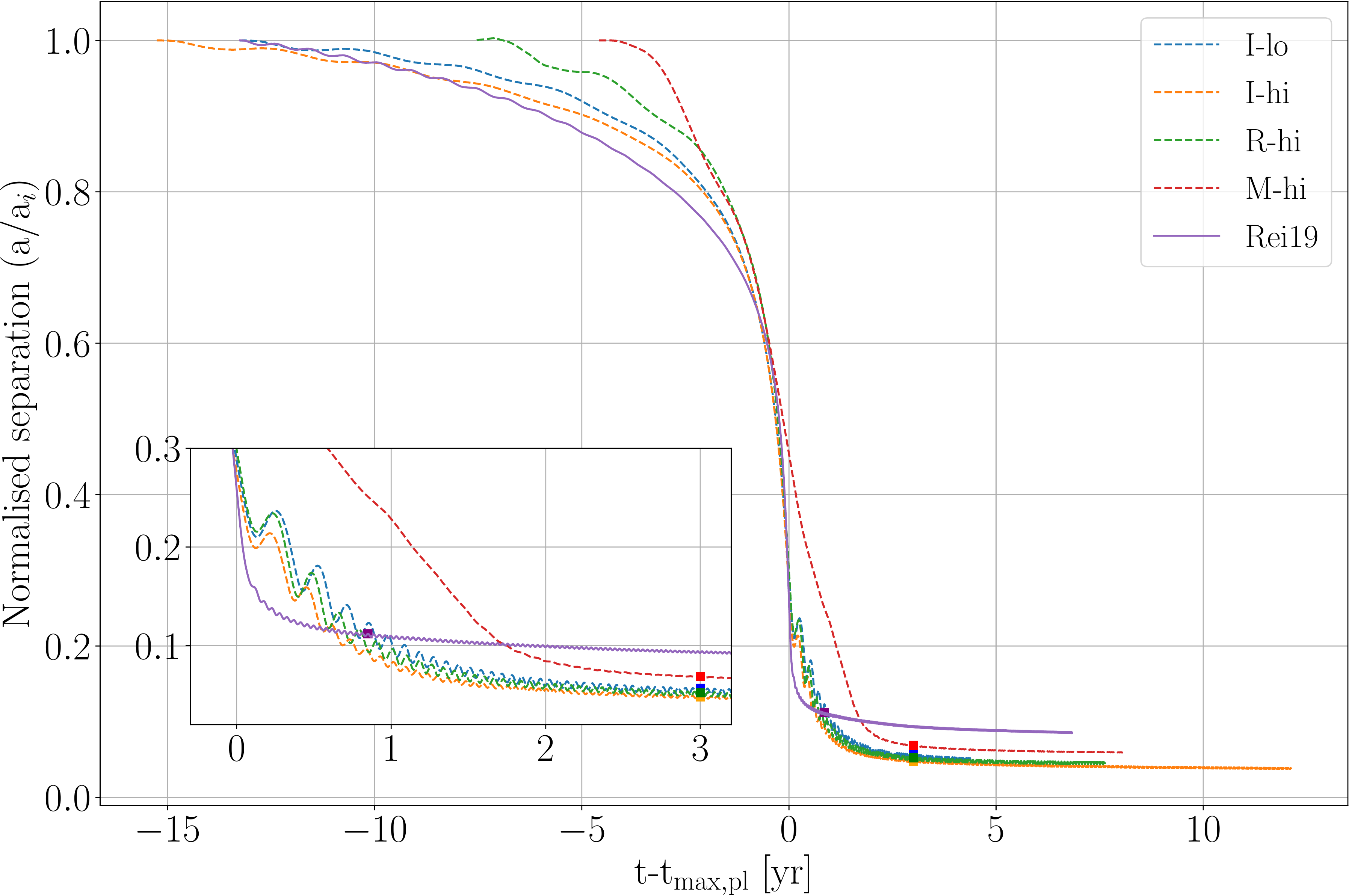} 
    \caption{Left panel: orbital evolution of the common envelope models. The black circles indicate the end of the plunge-in. The magenta square symbols on the I-hi and M-hi orbital evolution curves correspond to the times of the slices in Figure~\ref{fig:render_omega}. Right panel: orbital evolution of the models, including that of \citet{Reichardt2019}. The x-axis have been shifted so that $t=0$ %\mike{but you have $t-t_{pl}$ in your axes.}\miguel{Fixed}
    coincides with the steepest plunge-in moment, $t_{\rm max,pl}$ (Table~\ref{tab3:table}), in every model. The orbital separation is normalised to the initial separation between point mass particles. The square symbols show the times that are compared in Section~\ref{sec4.5:morph}.
    %\mike{The right panel adds very little information relative to the left panel. I would suggest indicating the start and end points with coloured markers placed on the separation vs. time curves in the left panel, and getting rid of the right panel. Same with the magenta squares.}\miguel{Done}
    }
    \label{fig:sep1}
\end{figure*}

Before we can describe the orbital evolution and compare simulations we must define a quantitative criterion to determine the time of the end of the dynamical part of the in-spiral. We define the dynamical in-spiral phase as the time at which $|\dot{a}P|/a>0.01$, where $a$ is the orbital separation and $P$ is the orbital period.
This criterion was used by \citet{Ivanova2016} as well as by \cite{Sand2020}; the former used a value of $0.1$ while the latter applied $0.01$. Because of the eccentricity that develops towards the end of the in-spiral, we apply a Savitzky–Golay filter to the point mass particles' separation. The filter applies a piecewise polynomial with a user-defined order, to a number of the data points within a window which is also determined by the user. We set a first order polynomial and a window length such that the residuals between the smoothed and the original data was at most the sum of the point mass particles' softening radii. The end of the plunge-in separation for all simulations, determined in this way, along with the separation at the steepest plunge-in point are listed in Table~\ref{tab3:table}, alongside the times when they happen. 
In Figure~\ref{fig:render_ins} we show density slices in the orbital and perpendicular planes for all our models taken at three times: (i) an arbitrary time shortly after the beginning of the simulation, when the dynamical in-spiral has not yet started, (ii) at a time close to when, visually, the orbital separation starts to decline more rapidly, and (iii) at the end of the in-spiral as quantified by the criterion above (called $t_{\rm pl,f}$ in Table~\ref{tab3:table}). For simulation M-hi there is no clear delineation between pre-plunge-in and plunge in, so for time (ii) we selected a somewhat arbitrary moment in the early in-spiral.

What is immediately evident in Figure~\ref{fig:render_ins} is that the envelope expansion and mass ejection are preferentially along the orbital plane. This ejection is primarily through the $L_{\rm 2}$ point. By the end of the plunge-in, a spiral structure is observed around the close binary, including a ring of material ejected at time $16.8$~yrs in the case of I-hi. Striking is that with the tabulated equation of state the distribution is more spherical and smooth. It is possible that as the envelope expands and recombines the recombination energy is delivered in a more spherically symmetric way and the mass is accelerated accordingly.

At the beginning of each simulation, the primary star fills its Roche lobe such that mass transfer starts immediately. The orbital period of the binary at the start of the simulations is $2.7$~years. Although there is some variability between models, the orbital separation decreases relatively slowly for the first few years, after which the plunge-in phase takes over. \review{This is followed by a relatively sudden slowing down of the in-spiral for all models.} A plot of the separation between core and companion as a function of time is shown in Figure~\ref{fig:sep1}.

Our baseline model, I-hi, spends $\sim$5~years in a pre-in-spiral phase. Eventually gas starts to outflow from the $L_2$ and $L_3$ Lagrangian points, at which point the in-spiral commences. Envelope gas expands at this point, some of which becomes unbound.For I-hi, the pre-plunge-in phase lasts visibly the longest of all models. For this model, the time of steepest plunge-in is at $t=14.2$ years from the start of the simulation. After the plunge-in, at about 17~yrs, the point mass particles move in an orbit with a mild eccentricity, $e\approx0.02$, and slowly decreasing average separation. The separation at the end of the plunge-in as defined above (16.8~yrs) is 32~\Rsun, while at the end of the simulation (27.4~yrs), the separation is 20~\Rsun. This is evident in Figure~\ref{fig:sep1}, right panel, where the sink separation curves are normalised by their initial values, and are shifted so that $t=0$ is at the time of steepest plunge-in, $t_{\rm max,pl}$  (Table~\ref{tab3:table}), for every model. Here, the slopes of the fast in-spiral phases are more easily compared, as is the end of the in-spiral (Fig.~\ref{fig:sep1}, right panel).

At low resolution the time before the plunge-in is shorter. This was also observed by \citet{Reichardt2019} and it is due to a higher mass transfer rate through $L_1$ that hastens the plunge-in. 
It is possible that at even higher resolution the pre-plunge-in time would be even longer as the early mass transfer rate would be lowered further. 
Under steady conditions (a star that is not naturally expanding or contracting over short timescales) it is possible that the pre in-spiral time would lengthen considerably. 
The TP-AGB star that we chose to simulate is, however, expanding  over short timescales, because of its thermal pulse, as described in Section~\ref{sec2:TP-AGB} and as can clearly be seen in Figure~\ref{fig:PD_rad_fracc}. In this situation it is likely that the early mass transfer at the time of first Roche lobe overflow, may be hastened by the natural expansion of the star. 
%\mike{Explain this expectation.} 
 The plunge-in slope and the post-plunge-in separation are very consistent between I-lo and I-hi (Figure~\ref{fig:sep1}, right panel), showing 
 %\mike{I don't think you discussed this before} \miguel{Fixed} 
 that resolution does not play a large role in these quantities. 
 
 The values of the separations for I-lo and I-hi at the end of plunge-in, $a_{\rm f,pl}$ in Table~\ref{tab3:table}, 
 are 54 and 32~\Rsun, respectively, a difference of 51 per cent.
 %taken either at the end of plunge-in or at the (arbitrary) end of the simulations 
 %\mike{I don't think you previously explained when you stop the simulations} \miguel{I fail to see the relevance in the time at which I stopped the simulations. If the systems were clearly on the plunge-in (even when there is no clear definition, we can say for sure the separations won't change too much after $t_f$ for each model), then we would have to justify why, but in this case we only need to state what is the stopping time for the simulations.}
 %vary by 26 and 51~per cent, respectively, but this variability is more due to the definition of the end of plunge-in 
 %\mike{what does this mean? Isn't the whole point of defining the end of the plunge-in to allow for a fair comparison across different simulations?} \miguel{This is just stating that there will be a variability between the separation at the end of the plunge-in and at the end of the simulations, regardless on how you chose those moments during the models' evolution} 
 %and to the different lengths of the two simulations 
 %\mike{In that case, I don't think you should mention the 51 per cent figure at all---it has little relevance} \miguel{I'm stating the difference between separations at different moments in time and comparing it at different resolutions. I think it is very relevant.}. 
 Aligning I-hi and I-lo at the steepest in-spiral point (Figure~\ref{fig:sep1}, right panel) and reading the separation at 4.4~years after that point, close to the end of the simulations, the two values are very similar: 24.2~\Rsun\ and 25.8~\Rsun, for I-hi and I-low, respectively. The eccentricity of the orbit after the plunge-in is comparatively larger for the low resolution simulation, I-lo (0.04) than for the high resolution one, I-hi (0.02). 

As we have discussed in Section~\ref{sec3.1:star_relaxation}, the R-hi initial stellar model is more extended and has higher overall particle speeds at its surface than the model I-hi (Figures~\ref{fig:PD_rad_fracc} and \ref{fig:vel_hist}). This may be responsible for a shorter Roche lobe overflow phase. The separation after the plunge-in, $a_{\rm pl,f}=44$~\Rsun\ (Table~\ref{tab3:table}), can be compared to the equivalent value for I-hi of 32~\Rsun.
The orbital separation 4.4~yrs after the time of maximum plunge-in is 25.3~\Rsun\ for R-hi and 25.8~\Rsun\ for I-hi. We therefore conclude that including radiation pressure in the equation of state does not lead to large differences.

The M-hi model has an in-spiral timescale shorter than for all the other simulations, including the one at lower resolution. The reason for this behaviour are  (i) the larger size of the star at the time of the start of the simulation, (ii) the slow radial expansion of the initial model and (iii) the delivery of recombination energy to the gas in the outer layers which promotes its expansion (and some early envelope unbinding - see Section~\ref{sec4.2:bound} for additional details). The exact role played by these three factors on the faster in-spiral on-set, in-spiral shape and final separation is not well quantified. However, we rerun simulation M-hi but this time starting from a stellar structure that was larger (285~\Rsun), but more stable and the in-spiral on-set, shape and final separation were remarkably similar to those associated with M-hi, showing that a star that is smaller, but slowly expanding results in a similar CE as one simulated using a stellar structure that is not expanding, but is already somewhat larger (see Appendix~\ref{app:alternative_Mhi} for more details). This demonstrates that the CE interaction characteristics are not strongly dependent on the fine details of the initial stellar structure. 

We note that M-hi does not just in-spiral sooner, but also has a completely different in-spiral shape compared to the other three simulations, with an overall shallower slope throughout (Figure~\ref{fig:sep1}, right panel). We will discuss this result in Section~\ref{sec5:discussion} where we will contextualise it with other work where similar comparisons are carried out.

The post in-spiral separation of M-hi is clearly larger as can be seen in Figure~\ref{fig:sep1}. The values at the end of the plunge-in is 35~\Rsun, 42~per cent larger, than the corresponding value for I-hi. Once I-hi and M-hi are aligned at the time of steepest in-spiral, the separation of M-hi 4.4~years after the steepest in-spiral time is 34.6~\Rsun\ compared to the equivalent value for I-hi of 24.2~\Rsun. As is also clear from Figure~\ref{fig:sep1}, right panel, M-hi stands out in having larger final separation compared to the other three simulations. 

\review{Since the orbits are still decreasing by the end of the simulation, we extrapolate the final separations using a exponential decay fit (Table~\ref{tab3:table}, column 7). The details of the extrapolation method are described on Appendix~\ref{app:extrapolation_finalsep}. The two values correspond to the upper and lower limits for the final separations. While these extrapolated values indicate that the orbits will further shrink, the differences are not significant to influence the envelope or considering a merger scenario. There is one caveat about this extrapolation technique. The number of oscillations in the orbital evolution after the plunge-in have to be large enough to find a decay factor that fits the decay of the orbit at later times, otherwise the upper limit of the extrapolation may be larger than the final separation in the simulation. This is the case for the orbital evolution in I-lo and Rei18. However, the rest of the AGB models have extrapolated (upper and lower) values that are below the values in column 6.} 

Lastly, the post-plunge-in orbit for the M-hi model does not present an eccentricity as large as the other models (0.005, Figure \ref{fig:sep1}, right panel and Table~\ref{tab3:table}). 

For all the simulations, the time between Roche lobe overflow and orbital stabilisation is at least one order of magnitude smaller than the thermal pulse of the TP-AGB star, from the moment the star reaches 260~\Rsun\ until it contracts to that same size, $400$ years later. {\it This suggests that the common envelope evolution can easily take place during a thermal pulse}. The only question is whether, in nature, the time between the on-set of Roche lobe overflow and the on-set of the in-spiral could be long enough that the CE would not take place, because the star contracts before in-spiral could start in earnest. After all, we do not have a good grasp on this pre-plunge-in time. We discuss this further in Section~\ref{sec5:discussion}. 
\begin{figure}
    \centering
    \includegraphics[width=\linewidth]{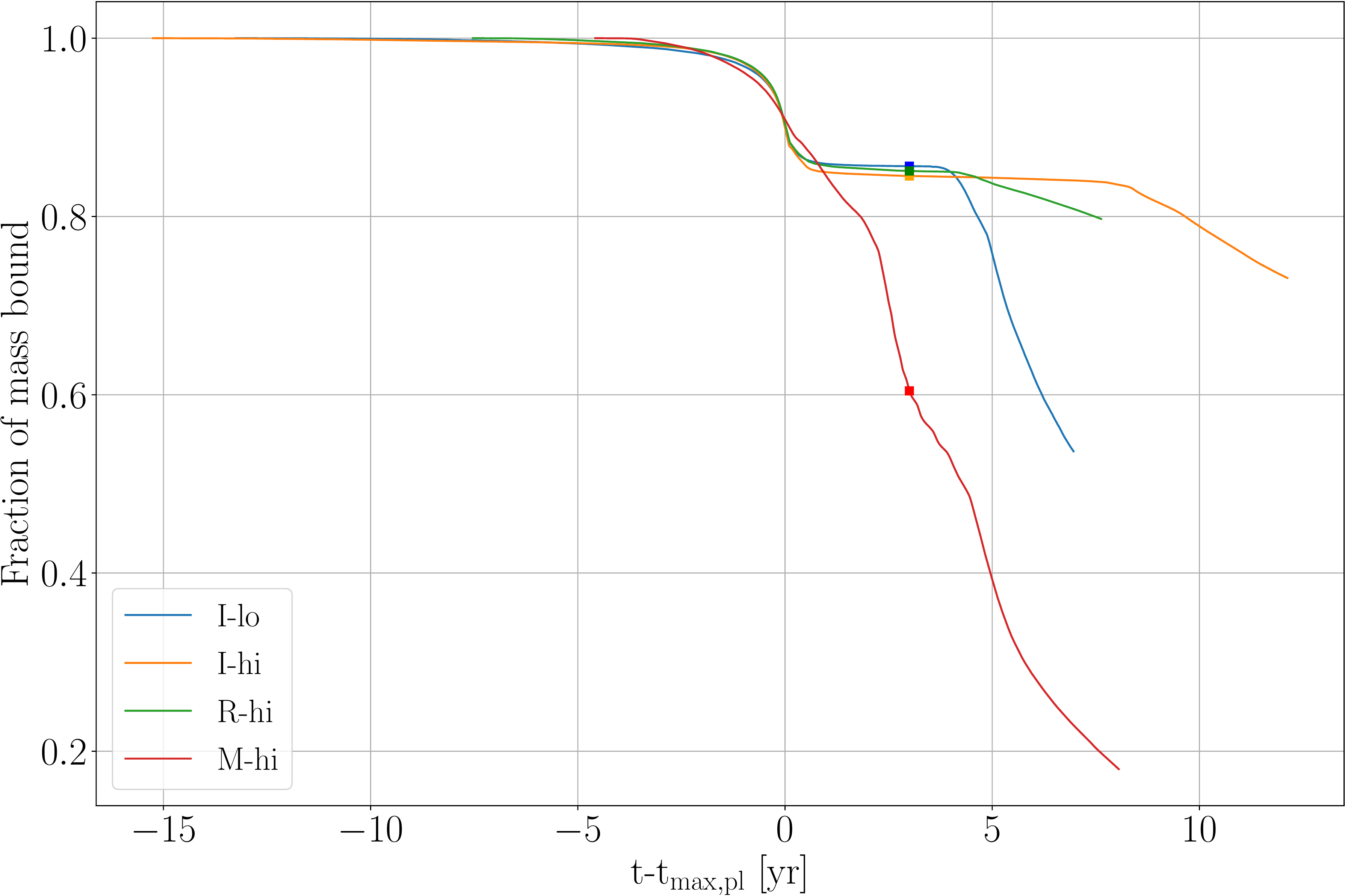}\
    \caption{Bound mass as a fraction of the envelope mass for all simulations, shifted so that $t=0$ corresponds to the time of plunge-in, $t_{\rm max,pl}$. The square symbols show the times compared in Section~\ref{sec4.5:morph}. For these plots, unbound mass is defined as $e_{\rm pot} + e_{\rm K} > 0$.}
    \label{fig:bound_mass}
\end{figure}

\begin{figure*}
    \centering
    \includegraphics[width=0.45\linewidth]{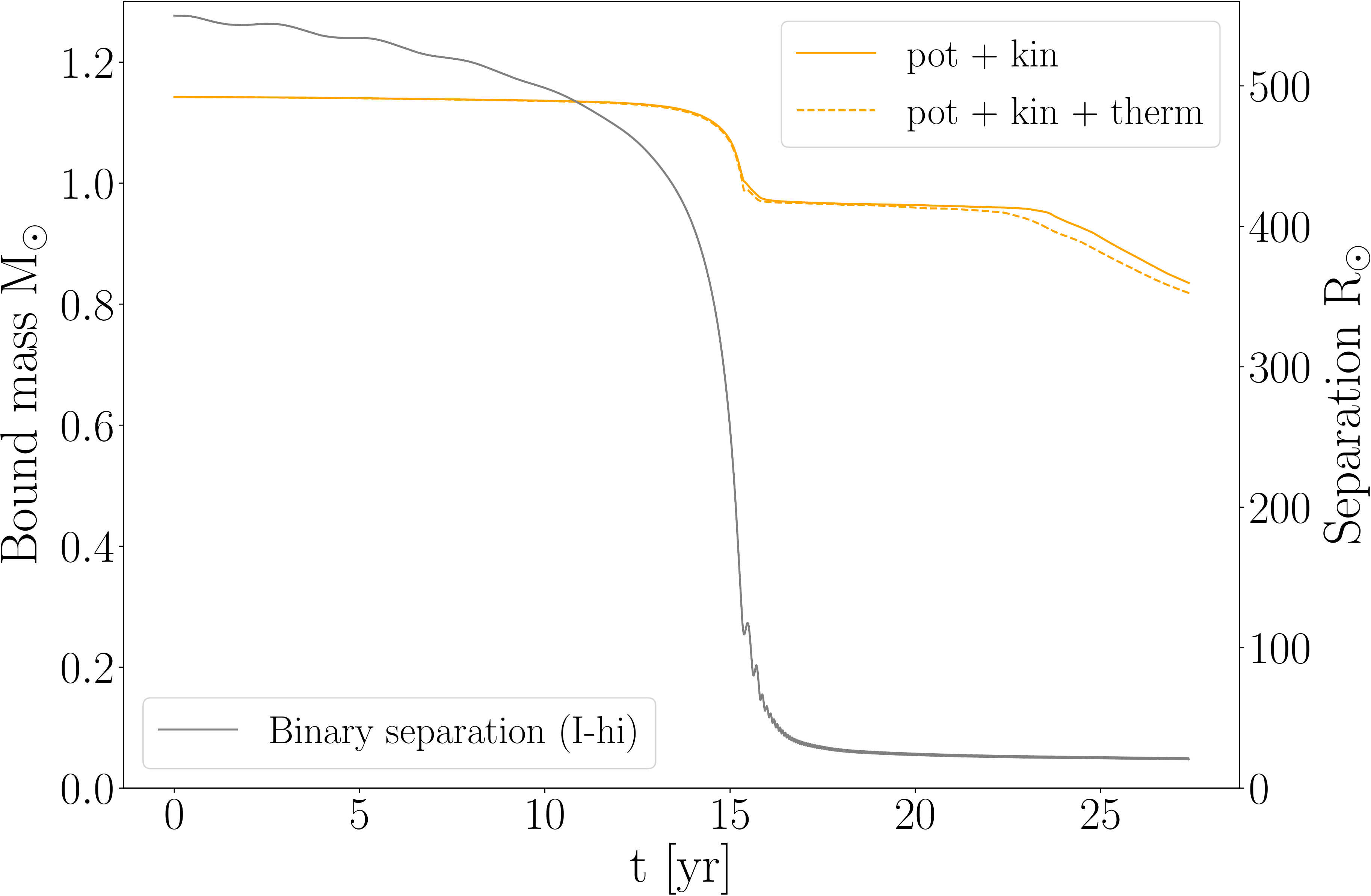}\ \ \ \ 
    \includegraphics[width=0.45\linewidth]{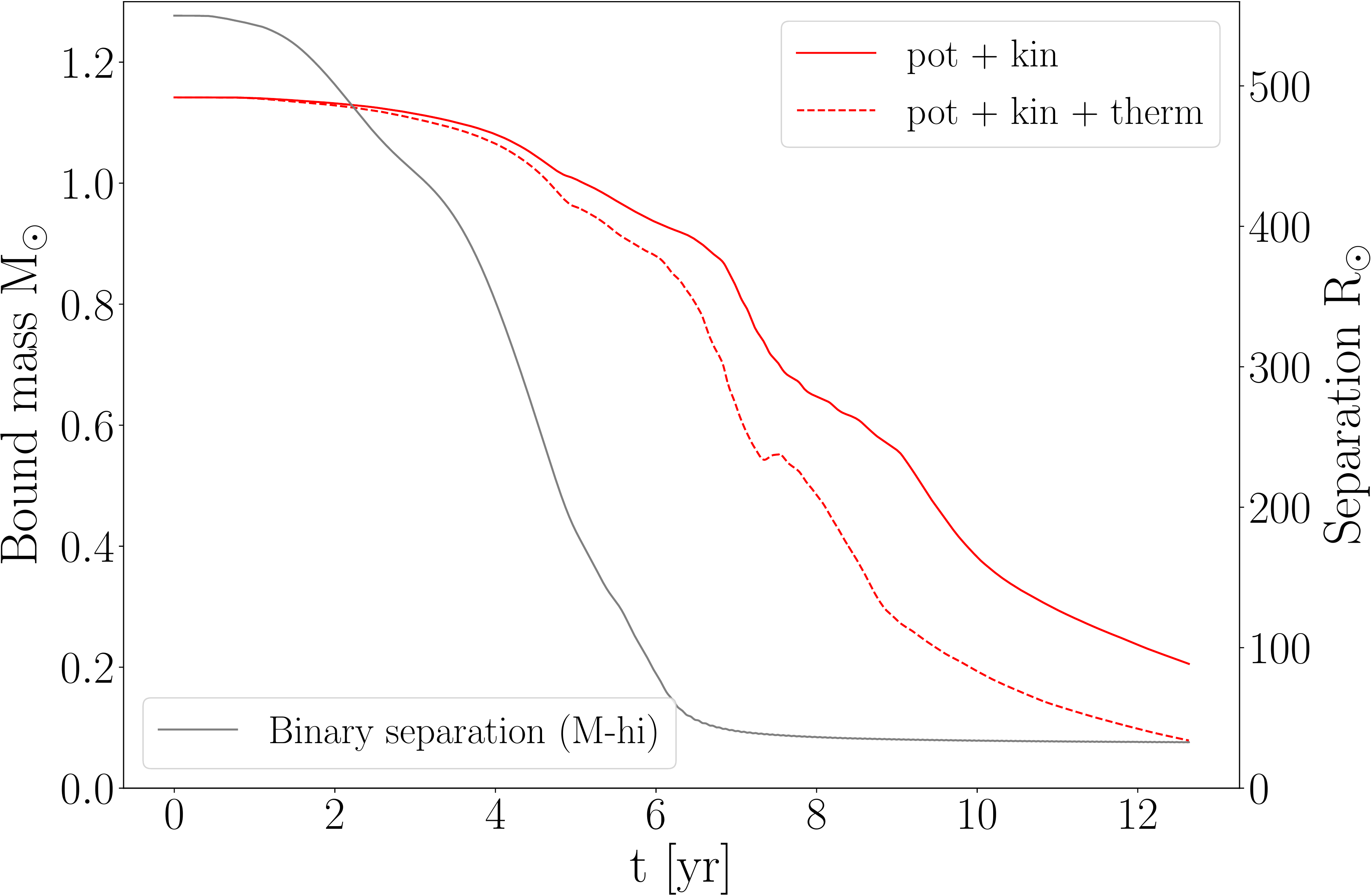}\
    \caption{Bound mass (coloured curves) and binary separation (black curves) plots of I-hi (left panel) and M-hi (right panel) models. We plot the bound mass using the mechanical (solid lines)  and thermal (dashed lines) criteria in each panel.}
    \label{fig:boundsep}
\end{figure*}

\subsection{Analysis of the unbound mass}
\label{sec4.2:bound}

In Figure~\ref{fig:bound_mass} we present the bound mass as a function of time for our four simulations as well as for that of \citet{Reichardt2019}, using the most stringent criterion for calculating unbound mass: $e_K + e_{\rm pot} \ge 0$, where $e_K$ and $e_{\rm pot}$ are the kinetic and potential energies of all the gas particles at each moment in time. \review{The potential energy for each SPH particle was calculated with respect to the rest of the SPH particles and each point mass particle.} As extensively discussed in several publications \citep[e.g.,][]{Iaconi2017,Staff2016}, including the thermal energy (and even more importantly the recombination energy) in this criterion assumes that the entire thermal energy payload of the stellar envelope (including the thermalised recombination energy) will, at some point, be transformed into bulk kinetic energy. Since this is not necessarily the case, it is good practice to retain a more stringent ``mechanical'' criterion as well as noting the amount of additional unbound gas when adopting a thermal energy criterion. 

In order to use a criterion that includes the thermal energy, we follow the procedure described below. 
The thermal energy calculation depends on the equation of state. For I-hi the thermal energy is provided by \phant\ in the form of the internal energy. 
For the \mesa\ and ideal plus radiation pressure equations of state, we calculate the thermal energy in the following way: 

\begin{equation}
    E_{ \rm th} = \frac32 \frac{k_{\rm B} T_i}{m_{\rm h} \mu} + \frac{a_{\rm rad} T_i^4}{\rho_{\rm i}}.
\end{equation}
where $k_{\rm B}$ is the Boltzmann constant, $m_{\rm h}$  is the mass of the hydrogen atom, $\mu$ is the mean molecular weight, $T_i$ is the temperature of the i-th particle, $\rho_i$ is its density and $a_{\rm rad}$ is the radiation constant. The temperature and mean molecular weights are output by the computation.

Figure~\ref{fig:boundsep} compares the unbound mass calculated with the criteria that omit, or include the thermal energy component,  for the I-hi and M-hi simulations. As expected, the mechanical criterion has less unbound gas than the thermal energy one. The bound mass at the end of I-hi is $0.83$ and $0.82$~\Msun, using the mechanical and thermal criteria, respectively. For M-hi, these values are $0.2$ and $0.08$~\Msun, respectively. Therefore, I-hi unbinds $27-28$~percent of the envelope, while M-hi unbinds $82-93$~percent. The curves in Figure~\ref{fig:boundsep} are all trending down, implying additional unbinding. However, as we are about to see, one has to be cautious about the later part of the unbinding.

\begin{figure*}
     \centering
     \begin{subfigure}[b]{0.45\textwidth}
         \centering
             \includegraphics[width=\textwidth]{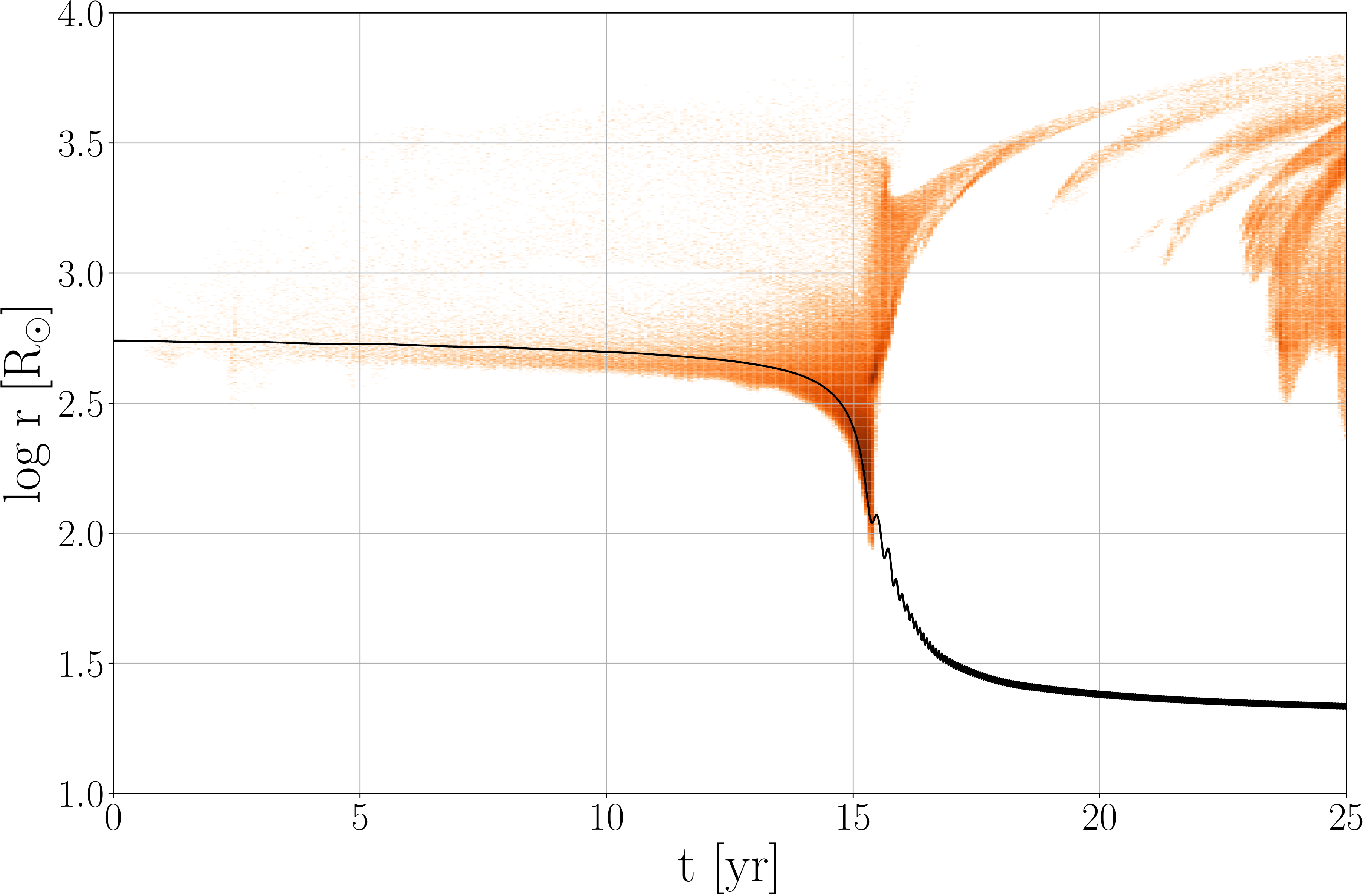}
         \caption{I-hi model, mechanical criterion.}
     \end{subfigure}
     \hfill
     \begin{subfigure}[b]{0.45\textwidth}
         \centering
         \includegraphics[width=\textwidth]{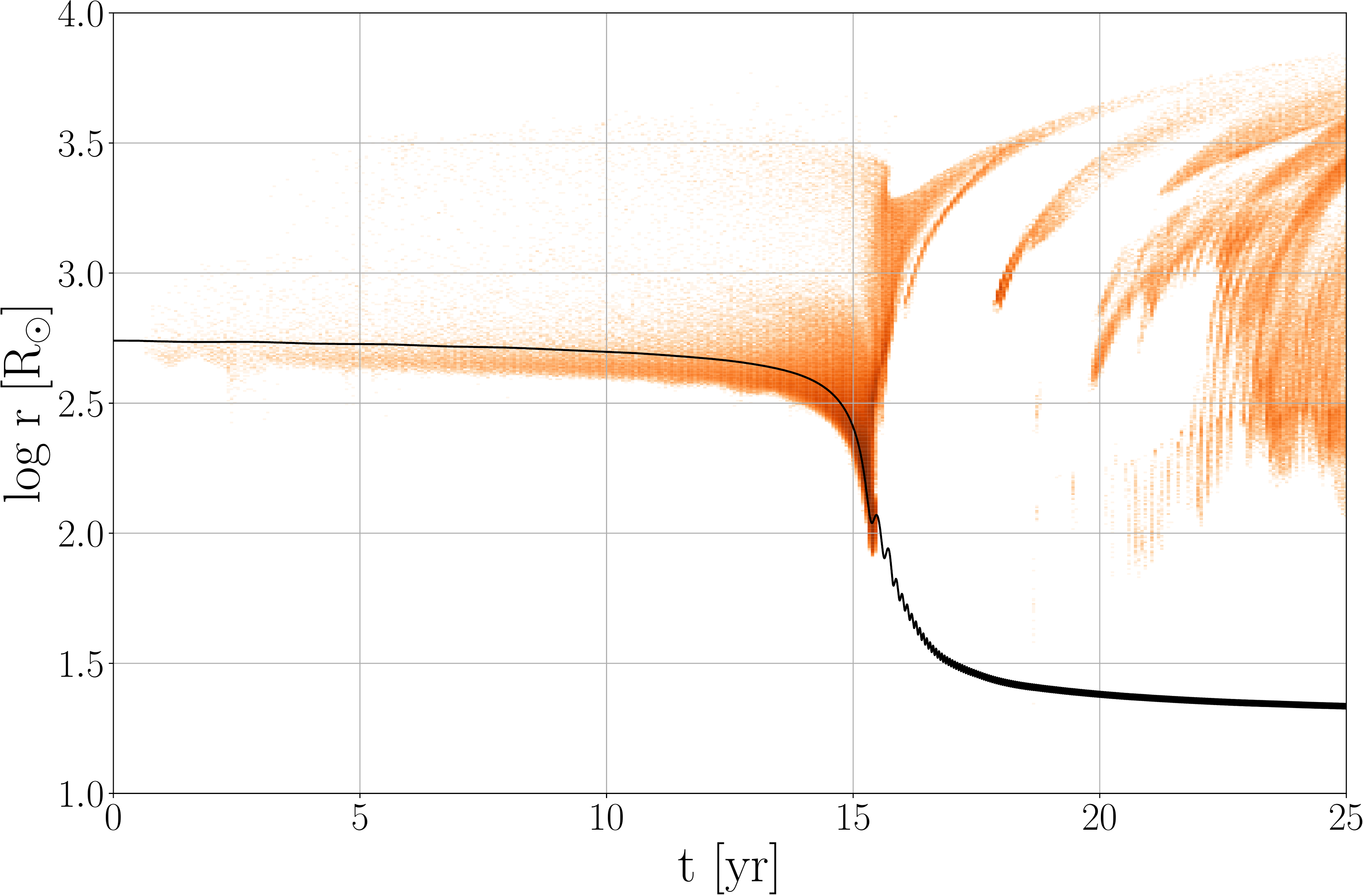}
         \caption{I-hi model, thermal criterion.}
         \end{subfigure}
     \hfill
     \begin{subfigure}[b]{0.45\textwidth}
         \centering
         \includegraphics[width=\textwidth]{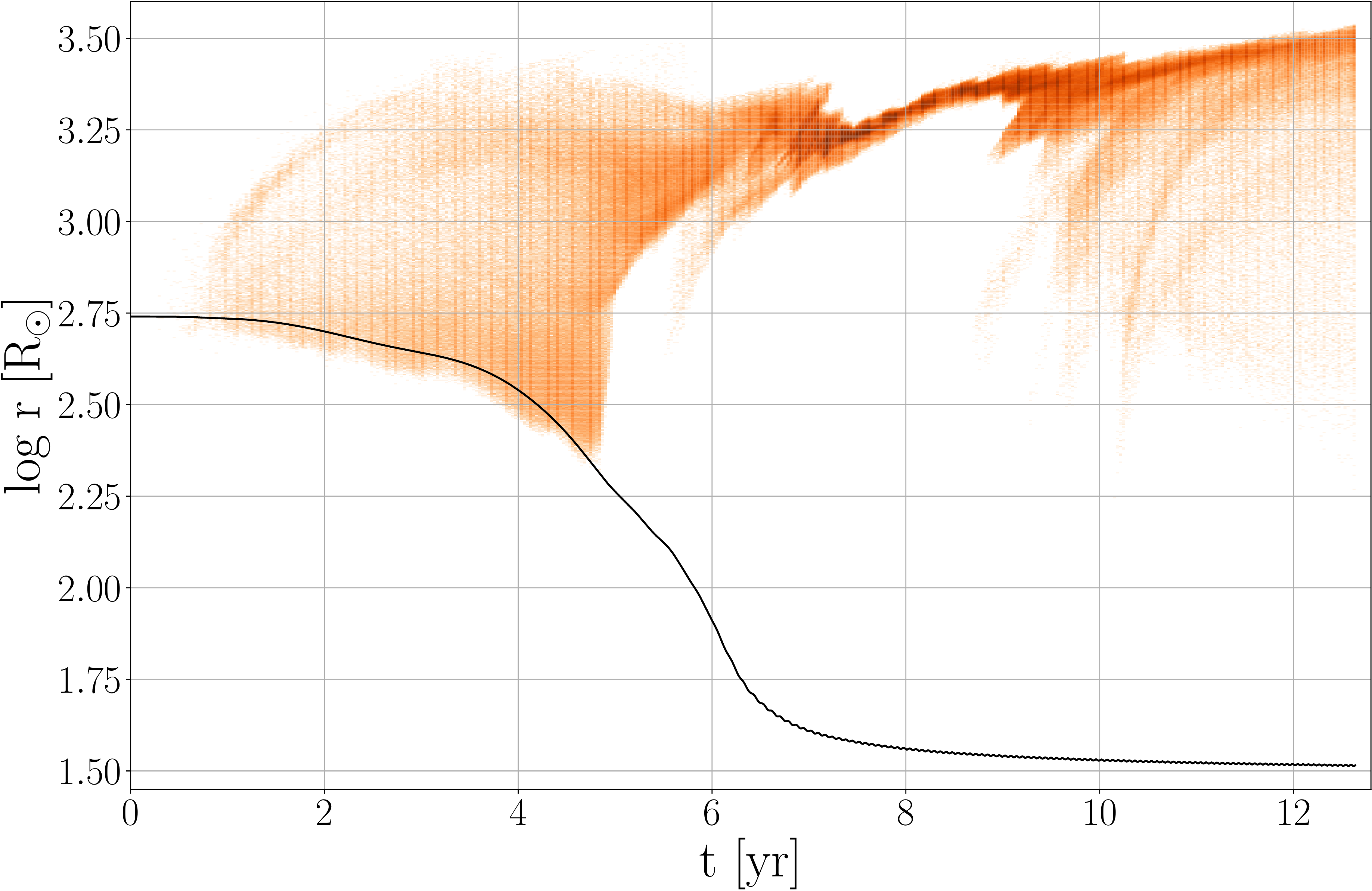}
         \caption{M-hi model, mechanical criterion.}
         \end{subfigure}
     \hfill
     \begin{subfigure}[b]{0.45\textwidth}
         \centering
         \includegraphics[width=\textwidth]{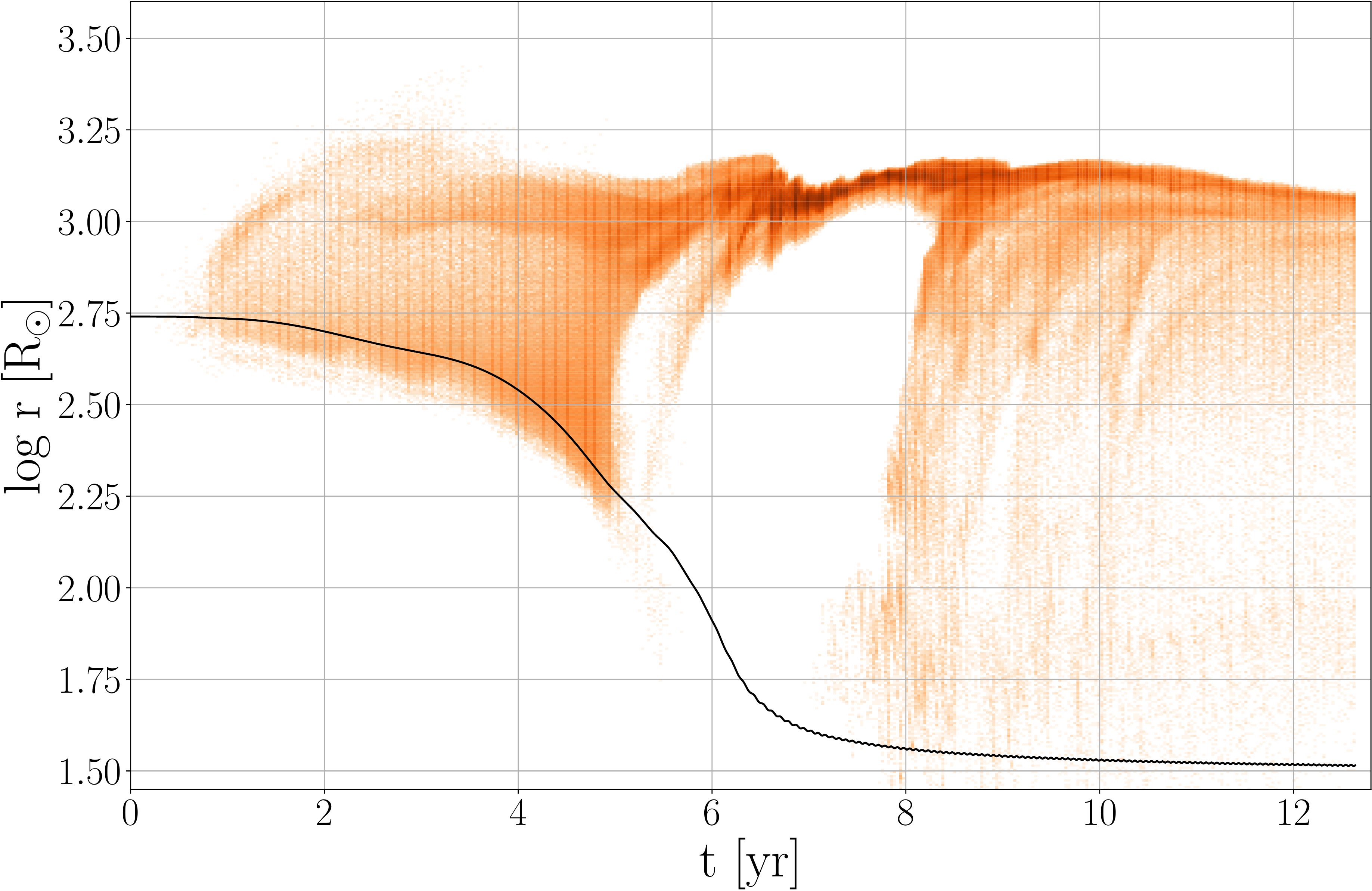}
         \caption{M-hi model, thermal criterion.}
         \end{subfigure}
        \caption{Newly unbound particles for the I-hi (top row) and M-hi (bottom row) models using the mechanical (left column) or thermal (right column) unbound mass criteria. The color grading indicates the amount of gas that is newly unbound, approximately every 5 days. The black curve represents the binary separation. The y-axis measures distance from the primary's point mass core particle.}
        \label{fig:new_UB}
\end{figure*}

Figure~\ref{fig:new_UB} shows four 2D histograms representing the radial position and time of {\it newly} unbound SPH particles for I-hi (top panels) and M-hi (bottom panels), using a heat map to represent the amount of mass that becomes unbound (in arbitrary units). 
We can divide the unbinding into four stages. The first stage is only clearly seen in M-hi: some particles become unbound starting at $\sim$1~yr, likely promoted by some early expansion and the delivery of recombination energy in those surface layers. This would promote additional expansion and precipitate the in-spiral, as explained in Section~\ref{sec4.1:orb}. 

The second phase, more prominent in I-hi, is due to the transfer of orbital energy to the gas. It appears as a band of particles right above the black curve that denotes the orbit. This unbinding starts before the plunge-in and it continues and intensifies in the early phase of the plunge-in, but it stops at approximately the time of maximum in-spiral at $t\sim 15$ years for I-hi and $t\sim 5$ years for M-hi. Gas unbound near the companion and moving outward transfers energy to gas further out leading to more unbinding -- this is seen as the curving streaks of unbound particles in  Figure~\ref{fig:new_UB}.

The third phase is denoted by a very localised region of unbound mass between $\sim$15~years ($\sim$5~years) and $\sim$22~years ($\sim$10~years) for the I-hi (M-hi) simulation in Figure~\ref{fig:new_UB}. Particles are no longer becoming unbound near the companion because the companion is closer to the core, where higher density, more bound gas resides. \review{Instead, the gas unbound in the previous phase collides and transfers some of its kinetic, and to a lesser degree thermal, energy to the bound gas in the upper layers of the envelope, effectively unbinding them.} The M-hi simulation differs from I-hi in this phase: the ``streak" of unbound gas is very dense (more unbound mass) with most of the mass being unbound here rather than in the previous phase as is the case for I-hi. This is likely due to the release of recombination energy.

The fourth and last phase occurs, for the I-hi simulation at $\sim22$~years and for M-hi at $\sim$10~years. Here the region of newly unbound particles extends again from deeper layers to near the surface. This fourth phase corresponds to the second drop in the bound mass for I-hi in Figure~\ref{fig:bound_mass} (left panel). For M-hi the decline in bound mass in the same Figure is more gradual and this fourth phase is not as distinct. The nature of this late unbinding is related to a decrease in resolution near the cores. This event is resolution dependent as can be seen by comparing I-hi and I-lo in Figure~\ref{fig:bound_mass} and was already described by \citet{Reichardt2020}. When the SPH particles' smoothing lengths become comparable to the softening length of the point mass particles, the gas particles fail to resolve the core \textemdash\ due to the modified point mass gravitational potential \textemdash\ and the pressure can become negative. This causes an acceleration of particles near the point masses, which at this point have a separation of $\lesssim 30$ \Rsun. These particles may or may not  become unbound, but even if not, they transfer their energy to upper layers of the envelope that, being only loosely bound, can become unbound. This resolution-dependent unbinding is far more obvious in ideal gas simulations and the extent to which the late unbinding in M-hi is resolution dependent has not yet been established.

The panels including thermal energy (right panels in Figure~\ref{fig:new_UB}) are similar, but in the fourth phase show additional regions where gas is becoming newly unbound. For I-hi this difference is marginal and shock heating may be the contributing agent. For M-hi the extra unbinding seen using the thermal criterion is more widespread and \review{likely indicates the release} of recombination energy (which is assumed to be immediately thermalised).

\begin{figure*}
    \centering
    \includegraphics[width=0.42\linewidth]{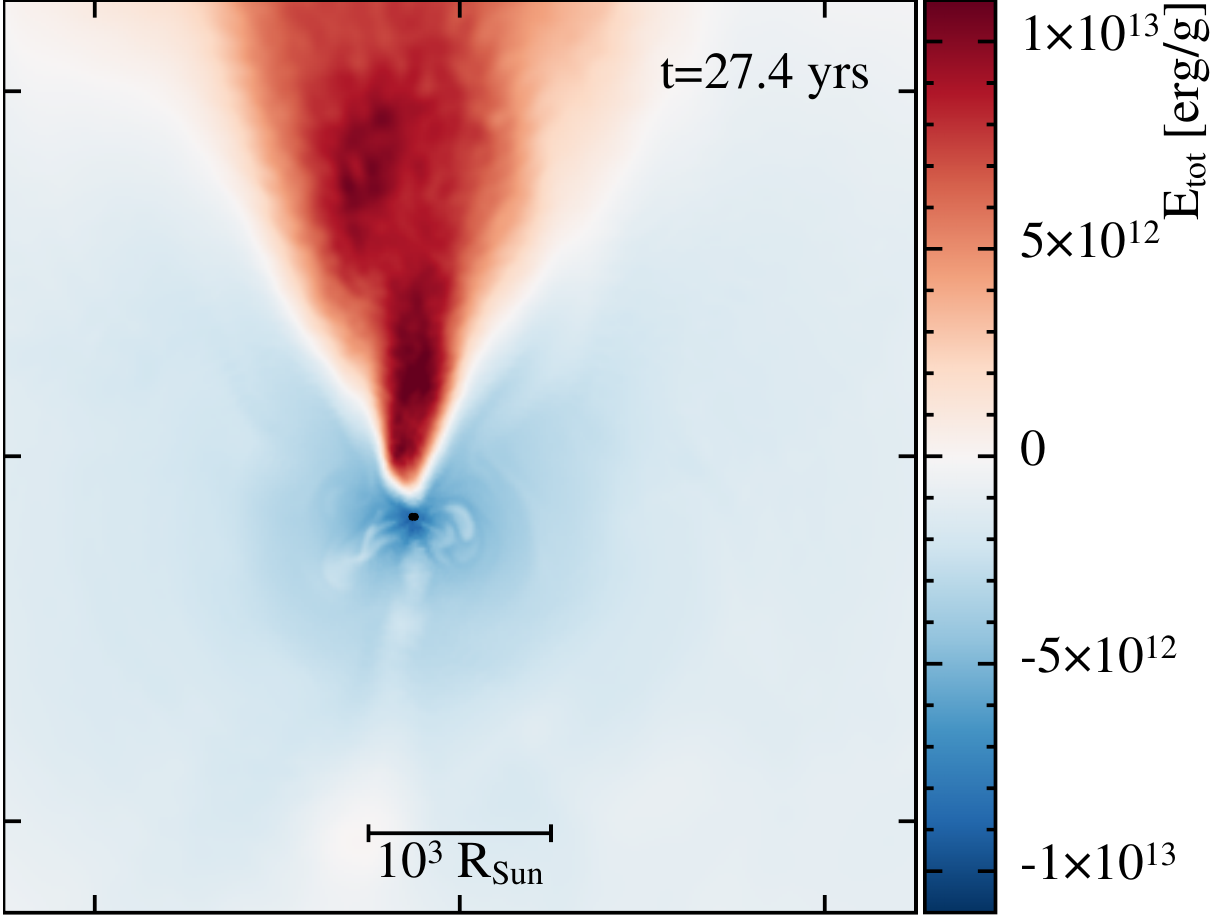}\ \ \ \ \ \ \ \ \
    \includegraphics[width=0.42\linewidth]{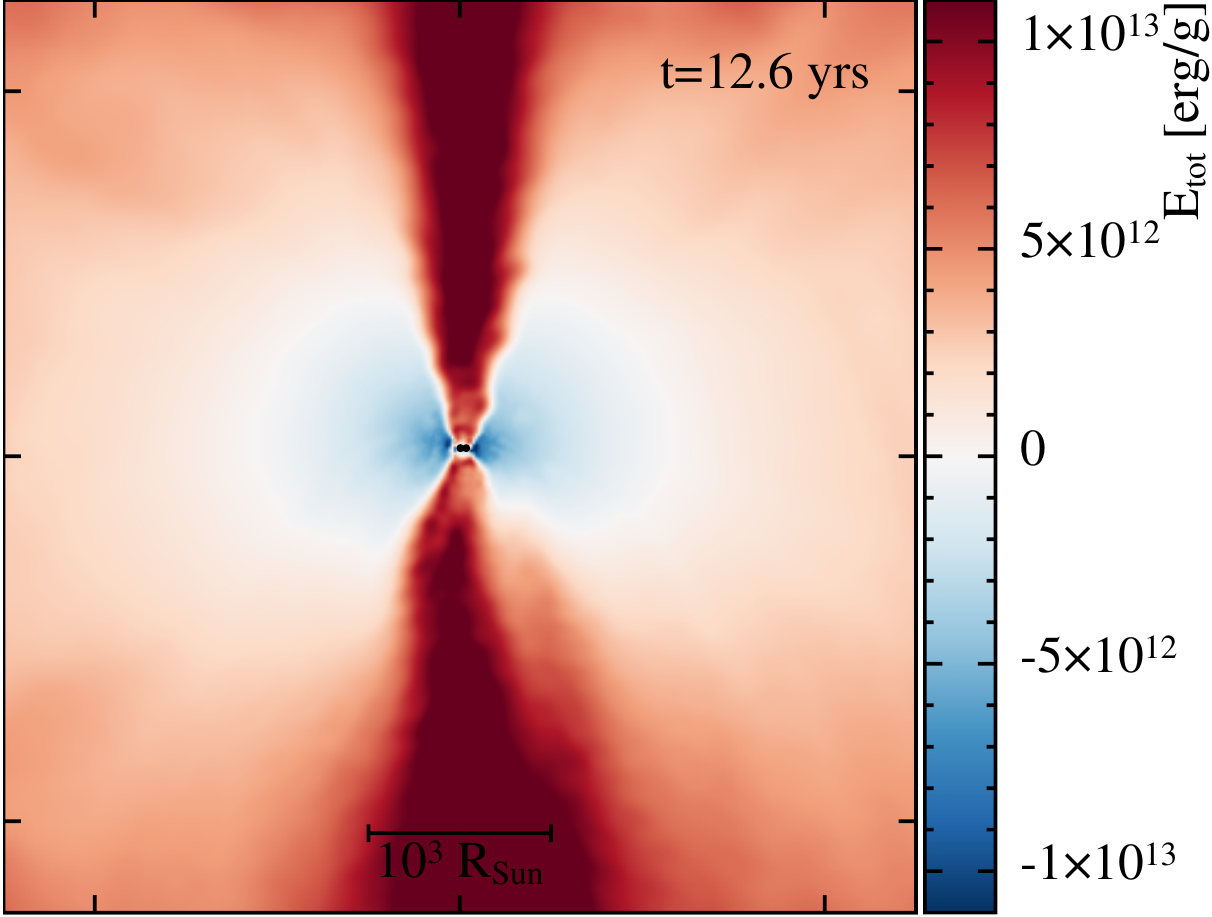}\
    \caption{Slices in the xz-plane for the I-hi (left) and M-hi (right) models at times 27.4 and 12.6 years, when the resolution-dependant unbinding is under way. The rendering correspond to the total specific energy of the gas, defined as the sum of the specific kinetic, potential, and thermal energies, with red and blue colours representing unbound and bound gas, respectively (using the thermal energy criterion).}
    \label{fig:art_ub}
\end{figure*}

We explore further the resolution-dependent unbinding in Figure~\ref{fig:art_ub}, where we show perpendicular slices of total energy (mechanical and thermal energy) at times when I-hi and M-hi display resolution-dependent unbinding. A bi-conic region of unbound gas becomes evident at the approximate time when the phase of additional, resolution-dependent unbinding starts. The geometry of the CE ejection is such that these regions have lower density than the density closer to the equatorial plane. This supports our interpretation that the particles near the core acquire higher kinetic energies but only  those in the lower density regions can  more readily escape as their path is not as impeded by high density overlaying particles.
As reference, the specific potential energy at the surface of the models at time zero is $4.1 \times 10^{12}$~erg~g$^{-1}$. 
The lack of one of the two cones in I-hi (Figure~\ref{fig:art_ub} - left panel) is due to the lack of perfect symmetry in the density distribution.

%\subsection{Resolution comparison} 
%\label{sec4.3:res}
%Ideal gas models, I-lo and I-hi, share almost all the same stabilisation and binary setup parameters. The only main difference is the point mass particle (Table \ref{tab1:table}), we used one with $0.5633$ and $0.5634$ \Msun for I-lo and I-hi, respectively. The reason for this difference is because, at the moment of mapping and stabilisation of the star, the models presented perturbations that propagated to the surface. Mapping the parameter space of both the softening length and the mass of the point particle, we found that these values suited each of the stellar profiles and minimize those perturbations on each case. 

%As we discussed on Section~\ref{sec:star_relaxation}, the expansion rate of I-lo and I-hi are the largest and lowest, respectively (Figure~\ref{fig:PD_rad_fracc}). These rates constrain the expansion of the original \mesa\ stellar evolution, which is convenient since the mass transfer phase has a strong dependance on $\frac{\Delta R}{R_1}^3$  

%\begin{itemize}

%    \item{I-lo vs I-hi CEE timescale comparison. ROLF phase longer at higher resolution, as expected.}
%    \item{Key events occurs with a ~2 yrs shift between simulations. Bound mass, final post-CE system and angular moment are similar, as expected. }
%    \item{"Artificial" unbound is smoother on I-hi}
%\end{itemize}

\subsection{Rotation and angular momentum}
\label{sec4.3:angmom}

The $z$ component of the initial angular momentum of the system with respect to the centre of mass is $9.55\times10^{52}$g~cm$^2$~s$^{-1}$ and it mostly belongs to the orbital angular momentum of the stellar cores. By the end of the interaction, most of the orbital angular momentum is transferred to the envelope, some of which becomes unbound  (Figure~\ref{fig:conservation_angmom}): $41\%$, $41\%$ and $88\%$ of the $z$-angular momentum unbound, for the I-hi, R-hi and M-hi, respectively (using the mechanical energy criterion). 
%These global values are consistent with the unbound mass discussed previously. However, the specific unbound of angular momentum is larger in I-hi and R-hi than in M-hi. 

If the envelope is only partly ejected by the dynamical plunge-in, the remaining angular momentum will play a key role in the adjustment of the remaining envelope into a new equilibrium configuration. Indicatively, if the remaining bound angular momentum at the end of the I-hi simulation (which unbinds only 27-28~per cent of the envelope)
%and M-hi, were evenly distributed in the original \mesa star
%\ryo{Why would you put it back into the original {\mesa} star? The star has lost quite a bit of mass already.} \mike{I don't understand what you mean by evenly distributing the bound angular momentum in the original star; can you clarify?},
were given to the initial I-hi model, and the star were to rotate as a solid body, its surface velocity would be $65$~km~s$^{-1}$, well above the local escape speed. This demonstrates that as CE mergers (that have not ejected a great amount of gas) attempt to readjust thermally, they have to contend with a great deal of angular momentum redistribution which will greatly impact the future structure of the star. %\mike{Can you clarify the point this sentence is making?} \miguel{It is a general estimation on how the remaining angular moment will be adjusted in the post-CE system. Angular momentum analysis is a complicated subject in literature, and in this case is not an exception so, for now, we only state the global values assuming a solid body in rotation, which can easily be compare with, let's say, the initial state of the primary star.}

\begin{figure*}
    \centering
    \includegraphics[width=0.9\linewidth]{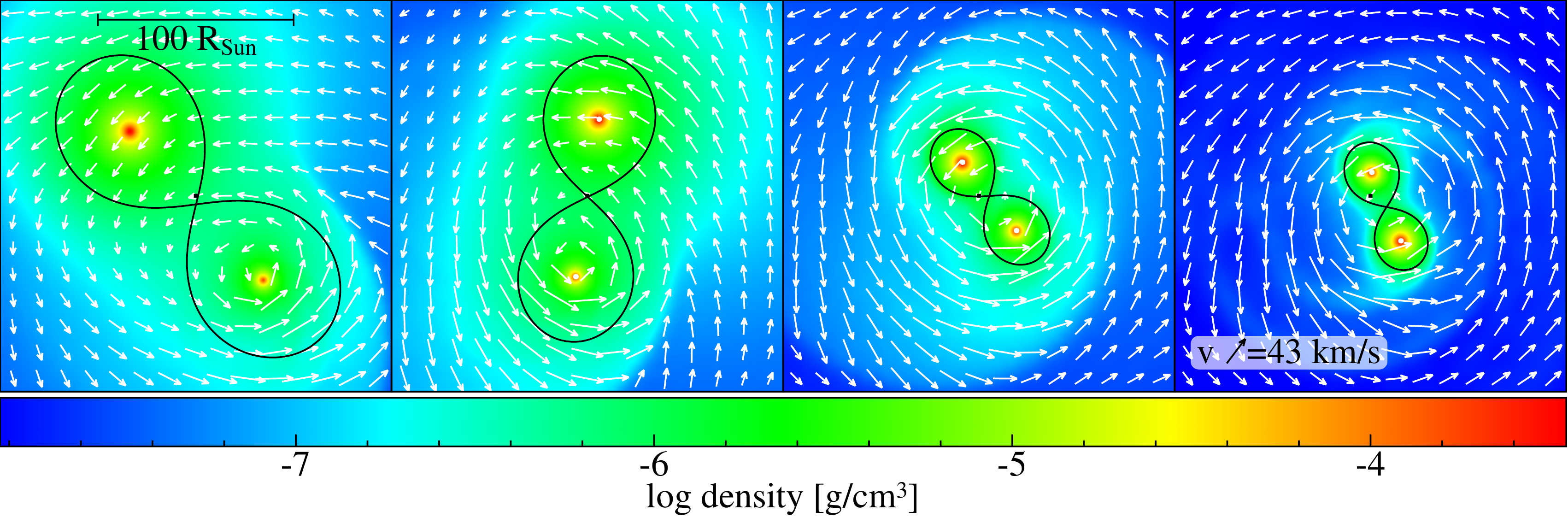}\\
    \includegraphics[width=0.9\linewidth]{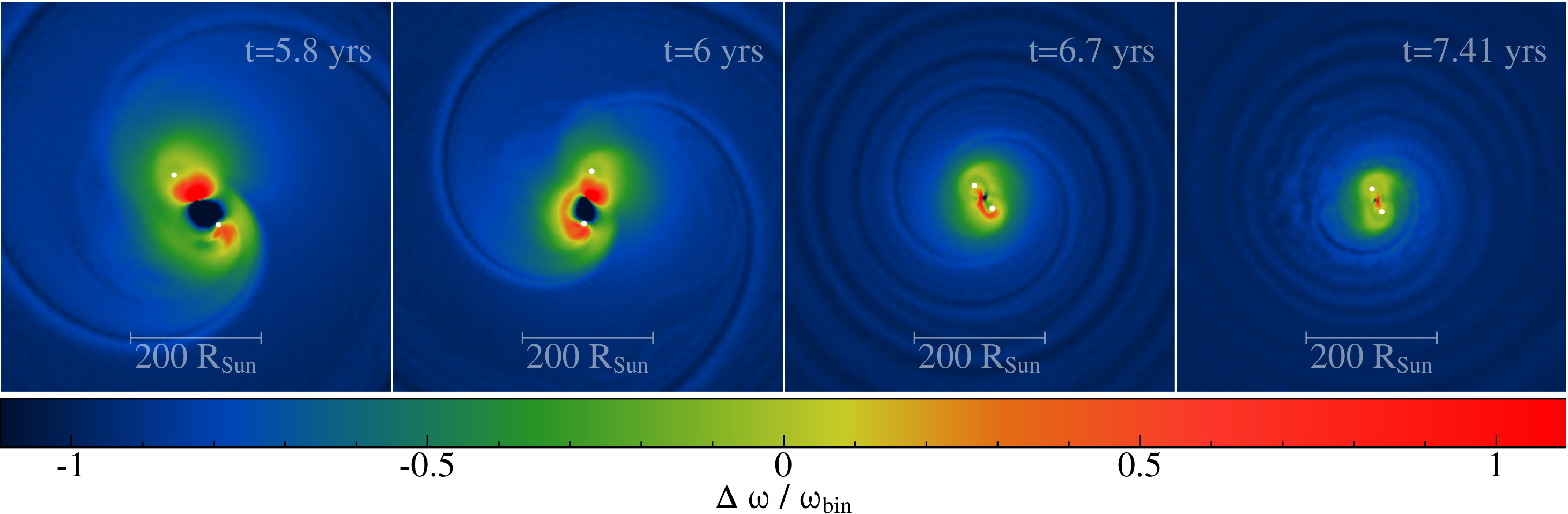}\\
    \caption{Slices of density including velocity arrows (top row) and `corotation' (bottom row) for the M-hi simulation. We show the Roche lobes corresponding only to the point mass particles system in the top panels. Corotation is defined as $\Delta \omega/\omega_{\rm bin}$, where $\Delta \omega \equiv \omega_{\rm bin} - \omega_{\rm gas}$ and $\omega_{\rm bin}$, $\omega_{\rm gas}$ are the angular velocity of the point mass particles and the of the gas, respectively, with respect to the centre of mass of the two point masses, only. The angular velocity, $\omega$, is defined as positive for anti-clockwise motion. The white dots represent the point masses. The times on the top panels correspond to the times labelled on the bottom panels. These times are also marked by magenta symbols in Figure~\ref{fig:sep1}. %\orsola{Miguel, we agreed to add a 4th time that is taken a little later, OR to delete the first column and add a time, to retain only three times.}\miguel{Done}
    }
    \label{fig:render_omega}
\end{figure*}

In Figure ~\ref{fig:render_omega} we analyse the gas density and velocity field (top panels), as well as the angular velocity (bottom panels) for M-hi. In the top panels, the Roche lobes are shown - the primary core is surrounded by higher density. At $t=5.8$ and $6$~years the gas in the primary's Roche lobe moves approximately in the same direction as the orbit as expected, while 
%The mass inside this lobe appears to rotate relative to a location close to the center of mass of the entire system (black square). T
the gas in the secondary's lobe orbits the point mass companion, as also expected after entering the lobe via L1.
%but a large portion is being ejected %\mike{Why is the gas in the secondary Roche lobe rotating w.r.t. the secondary but not for the primary? And how does one see that a large amount of this gas is leaving the secondary Roche lobe, as you suggest?}. \miguel{For question 1: The gas in the primary lobe moves along primary sink because this was originally on equilibrium with it. This system was never in co-rotation, therefore the gas does not rotate w.r.t the primary. On the other hand, when the gas enters the secondary lobe, it enters through L1 and rotates due to the angular momentum it has w.r.t. to the companion. For question 2: Figure \ref{fig:new_UB} has a small unbound region around 6yrs and $10^3$\Rsun that becomes thicker at 7 yrs and $10^3.2$\Rsun. This region contains a parcel of extremely unbound gas that was ejected in the first three panels of Figure \ref{fig:render_omega}.} 
%this time the binary is still deep in the plunge-in phase and the ejected gas takes away the angular momentum of the secondary point mass, reducing further the binary orbit. 
The corresponding bottom panels show an angular velocity contrast pattern that is in line with what we see in the top panels.
%high contrast in the angular velocity between the orbit, due to the roughly uniform flow near the rotation reference frame %\mike{I don't really understand this sentence...}\miguel{If the gas goes in the same direction near the rotation reference point, you will see in the $\omega$ render plot two very negative and very positive regions, since they both rotate in the opposite direction w.r.t this reference point.}. 
At $6.7$~years, right at the end of the plunge-in, the region within and around the lobes is rotating with the orbit and is close to corotation (green colour). Any remaining patterns of differential rotation are quickly disappearing.
%The center of rotation of the secondary lobe is shifting towards L1 and the bulk velocity inside the primary lobe is becoming similar to the one in the secondary %\mike{What do you mean by bulk velocity?}\miguel{The average flow velocity in that region.}. 
%The bottom central panels have a region of high angular velocity connecting both lobes, which is consistent with the shifting center of rotation in the top panels 
%\mike{I don't quite understand this sentence.}\miguel{When there is a red region connecting the lobes, as opposed to the previous panels, it means that there is a parcel of gas rotating with the same omega. This coincides with the flux of gas rotating together w.r.t. to a point near the center of mass of the whole system in the upper panels. Next sentence is new and illustrates this comment}. 
%This indicates that the gas in the central region is now rotating with the respect of a local center of mass, closed in distance but not necessarily the center of mass of the entire binary. 
By $7.41$ years the gas in and around the orbit is mostly in corotation. We conclude that the end of the in-spiral in these simulations is primarily the result of the inner gaseous region being brought into corotation by the binary cores.

\subsection{The morphology of the ejected envelope}
\label{sec4.5:morph}
\begin{figure*}
     \centering
     \begin{subfigure}[b]{\textwidth}
         \centering
         \includegraphics[width=0.88\textwidth]{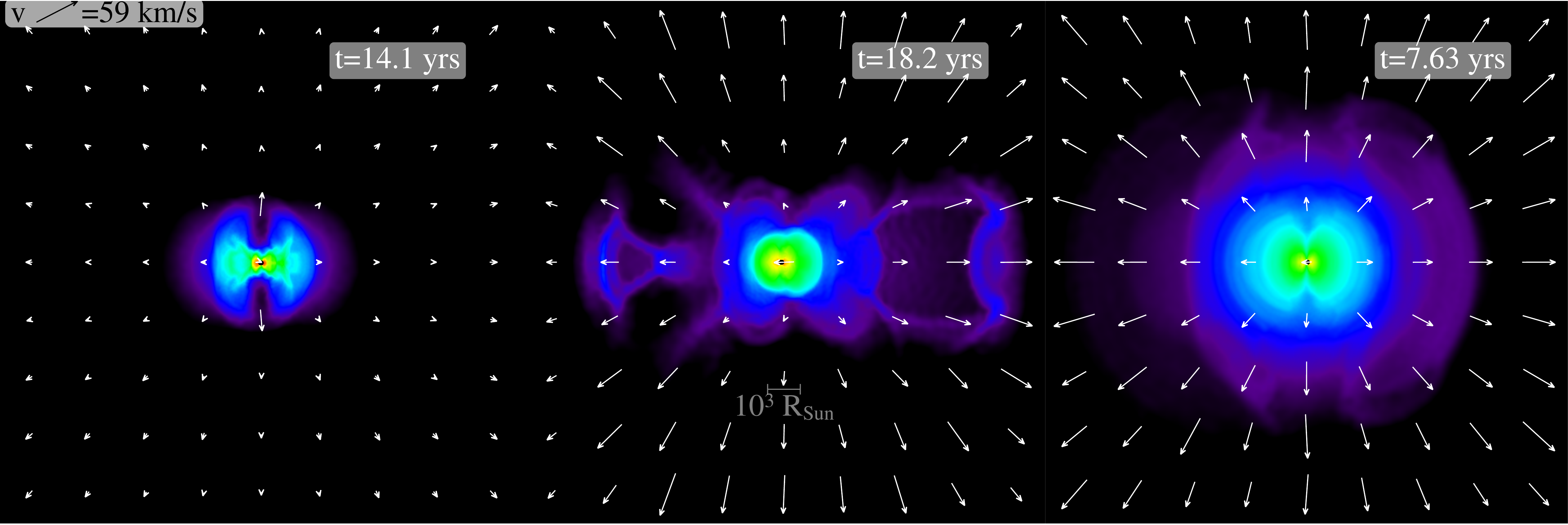}\\
         \includegraphics[width=0.88\textwidth]{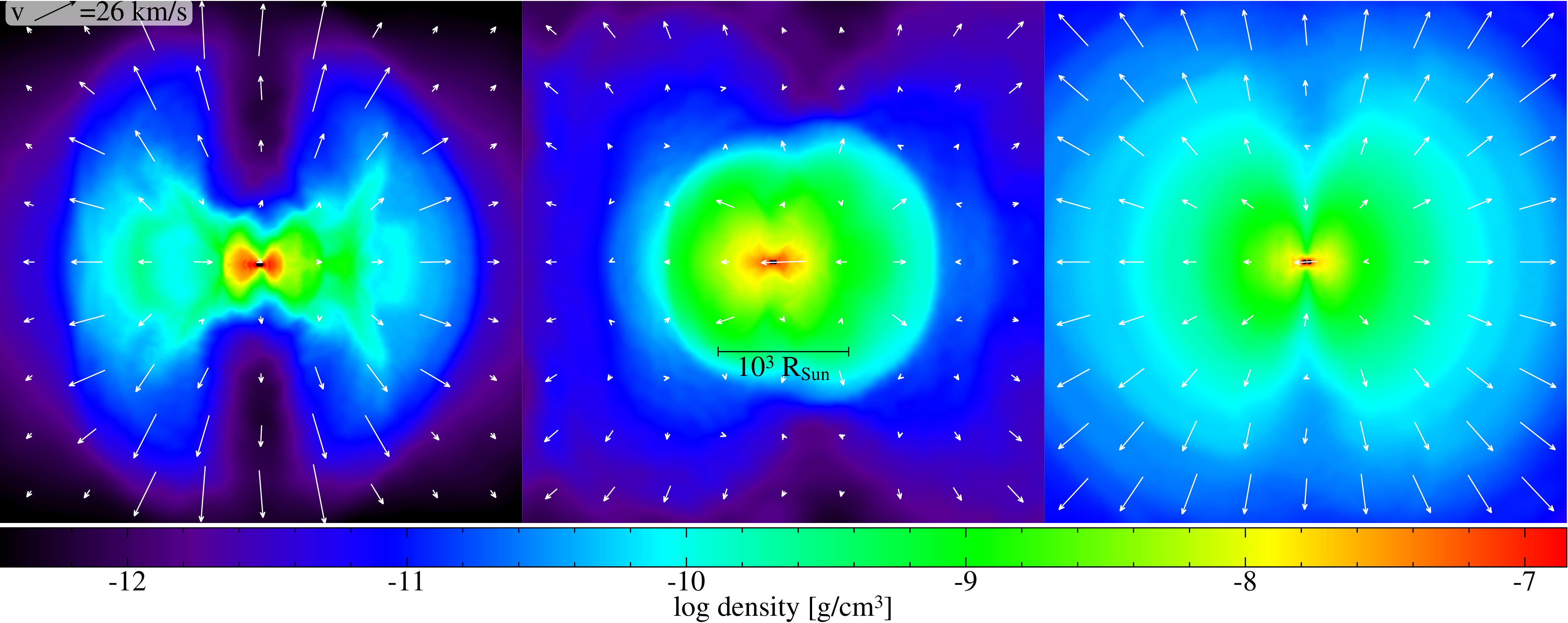}\\
     \end{subfigure}
     \vspace{0.3in}
     \begin{subfigure}[b]{\textwidth}
         \centering
         \includegraphics[width=0.88\textwidth]{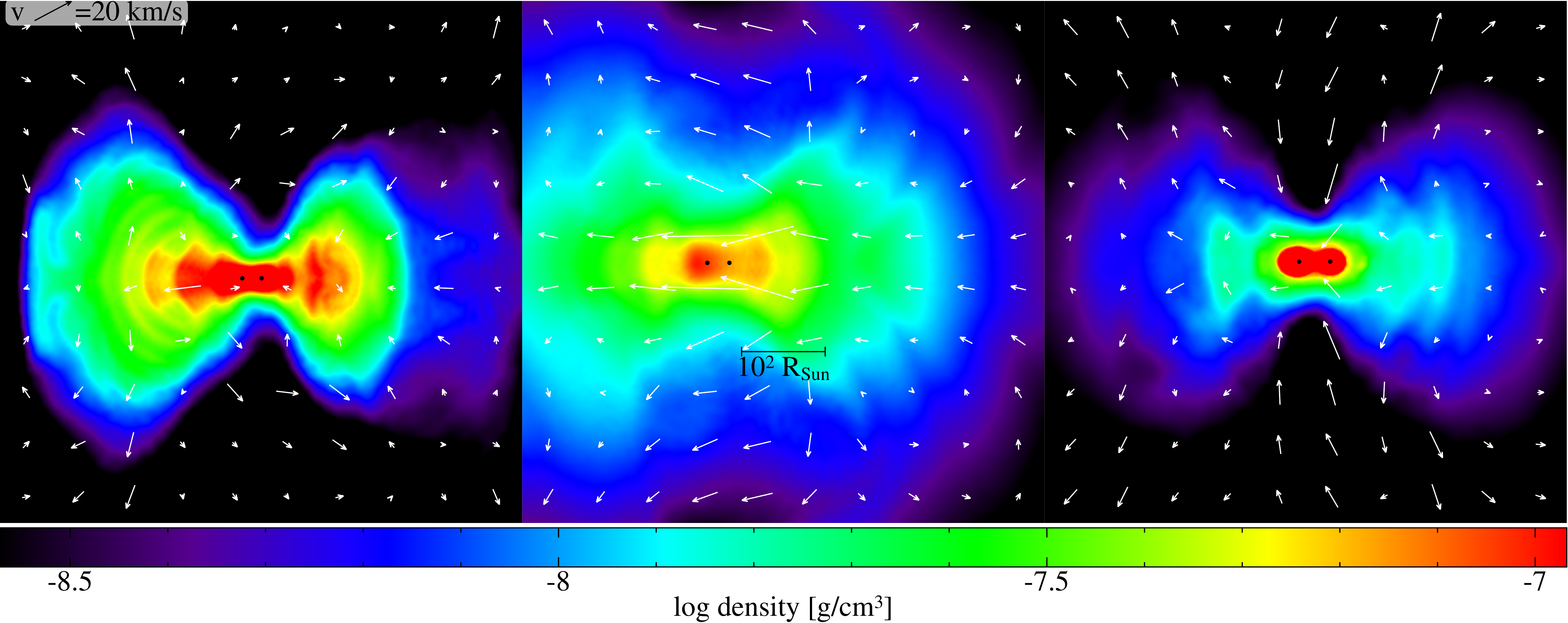}
         \end{subfigure}
\caption{Density slices of Rei19 (left column), I-hi (central column) and M-hi (right column) at three zoom levels in the x-z plane: top and middle rows show two zoom levels using the same density colour scale, while the bottom row zooms in further using a colour scale that shows only the densest parts of the structure. The first row has a box size of $1.6 \times 10^4$ \Rsun\  per side. The second row has sides of 4000\Rsun on each box. The bottom row is has dimensions of $625\time450$ \Rsun.}
\label{fig:PN}
\end{figure*}

Approximately 20~per cent of all planetary nebulae (PNe) have a surviving post CE binary in the middle \citep[e.g.,][]{Jacoby2021}, indicating that the nebulae are the ejected envelopes. The great range of post-CE PN shapes is in stark contrast to the relative homogeneity of the post-CE binaries at the centres of the nebulae. This suggests that \review{post-CE PN morphology is very sensitive to small changes in stellar and binary parameters.}

\citet{GarciaSegura2018}, \citet{Frank2018} and \citet{Zou2020} were first to simulate PN formation in the aftermath of a CE interactions. However, they used, as initial conditions, CE simulations with RGB, instead of AGB stars; in addition, those CE simulations had only partly ejected their envelopes. One of the aims of this study is to generate an AGB CE simulation that fully ejects the envelope to be used as input for a new PN simulation. Here, we therefore compare the I-hi and M-hi simulations' gas distributions with that of the simulation of \citet{Reichardt2019} that was already used to simulate the PN formation by \citet{Frank2018} and \citet{Zou2020}. We use this comparison to predict whether our new simulation is expected to result in substantially different nebular shapes. 

%The importance of studying post common envelope systems cannot be underestimated. Depending on the energy released during the interaction we can end up having a merger of the stellar cores or a close binary stars. This task becomes more complicated when we want to compare post-CE systems with widely different initial conditions; therefore any contrast method has to take this into consideration. In this section we propose a system to compare \cite{Reichardt2019} common envelope simulation, particularly the time frame used by \cite{Zou2020} to analyse dusty winds on planetary nebulae implementing the grid code \astrobear.

We start by selecting a post-CE time for I-hi and M-hi at which we can compare the CE structure with the simulation of \citet{Reichardt2019} adopted by \citet{Zou2020}. 
%we need to select a time that is dynamically equivalent to the age at which the simulation of \citet{Reichardt2019} was used as input for the nebular analysis. We state that 
We do not know the time after the CE ejection when the fast, post-AGB wind that ploughs into the CE gas should start blowing. We can presume that the fast wind will start when the phtospheric temperature reaches a certain value ($\sim$20\,000~K). This time, in turn, depends on the adjustment of the core once the envelope has departed. None of these processes are accurately modelled by the CE simulation. \citet{Zou2020} therefore used a CE simulation that was stopped about one year after the time of maximum in-spiral. 
%Absolute physical time in each of the three simulations is not comparable. First, the three simulations are started with slightly different degrees of Roche lobe filling and have therefore different pre in-spiral times. Second, the RGB and AGB stars have different dynamical times and therefore different inherent timescales. 
%We therefore, s

Somewhat arbitrarily, we select the dynamical time of the primary star prior to the start of the CE interaction as a ``yardstick''. We then assume that the CE ejection takes place at the time of maximum in-spiral and call that time-zero: $t_0=t_{\rm max, pl}$ (Table~\ref{tab3:table}). The \citet{Reichardt2019}  simulation was used to generate the PN at 14.1~years of simulation time and 0.8~years after $t_0$. With a dynamical time for that star of $\tau_{\rm dyn, RGB}$ = 11~days, the simulation was stopped $26 \times \tau_{\rm dyn, RGB}$ after $t_0$. 
Using the same factor, together with the AGB star dynamical time of $\tau_{\rm dyn, AGB}=41$~days, we stop the simulation for the purpouse of comparison, 2.9~years after $t_0$, or $18.2$ and $7.58$~yr from the start of the I-hi and M-hi simulations, respectively.
These times are marked with square symbols in Figures~\ref{fig:sep1} (right panel) and \ref{fig:bound_mass}.

We finally compare the three simulations in Figure~\ref{fig:PN}. We observe that our AGB CE simulations lead to a somewhat more extended and more spherical post-CE gas distribution (top row). 

At higher magnification (Figure~\ref{fig:PN}, middle row) all three simulations show a peanut shaped overdensity, with the simulation of \citet{Reichardt2019} being the most pronounced. This shape leaves a reasonably evacuated corridor in the vertical direction, which is responsible for hydrodynamically collimating the fast wind \citep{Zou2020}. In Figure~\ref{fig:PN}, bottom row, we zoom further into the centre of the gas distribution and change the density colour bar to highlight the structure there. Clearly, at the highest densities ($\log \rho /{\rm g~cm}^{-3} = 7-7.3$) the M-hi is far more spherical than the two ideal gas simulations. 

The velocity field shows some striking differences between the three simulations. Overall, the M-hi simulation seems to show faster outflow velocities with an inflow at the poles at smaller scales (Figure~\ref{fig:PN}, lower row). The RGB simulation of \citet{Reichardt2019} has fast outflow velocities (including at the poles) only at intermediate scales (Figure~\ref{fig:PN}, middle row), while the I-hi AGB simulation has outflow velocities comparable with M-hi at large scale, but at smaller scale the velocity field is slow and confused.  Overall, the simulation that unbinds the envelope  (M-hi) \review{appears to have a faster outflow}.

In conclusion \review{we propose that M-hi and other simulations like it, that include recombination energy, are the most realistic and hence the most appropriate simulations from which to derive PN shapes. It will be then interesting to see whether the strong collimation observed by \citet{Zou2020} is still a feature, something that we leave for future work.}

\subsection{Dust and the expanding common envelope}
\label{ssec:dust}

The dust formation that is likely to happen in the dense, yet expanding and cooling common envelope has never been calculated consistently. \citet{Iaconi2019} and \citet{Iaconi2020} have, however, post-processed the dust formation in the RGB, CE simulations of \citet{Reichardt2019} and found that a CE event like that would produce approximately 10$^{-3}$~\Msun, of dust independently of whether the dust is carbon- or oxygen-rich.

\begin{figure*}
    \centering
    \includegraphics[width=0.65\linewidth]{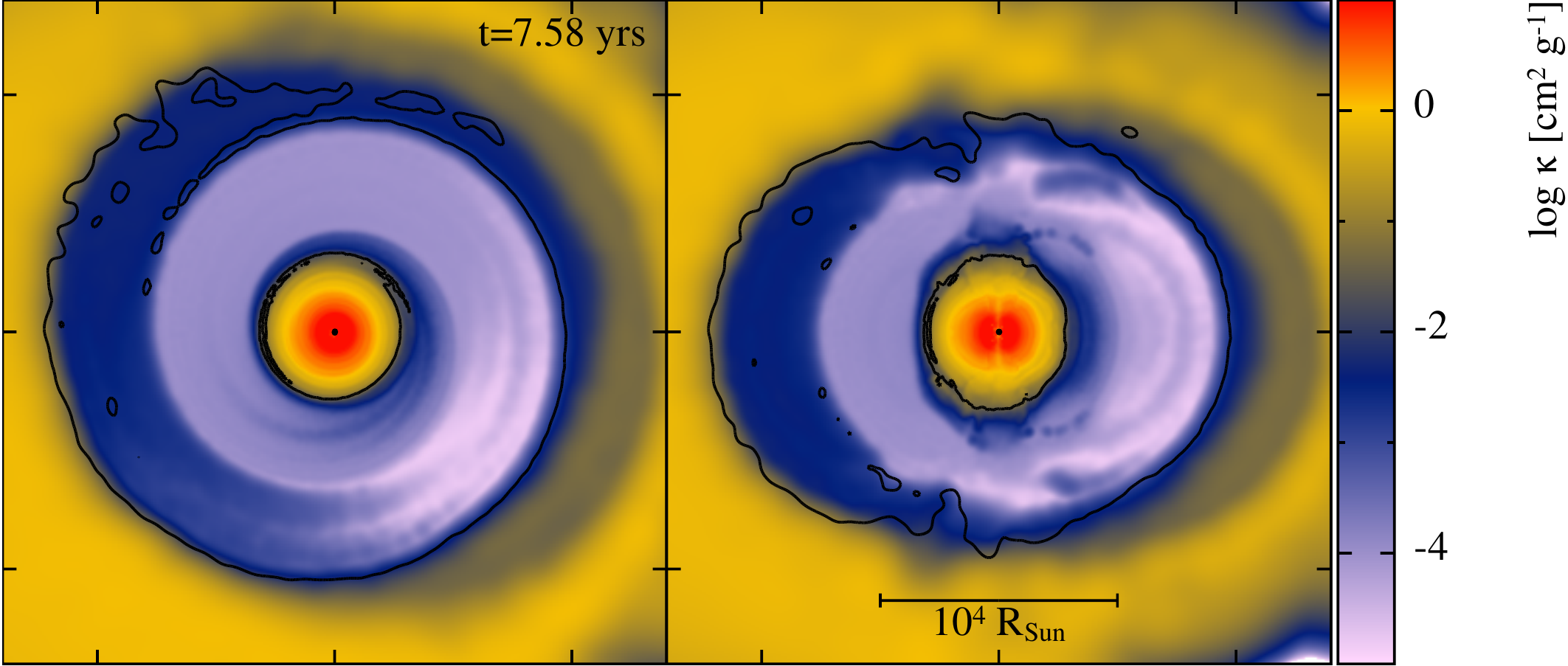}\\
    \includegraphics[width=0.65\linewidth]{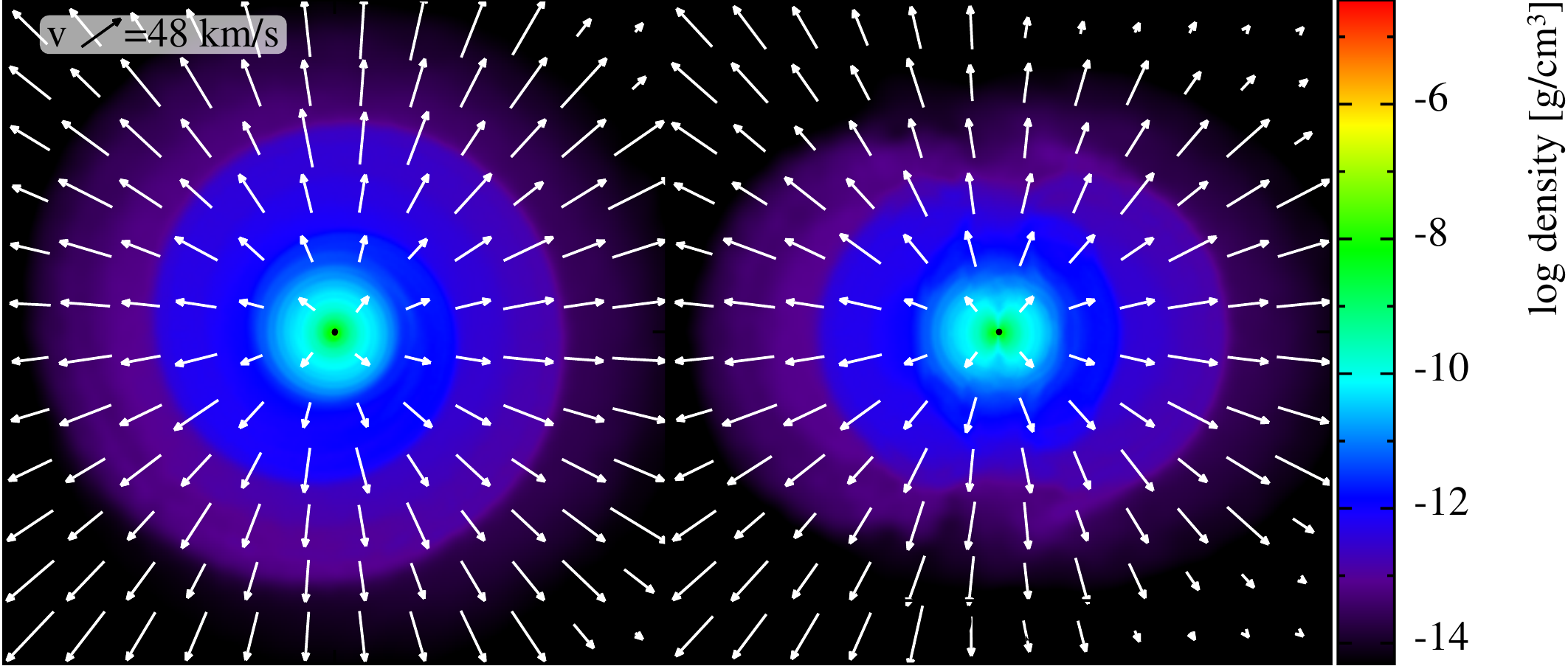}
    \caption{Top row: opacity slices in the orbital (left panel) and perpendicular (right panel) directions for the simulation M-hi. Bottom row: equivalent slices of density with, overlaid, projected velocity arrows.}
    \label{fig:kappa}
\end{figure*}

For AGB donor stars, the formation of dust in the expanding CE is likely to play an even greater role than for RGB giants, impacting the objects' optical properties as well as the dynamics of the expansion. In our simulations we have not considered dust formation. However, we can infer where the dust particles would likely form based on the gas temperature and density from which our tabulated equation of state tables provide knowledge of the expected opacity. In Figure~\ref{fig:kappa} we show horizontal and vertical slices of the opacity for M-hi at $t=7.58$ years. This figure is similar to one by \citet[][figure 10]{Reichardt2020} for their RGB star. The two black contours represent the recombination temperature for hydrogen ($6000$~K, inner contour) and an approximate temperature for dust condensation ($1500$~K, outer contour). Inside the $6000$~K contour, hydrogen is ionised and the opacity is high and dominated by electron scattering. Outside of the $T=6000$~K region, hydrogen is recombined and the medium becomes transparent. However, as we move outwards, the temperature decreases and when it reaches $\sim$1500~K, molecular and dust opacity contribute to increasing the overall opacity once again.

Finally we observe that in the 0.88~\Msun\ RGB star of \citet{Reichardt2019}, the high opacity shell is concentrated around the equator, with lower opacity along the polar direction. \review{On the other hand}, their 2~\Msun\ RGB giant model shows higher opacity in the direction of the polar caps, with lower opacity at the equator. In our case, M-hi shows a thick shell of higher opacity all around the star, with a somewhat oblate shape similar to the 1-\Msun\ simulation of \citet{Reichardt2019}.  In future work we will calculate the dust mass associated with this CE and infer the dust driving properties of the dusty envelope.

\section{Discussion}
\label{sec5:discussion}

We have presented four simulations of a common envelope interaction using a TP-AGB star donor. We have used the 1D, stellar structure and evolution code \mesa\ to calculate a 2 \Msun, ZAMS star that we have evolved until the seventh AGB thermal pulse, when it had a radius of 260~\Rsun\ and a mass of 1.7~\Msun. We have implemented two different relaxation procedures, one used by \cite{Ohlmann2017} and the other designed by \cite{Lau2022}, to map and relax the stellar model in the 3D smoothed particle hydrodynamics (SPH) code \phant. \review{The relaxed star was set up} with a $0.6$-\Msun\ companion at an orbital distance of 550~\Rsun, such that the primary was Roche lobe-filling. Moreover, we have simulated the system using three different equations of state: ideal gas, ideal gas plus radiation and a tabulated form that takes recombination energy into account. All simulations were run with a resolution of $1.3\times10^6$ SPH particles. For the ideal gas model, we also run a simulation at low resolution, with $1.37\times10^5$ SPH particles.

The final orbital separation ranges between $20$ (I-hi) and $31$~\Rsun\ (M-hi), depending on the equation of state. The adoption of a tabulated equation of state results in a final separation that is $55$~per cent larger than for the simulation with an ideal gas equation of state.  \citet{Reichardt2020} found almost no difference in the final separations between the two equations of state for their $0.88$~\Msun, RGB simulation and for their $2$~\Msun, RGB simulation. \citet{Sand2020} found instead an increase between $11$ and 25~per cent (depending on companion mass) in final separation when using a tabulated equation of state in their 1.2~\Msun\ AGB star. 

In addition to the final separation, the shape of the inspiral curve in the separation vs. time plot (Figure~\ref{fig:sep1}), changes dramatically between the simulations I-hi and M-hi: not only does the M-hi simulation in-spiral sooner, but the slope of the in-spiral is overall less steep. \citet{Reichardt2020} observed virtually no difference in the in-spiral shape between ideal gas and tabulated equation of state simulations. \cite{Sand2020}, on the other hand, found that the early in-spiral is similar for the two equations of state, but that the late in-spiral is different, with the tabulated equation of state resulting in a less steep curve, as we also find.

We interpret these differences as due to early expansion of the stellar structure because of a slight lack of hydrostatic equilibrium, as explained in Section~\ref{sec4.2:bound} {\it and} to the recombination energy delivered upon early expansion aiding that expansion. Our starting the simulations from the time of Roche lobe overflow, affords more time before the in-spiral than for the simulations of \citet{Reichardt2020}, which were started on the surface. During this time the M-hi star ``expresses" its tabulated equation of state by enacting additional expansion driven by the outer layers recombining (Section~\ref{sec4.2:bound} and Figure~\ref{fig:new_UB}). This leads to a relatively faster in-spiral on-set, compared to I-hi. As the companion in-spirals and further recombination energy is delivered, a gentler inspiral slope is observed, as the densities encountered are lower than for I-hi. Finally, we see a wider separation, as the star has been effectively ``helping" with its own unbinding.

\citet{Sand2020} started their simulations with the companion farther away from the primary surface, but closer than Roche lobe overflow, such that they  observed more similarity in the in-spiral between ideal and tabulated equations of state, but the shape of the in-spiral diverged later on with the tabulated equation of state giving rise to a gentler slope, in a manner not dissimilar to us\footnote{This is generally true for their 6 simulations, but there is some variability across the three companion masses that complicated the analysis further.}.
The simulations of \citet{Reichardt2019} were instead started with the companions on the surface of the stellar structures, such that the in-spiral was initiated immediately, giving no time to the star to expand and for the recombination energy to play a role. Their envelope ejection took place afterwards. This likely explained the similarity of their in-spirals and final separations. 
{\it We conclude preliminarily that the equation of state does play a role on the envelope dynamics and even on the final separation}, contrary to what was concluded by \citet{Reichardt2020}. 

Our final orbital separations (20-31~\Rsun) are wider than encountered in the bulk of observed post-CE central binaries in PN \citep{Iaconi2019b}. This is so also for the 1~\Msun, AGB star simulations of \citet[][22-69~\Rsun, depending on the companion's mass]{Sand2020} and even for the 1.78~\Msun, AGB star simulation with a 0.98~\Msun\ companion by \citet[][15.5~\Rsun, and we note they used an ideal gas equation of state, suggesting that the separations would be even larger had they adopted a tabulated equation of state]{Chamandy2020}. 

The in-spiral itself is very brief, taking place on the stellar dynamical timescale and about an order of magnitude shorter than the thermal pulse. However, we could ask whether the period of Roche lobe overflow preceding the in-spiral, which could be longer than we have modelled, may be long enough that an in-spiral may not happen at all before the star contracts at the end of the thermal pulse. We propose here that the information gathered so far indicates that once a thermal pulse leads to Roche lobe overflow the expanding envelope of the plusating star will force a CE in-spiral. The natural expansion of the thermally pulsating AGB star radius is not dissimilar to our gently expanding structures (that expand due to a slight violation of hydrostatic equilibrium; Figure~\ref{fig:PD_rad_fracc}). Additional simulations with increasing accuracy of the stellar structure will be needed to finalise this conclusion.

In Figure~\ref{fig:boundsep}, it is clear that even adopting the most stringent, mechanical criterion, M-hi unbinds a great deal more envelope than I-hi or R-hi. Including thermal energy in the criterion, results in an even greater amount of unbound gas. On the other hand, the inclusion of thermal energy makes almost no difference to the unbound mass calculated for I-hi. \review{This is likely because there is much less marginally bound material in the I-hi model that would be marked as unbound with the addition of a small amount of thermal energy}.
The overall trend in the bound gas curve over time for M-hi indicates that the entire envelope could easily be unbound in a short time. Interestingly, this trend is very different to what was observed by \citet{Reichardt2020} in their RGB star simulations: their 0.88-\Msun\ simulation with ideal gas and tabulated equation of state behaves similarly to ours. However, their 1.8~\Msun\ simulation unbound very little gas, even when using a tabulated equation of state showing that their recombination energy had not been fully released by the end of that simulation. 

There is relatively little difference between I-hi and R-hi, where the latter includes radiation energy. While this is not unexpected for low mass giants, we note here some subtle differences. R-hi starts to in-spiral sooner than I-hi, likely due to the same reasons argued above, namely the initial model was ever so sightly more out of hydrostatic equilibrium than I-hi. The shape of the in-spirals, however, are very similar. The separation at the end of the plunge-in, $a_{\rm f, pl}$, and the separation at the end of the simulations, $a_{\rm f}$, are both larger for R-hi (44 and 27~\Rsun) than for I-hi (32 and 20~\Rsun) and take place 1.2 years after the time of maximum plunge in, $t_{\rm max, pl}$, compared to 1.5~years for I-hi. The unbound mass at the end of the in-spiral (which excludes the resolution-dependent unbinding at the end) is almost identical for I-hi and R-hi.

Towards the end of all simulations, but particularly in I-lo/I-hi and R-hi, there appears to be a second envelope unbinding phase (Figure~\ref{fig:bound_mass}), no matter what criterion is used (Figure~\ref{fig:new_UB}).
\citet{Reichardt2019} described this resolution-dependent unbinding as caused by the decrease in density (and concomitant increase in the local SPH particles smoothing lengths) in the proximity of the core. When the local SPH particles no longer ``resolve" the core, the pressure gradient inverts and they evacuate the core region. When they confer their kinetic energy to surrounding particles, these particles may become unbound.
Figure~\ref{fig:art_ub} indicates that this late, resolution-dependent ejection occurs in the direction of z-axis. This zone has a lower density and would facilitate the escape of particles.     
For M-hi, this phase of resolution-dependent unbinding is less obvious from Figure~\ref{fig:bound_mass}, but we determined that it does happen at about 10~years from the start of the simulation. While this reduces the fraction of unbound envelope that we can measure with certainty, the gas unbinding ratio observed in Figure~\ref{fig:bound_mass} for M-hi still indicates that the envelope will soon become fully unbound: \review{$15.6$ and $16.7$ years with the final mass-loss rate for mechanical and thermal criteria, respectively}.
%\ryo{"mass loss ratio" -> do you mean, "unbound mass fraction"?}

Morphologically and kinematically, the ejected CE of this AGB star is \review{somewhat} different from that of the RGB star of \citet[][Figure~\ref{fig:PN}]{Reichardt2019}. The innermost structure of our M-hi simulation is \review{less elongated} than the equivalent region in the RGB simulation of \citet{Reichardt2019} - this inner mass distribution influences greatly the nascent PN. In future work we will use the ejected CE of M-hi to simulate the PN and will further analyse differences at that time. The opacity within the envelope also indicates that dust formation would take place. This would likely result in radiation driving and unbinding further envelope gas.  

Finally, some considerations on the common envelope unbinding efficiency. Following \citet{Lau2022} we can determine the common envelope efficiency parameter based on our three high resolution simulations, which have individual values for the orbital energy delivered, $\Delta E_{\rm orb}$ \review{(where we use the average of the apastron and periastron fitted values from Table~\ref{tab3:table} to calculate it). We also use a definition of the binding energy that includes thermal energy. If thermal energy were omitted, the values of $\alpha$ would increase by a factor of two.} 
%For each of the three simulations, we use two different definitions for the binding energy: $E_{\rm bin} = E_{\rm grav}$, or $E_{\rm bin} = E_{\rm grav} + E_{\rm th}$. 
In Table~\ref{tab:alpha} we list
%therefore list $\alpha_{\rm grav}$ and $\alpha_{\rm th}$  
$\alpha$ 
%for each of the two binding energy definitions and 
for each of the three high resolution simulations. The values of $\alpha$ larger than unity tell us that the relatively meager orbital shrinkage \review{in the simulations} did not deliver enough energy to unbind the envelope, according to the two binding energy definitions. \review{The reason why the in-spiral stalls before delivering sufficient energy to unbind the envelope is that our simulations also contain recombination energy, which aids the envelope ejection and which is not accounted for in the formalism.}

\citet{Lau2022} used \review{three} definitions of the binding energy. \review{One without thermal energy, only includes gravitational potential energy. The second, is the same we have used, while the third includes recombination energy}
%\ryo{Again, "potential energy" should be replaced with "binding energy".} 
($E_{\rm bin} = E_{\rm grav} + E_{\rm th} + E_{\rm rec}$).
%, or in other words, a binding energy definition that includes gravitational potential energy, thermal and recombination energies.
\review{Adopting a binding energy definition that excludes thermal energy would increase $\alpha$ by a factor of two. We cannot use the third definition here: the inclusion of recombination energy in the definition of binding energy leads to a formally unbound envelope and a negative value of $\alpha$.} The negative value of $\alpha$ simply states that given that the binding energy including recombination is negative (the envelope is formally unbound) the orbit needs to expand instead of contracting. Clearly this means that the alpha formalism does not work here. The reason is that while the envelope might be formally unbound at time zero, when the in-spiral commences, the thermal and recombination energies have not yet turned into bulk kinetic energy so the companion can still feel a drag force, spiral in, and deliver orbital energy. This formalism does not include this time sequence and fails.

\begin{table}
    \centering
    \begin{tabular}{lc}
    \hline
          %Sim (EoS) & $\alpha_{\rm grav}$&$\alpha_{\rm th}$&$\alpha_{\rm int}$\\
          %Sim (EoS) & $\alpha_{\rm grav}$&$\alpha_{\rm th}$\\
          Sim (EoS) & $\alpha$\\
          \hline
         %I-hi (Ideal)   & 2.08 & 1.08 & --0.037 \\
         %R-hi (Gas+rad) & 2.94 & 1.52 & --0.052\\
         %M-hi (Full)    & 3.47 & 1.79 & --0.061\\
         %I-hi (Ideal)   & 2.08 & 1.08  \\
         %R-hi (Gas+rad) & 2.94 & 1.52 \\
         %-hi (Full)    & 3.47 & 1.79 \\
         I-hi (Ideal)   &  1.03  \\
         R-hi (Gas+rad) &  1.24 \\
         M-hi (Full)    &  1.76 \\
         \hline
    \end{tabular}
    \caption{The common envelope efficiency parameters derived from the three high resolution simulations. \review{The binding energy includes thermal energy. The delivered orbital energy uses the mean of periastron and apastron final orbital separations as listed in Table~\ref{tab3:table}. }
    %using three definitions of the envelope binding energy that include gravitational energy only ($\alpha_{\rm grav}$), or that include gravitational and thermal energy ($\alpha_{\rm th}$).
    }
    \label{tab:alpha}
\end{table}

\section{Conclusions}
\label{sec6:conclusions}
We simulated the common envelope interaction between a 1.7~\Msun, thermally pulsating AGB star with a radius of 250~\Rsun, and a 0.6-\Msun\ companion started approximately at the time of Roche lobe overflow using an ideal gas, ideal gas plus radiation and tabulated equations of state. Our conclusions are:

\begin{enumerate}
    \item The thermalisation of recombination energy released during common envelope expansion is sufficient to unbind most of the envelope. \review{This is consistent with previous simulations that include recombination energy \citep{Nandez2016,Sand2020,Ivanova2016,Lau2022}. We assume the gas opacity to be sufficiently high to absorb the photons emitted from the recombination process before they can escape the envelope. While this may not be true for all of the hydrogen recombination photons (albeit some fraction of them may be thermalised), there is reason to believe that the entirety of helium recombination energy can be transferred to the envelope through this process \citep{Ivanova2018,Lau2022b}}.
    
    \item The inclusion of recombination energy leads to a dynamical, but shallower in-spiral and a wider final separation, \review{by about 58~per cent (or 66~per cent using the lower limit extrapolated values). This is a similar trend to that obtained by \citet{Sand2020}, who modelled a low mass AGB star and three companions, where the final separations were larger by between 20 and 30 per cent with a tabulated EoS, and by \citet{Lau2022} who modelled a 12~\Msun\ red supergiant and obtained separations that were wider by 16\% when recombination energy was included. \citet{Reichardt2020} modelled two RGB donor stars and obtained separations that were wider for their lower binding energy star (by 16 per cent), but \review{ever so slightly} smaller for their more bound RGB star (by 4 per cent). 
    Clearly this conclusion does depend on the specific parameters of the interaction, but the reason for a wider separation when recombination energy is included is likely to be that recombination energy expands the stellar envelope more and earlier than is the case when using an ideal gas EoS, which contributes to less in-spiral.} 
    \item Thermal pulses are likely to trigger a common envelope in-spiral, so long that the primary giant overfills its Roche lobe when the star is in the expanding phase of the thermal pulse. \review{Doubling the resolution just increases the RLOF timescale by 18~per cent or by 2 years. While a third simulation with another increment in resolution is needed to do a proper convergence test, the RLOF timescale is still under one order of magnitude smaller that the duration of the pulse. In addition, in Nature, the star is expanding between 250 and 300~\Rsun\ in less than 100 years, forcing a CE even if, under a constant radius, the stable mass transfer would last longer.} {\it We conclude that thermal pulses increase the ability of AGB stars to capture companions into common envelopes}, thereby increasing the relative number of post-AGB, post-CE systems relative to a situation where the thermal pulse does not cause a CE to take place.
    \item \review{At the large scale,} the CE shape shortly after the in-spiral is \review{somewhat} more spherically distributed for AGB CEs than for RGB CEs and even more so for AGB CEs calculated with the inclusion of recombination energy.
    \item While formally calculating the CE efficiency fails, we note that the wide final separations obtained in our AGB CE simulation are in disagreement with the smaller separations seen in observations \citep{Iaconi2019b}. This leaves wide open the  question of how AGB CE result in final separations of the order of a few solar radii.

\end{enumerate}

\section*{Acknowledgements}
M. G. B. acknowledges funding support from Macquarie University through the International Macquarie University Research Excellence Scholarship (‘iMQRES’) and to Ilya Mandel for his insightful commentaries during the course of this research. OD acknowledges funding from Australian Research Council Discovery Project, DP210101094. The authors thank to the Referee for the constructive comments and recommendations which helped to improve the quality of the paper. Parts of this research work were performed on the Gadi supercomputer of the National Computational Infrastructure (NCI), which is supported by the Australian Government. This research has made use of the NASA’s Astrophysics Data System (ADS) and NASA Exoplanet Archive, which is operated by the California Institute of Technology, under contract with the National Aeronautics and Space Administration under the Exoplanet Exploration Program. 

This is a pre-copyedited, author-produced PDF of an article accepted for publication in Monthly Notices of the Royal Astronomical Society following peer review. The version of record is available online at: \url{https://doi.org/10.1093/mnras/stac2301}.

\section*{Data availability}
The data underlying this article will be shared on reasonable request to the corresponding author.

\bibliographystyle{mnras}
\bibliography{references,bibliography2} 
\appendix

\section{Stellar relaxation}
\label{app:stellar_relaxation}

In this Appendix we provide additional details of the stellar relaxation procedures adopted in this paper and of the quantification of the hydrostatic stability achieved.

\begin{figure*}
    \centering
    \includegraphics[width=\textwidth]{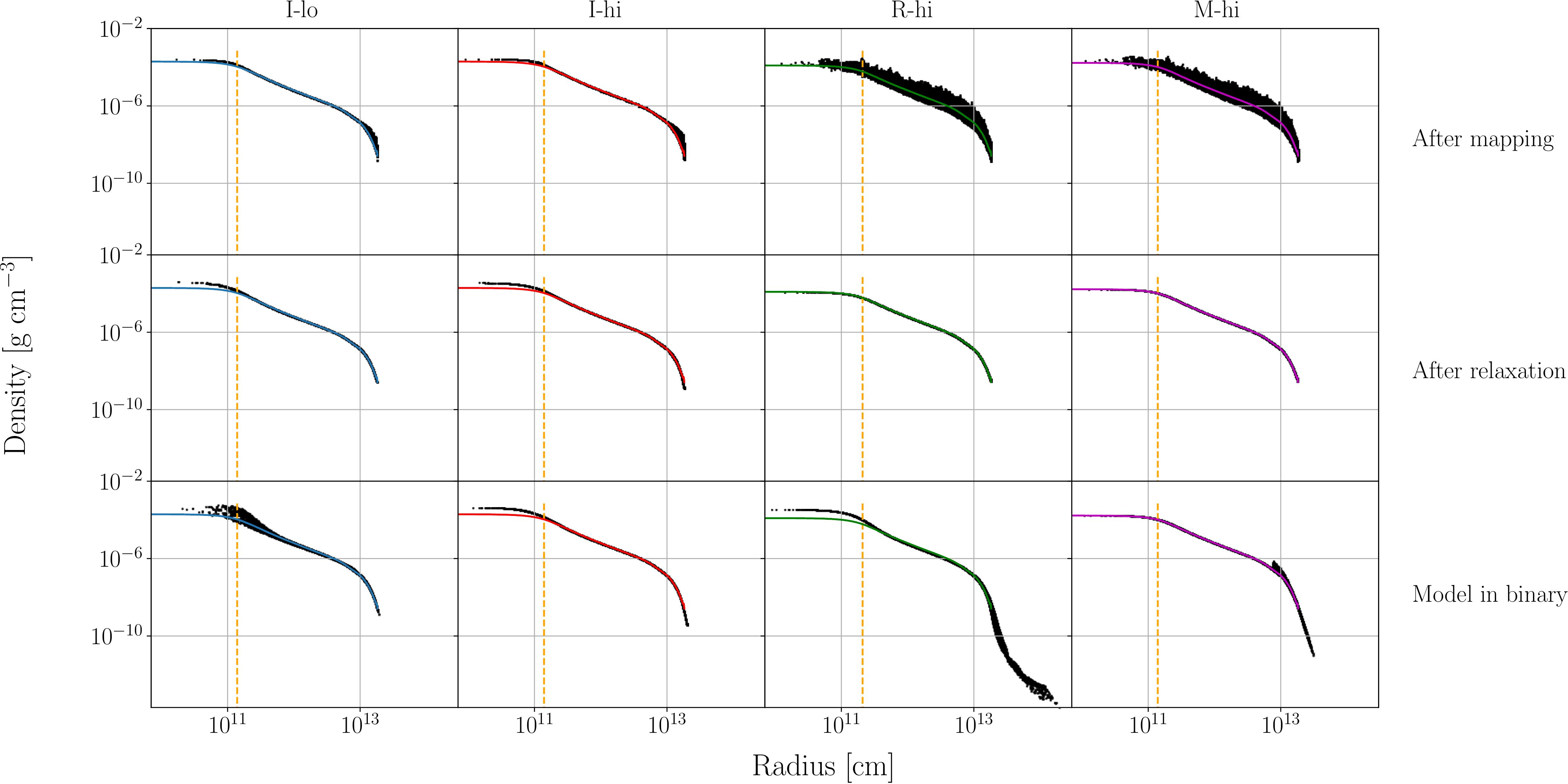}
    \caption{Density profiles of the four stellar models at various phases of setup. The first column (blue) is for I-lo, the second (red) for I-hi, the third (green) for R-hi and the last column (magenta) for M-hi. The coloured profiles are the cored \mesa\ profiles used in the initial mapping. The vertical dashed lines show the softening length, $r_{\rm soft}$, of the core point mass particles. The top row is right after mapping, the middle row is after damping (for I-lo and I-hi) or {\sc relax-o-matic} (for R-hi and M-hi) and the last row is at the end of the 5 dynamical times ($\approx 0.56$ years).}
    \label{fig:rho-R}
\end{figure*}

To assess the fidelity of our stellar structures we plot, in Figure~\ref{fig:rho-R} the density profiles of the four models, taken right after mapping, after the damping (for the I-lo and I-hi models) or after the {\sc relax-o-matic} for the R-hi and M-hi models), time and after evolving with no damping. 

The last part of the relaxation procedure, however, sees some particles moving outward in both the R-hi and M-hi models. Also, both I-hi and R-hi have a slightly higher central density than the cored \mesa\ models.

To gauge the stability of the 3D hydrostatic equilibrium of the stellar structures we checked the distribution of the SPH particle velocities at the time after relaxation, when each profile was used as input to the binary simulation. From there it is clear that only 1~per cent of all particles are at velocities above 7.5~km~s$^{-1}$ for I-lo, R-hi and M-hi, while for I-hi it is 0.50~km~s$^{-1}$. While this is a reasonable number, and the best we can obtain with the current method, we should be cognisant that at 7.5 km~s$^{-1}$ a particle would cross the stellar diameter in 1.5 years, a time that is comparable with the simulation time.

    \label{fig:vel_hist}
%\end{figure*}

\section{Conservation properties of our simulations}
\label{app:conservation_properties}

\begin{figure*}
    \centering
    \includegraphics[width=0.45\linewidth]{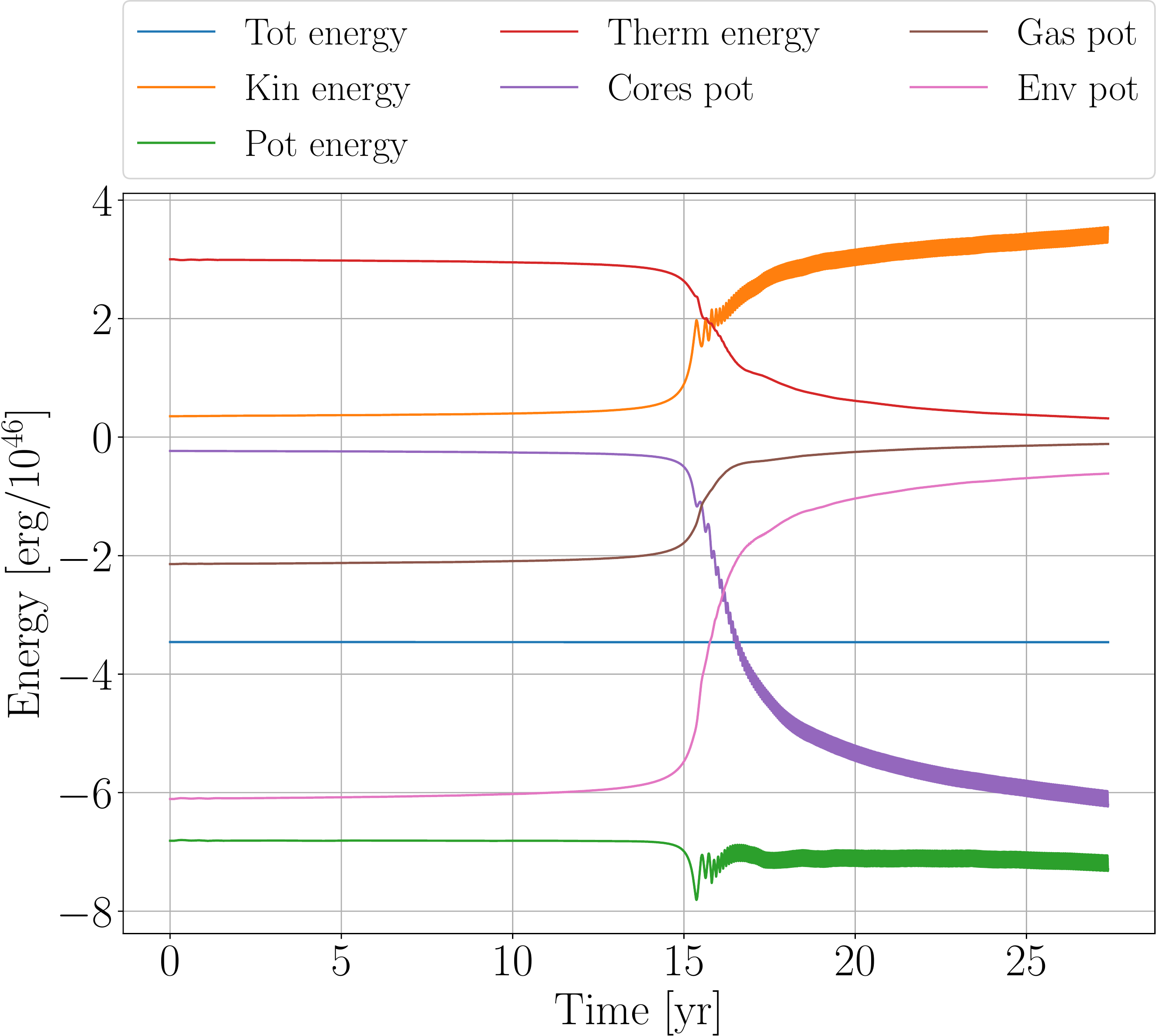}\ \ \ \ \ \ \ \ \ \
    \includegraphics[width=0.45\linewidth]{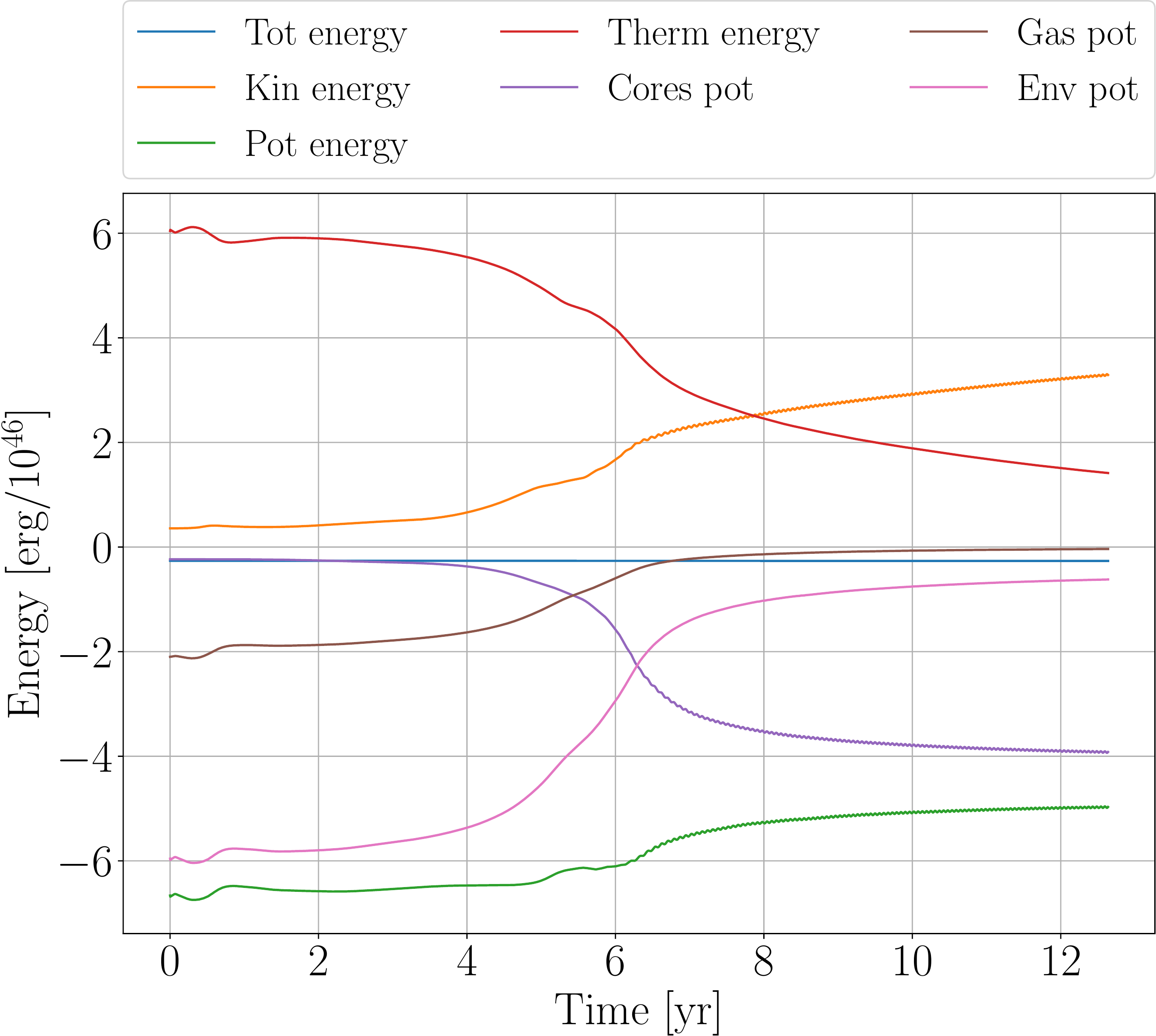}
    \caption{Energy evolution for the I-hi (left) and M-hi (right) simulations. The plots show total energy (Tot energy), kinetic energy (Kin energy), potential energy (Pot energy), thermal energy (Therm energy), orbital energy of the point mass particles (Cores pot), potential energy of the gas without point mass particles (Gas pot) and potential energy of the gas with the point mass particles (Env pot). } \label{fig:energy_conservation}
\end{figure*}

\begin{figure*}
    \centering
    \includegraphics[width=0.48\textwidth]{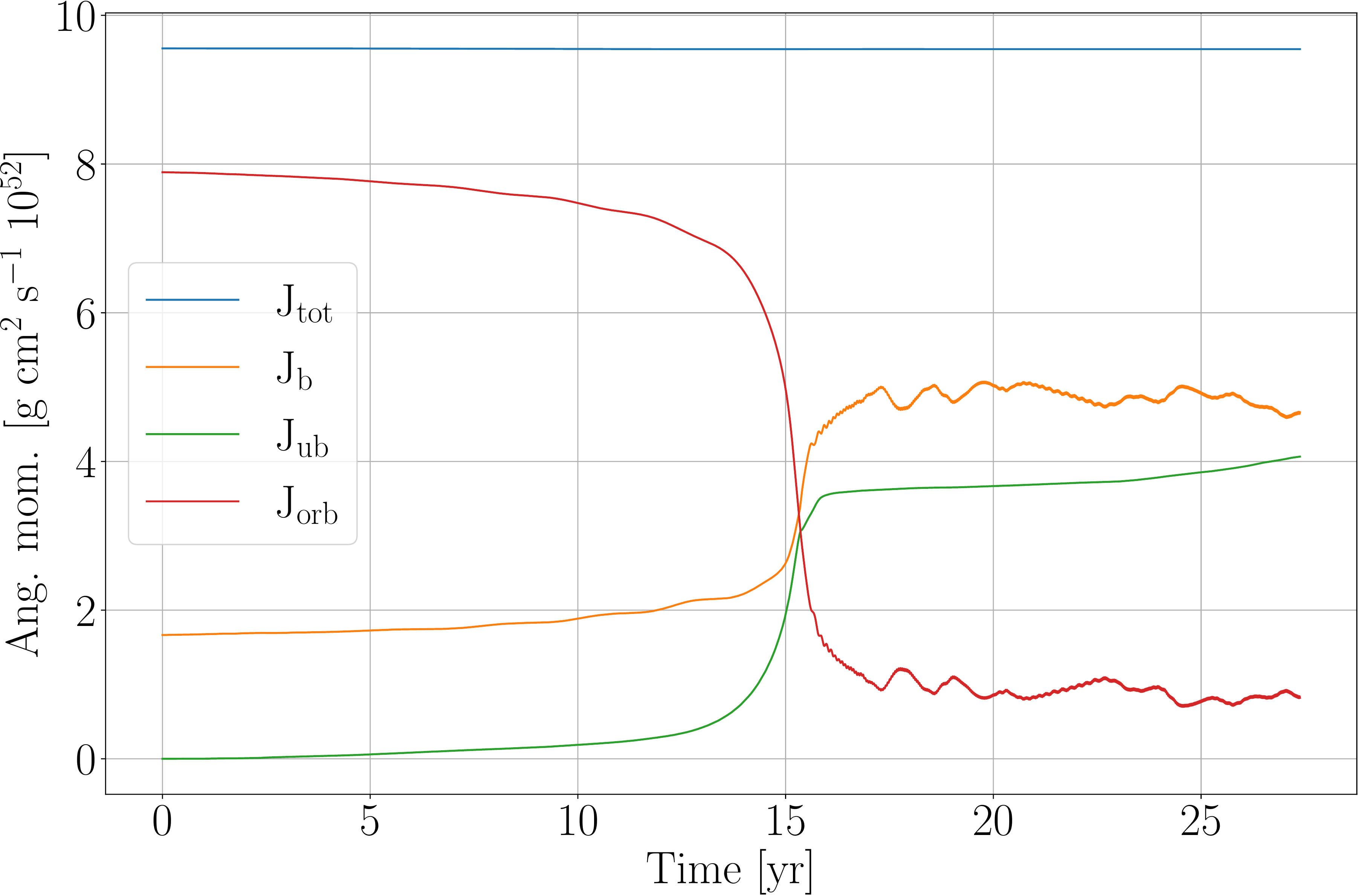}\ \ \ \ \ \
    \includegraphics[width=0.48\textwidth]{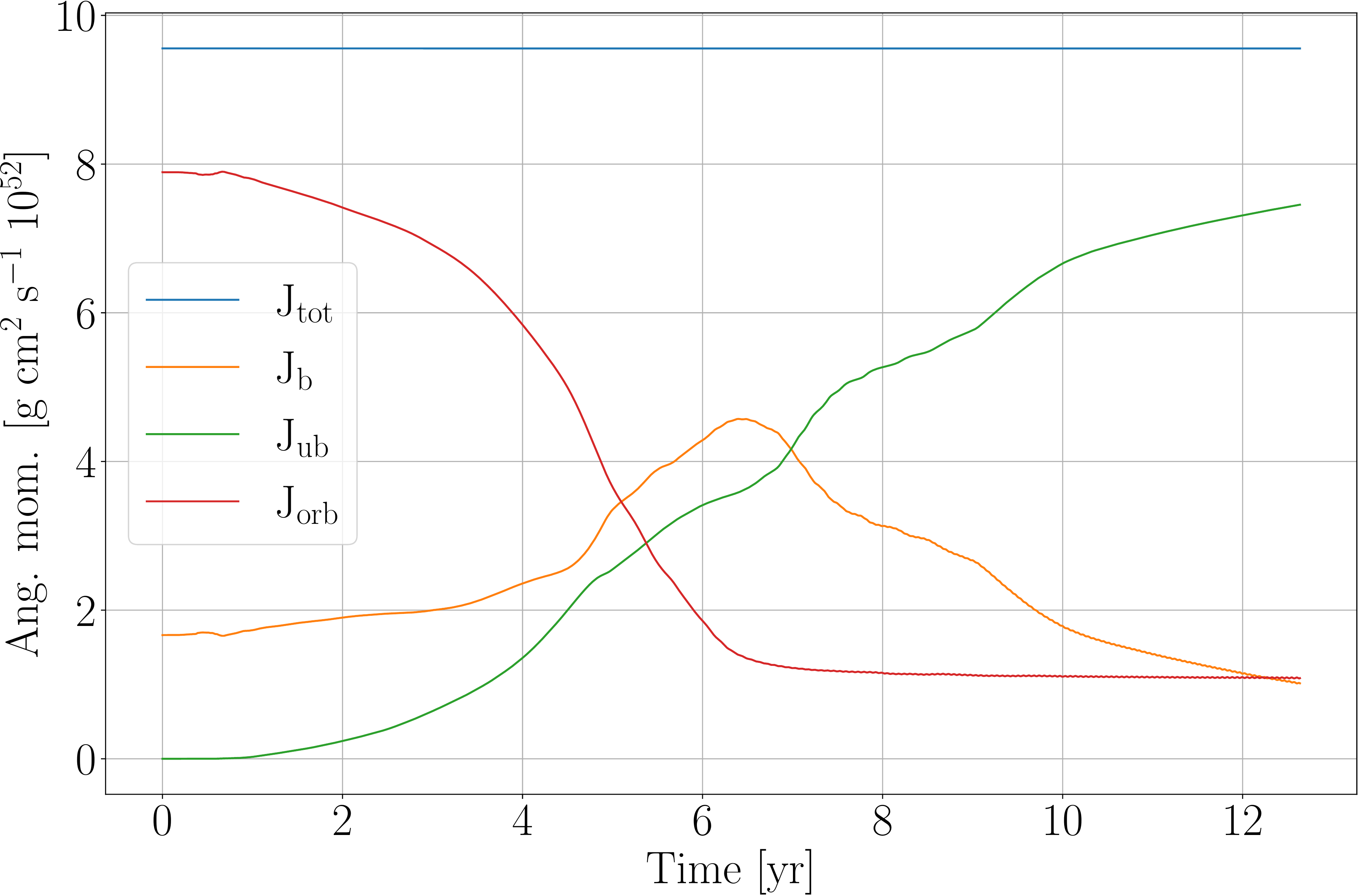}\
    \caption{Evolution of the $z$ component of the angular momentum for the I-hi (left) and M-hi (right). In both cases, the total angular momentum, $J_{\rm tot}$, is conserved. $J_{\rm orb}$, $J_{\rm b}$ and $J_{\rm unb}$ are the orbital, bound and unbound (using mechanic criterion for all curves) angular momenta, respectively.} \label{fig:angular_momentum_conservation}
    \label{fig:conservation_angmom}
\end{figure*}

Figures~\ref{fig:energy_conservation} and \ref{fig:conservation_angmom} show the energy and angular momentum conservation properties of the simulations. The conservation of energy and angular momentum are excellent, with energy conserved to 0.05~per cent, and angular momentum conserved to 0.1~per cent in I-hi and conservation in the other simulations being similarly good. The energy and angular momentum exchanges between the different components of our systems are typical for CE simulations \citep[e.g.,][]{Iaconi2017,Sandquist1998}. The simulation M-hi unbinds almost the entire envelope, justifying the observed differences between the left and right panels of Figures~\ref{fig:energy_conservation} and \ref{fig:conservation_angmom}.

\section{Sensitivity of the in-spiral to initial stellar structure}
\label{app:alternative_Mhi}

To test whether the speed of in-spiral on-set, in-spiral shape and final separation were due to the fact that the initial stellar model in M-hi was somewhat expanding, we calculated an additional model but where the initial stellar model was taken at a later time, 2.5~years after application of the {\sc relax-o-matic} relaxation technique. At this time, the stellar radius had expanded and then contracted, to reach a stable size of 285~\Rsun. 

\begin{figure}
    \centering
    \includegraphics[width=\linewidth]{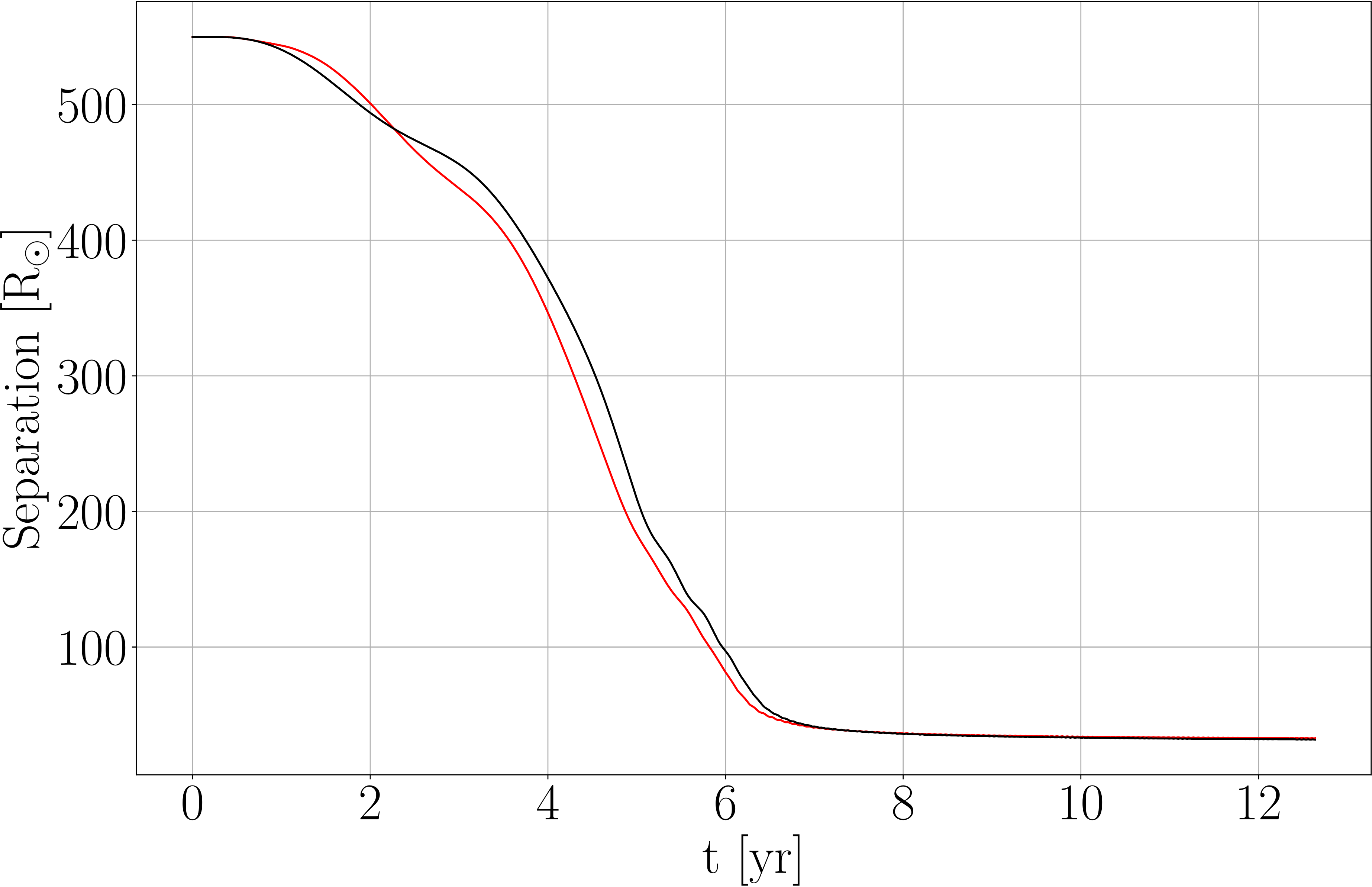}\ \ \
\caption{A comparison of M-hi (red) and an identical simulation started with a stellar model that was evolved in isolation in the \phant\ computational domain for an additional period of time, such that it reached a more stable, albeit larger radius of 285~\Rsun (black).} 

\label{fig:alternative_Mhi}
\end{figure}

In Figure~\ref{fig:alternative_Mhi} we see that the orbital separation over time for M-hi and for the additional simulation are extremely similar. M-hi takes ever so sightly longer to inspiral, but overall the two simulations have a similar in-spiral shape and final separation. This demonstrates that the expansion of the single star used for the M-hi simulation ($R=260$~\Rsun, but expanding to 300~\Rsun in $\sim$1~year; Figure~\ref{fig:PD_rad_fracc})  is not inducing a systematically different behaviour compared to a more stable, albeit slightly larger (285~\Rsun) stellar structure.

\section{Extrapolation of the final separation}
\label{app:extrapolation_finalsep}

\begin{figure}
    \centering
    \includegraphics[width=\linewidth]{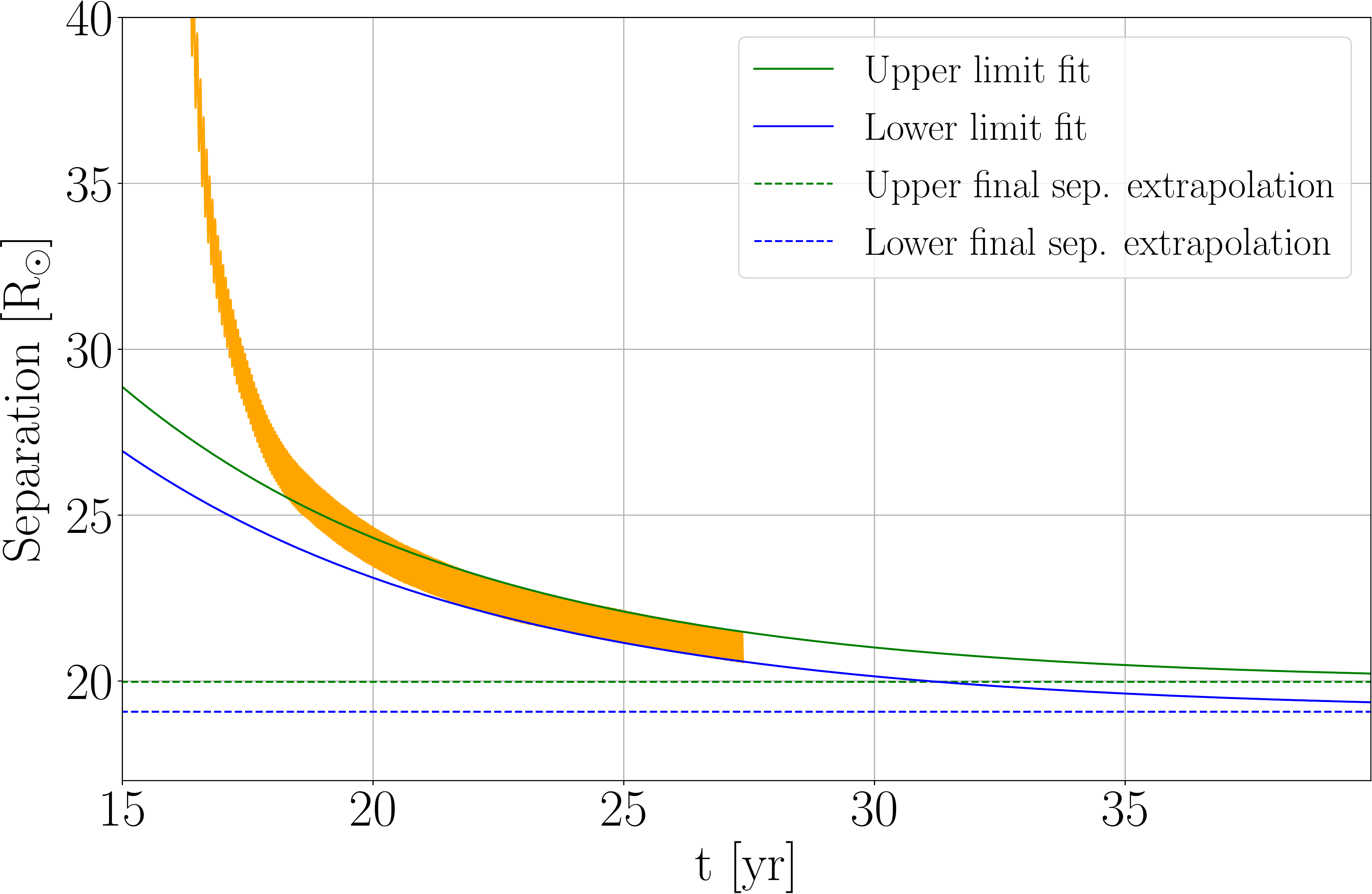}
    \caption{Extrapolation of the orbital separation for simulation I-hi (orange curve). The green and blue solid curves are fits to the apastron and periastron separations, respectively, using the method described in appendix~\ref{app:extrapolation_finalsep}. These curves have two different asymptotic values ($\omega$ in Equation~\ref{eq:fit_function}), indicated by the dashed lines.}
    \label{fig:fit_final_sep}
    
\end{figure}

In order to obtain a value of the final orbital separation that would more closely resemble that which we would obtain by running the simulation further, we extrapolated the apastron and periastron orbital distances observed at the end of the simulations. To do so, we used the \texttt{curve\_fit} function of the {\sc scipy} module in Python 3 \citep{scipy2020} to carry out a non-linear fit. For the fitting curve, we used an exponential function, $s(t)$:

\begin{equation}
    s(t) = \gamma e^{-\kappa t} + \omega,
    \label{eq:fit_function}
\end{equation}

\noindent where $\gamma$, $\kappa$ and $\omega$ are the fitting parameters. Particularly, $\kappa$ and $\omega$ are the decay factor and the extrapolated final separation. We decided to fit perastron and apastron separations separately. For each fit, we select the last $n$ data points, such that $n$ is the number of points that minimizes the element $\sigma_{\kappa,\kappa}$ in the covariance matrix $C(\gamma,\kappa,\omega)$:

\begin{equation}
C(\gamma,\kappa,\omega)=
\begin{bmatrix}
    \sigma_{\gamma,\gamma} & \sigma_{\gamma,\kappa} & \sigma_{\gamma,\omega} \\
    \sigma_{\kappa,\gamma} & \sigma_{\kappa,\kappa} & \sigma_{\kappa,\omega} \\
    \sigma_{\omega,\gamma} & \sigma_{\omega,\kappa} & \sigma_{\omega,\omega} \\
\end{bmatrix},
\end{equation}

\noindent where the diagonal elements are the variance values for each of the three free parameters, compared to the data points. The result of the fitting for simulation I-hi is shown in Figure~\ref{fig:fit_final_sep}. 

% Don't change these lines
\bsp	% typesetting comment
\label{lastpage}
\end{document}